\documentclass[10pt,letterpaper,twocolumn,english,final,floatfix,showpacs,showkeys,citesort,preprintnumbers,prd,nofootinbib,superscriptaddress]{revtex4-1}
\usepackage[T1]{fontenc}
\usepackage[latin1]{inputenc}
\usepackage{color}
\usepackage{babel}
\usepackage{float}
\usepackage{calc}
\usepackage{amsmath}
\usepackage{amssymb}
\usepackage{graphicx}
\usepackage{esint}
\usepackage[unicode=true,pdfusetitle,
 bookmarks=true,bookmarksnumbered=false,bookmarksopen=false,
 breaklinks=false,pdfborder={0 0 1},colorlinks=false]
 {hyperref}

\usepackage{paralist}

\makeatletter

\pdfpageheight\paperheight
\pdfpagewidth\paperwidth

 \@ifundefined{textcolor}{}
 {%
   \definecolor{BLACK}{gray}{0}
   \definecolor{WHITE}{gray}{1}
   \definecolor{RED}{rgb}{1,0,0}
   \definecolor{GREEN}{rgb}{0,1,0}
   \definecolor{BLUE}{rgb}{0,0,1}
   \definecolor{CYAN}{cmyk}{1,0,0,0}
   \definecolor{MAGENTA}{cmyk}{0,1,0,0}
   \definecolor{YELLOW}{cmyk}{0,0,1,0}
 }
\newenvironment{lyxlist}[1]
{\begin{list}{}
{\settowidth{\labelwidth}{#1}
 \setlength{\leftmargin}{\labelwidth}
 \addtolength{\leftmargin}{\labelsep}
 }}
{\end{list}}

\@ifundefined{definecolor}
 {\usepackage{color}}{}
\usepackage{babel}

\newcommand{\msbar}{\overline{\rm MS}}
\newcommand{\FFSb}[1]{\ensuremath{\mathrm{S}^{(#1)}}}
\newcommand{\FFS}[2]{\ensuremath{\mathrm{S}^{(#1,#2)}}}
\newcommand{\mum}[1]{\ensuremath{\mu_M^{(#1)}}}
\newcommand{\mut}[1]{\ensuremath{\mu_S^{(#1)}}}
\newcommand{\ord}{{\cal O}}
\newcommand{\nf}{{\ensuremath{N_F}}}
\newcommand{\nr}{{\ensuremath{N_R}}}
\newcommand{\alphas}{{\alpha_{S}}}

\@ifundefined{showcaptionsetup}{}{%
 \PassOptionsToPackage{caption=false}{subfig}}
\usepackage{subfig}
\makeatother

\begin{document}

\preprint{
\vbox{
\null \vspace{0.3in}
\hbox{SMU-HEP-13-20}
\hbox{LPSC-13-104}
\hbox{LTH 977}
\hbox{LPN13-024}
\hbox{IPPP/13/22}
}
}

\title{\null \vspace{0.5in}
A Hybrid Scheme for Heavy Flavors: Merging the FFNS and VFNS}

\author{A.~Kusina}
\thanks{akusina@smu.edu}
\affiliation{Southern Methodist University, Dallas, TX 75275, USA}

\author{F.~I.~Olness}
\thanks{olness@smu.edu}
\affiliation{Southern Methodist University, Dallas, TX 75275, USA}

\author{I.~Schienbein}
\thanks{schien@lpsc.in2p3.fr}
\affiliation{Laboratoire de Physique Subatomique et de Cosmologie, Universit\'e
Joseph Fourier/CNRS-IN2P3/INPG, \\
 53 Avenue des Martyrs, 38026 Grenoble, France}

\author{T.~Je\v{z}o}
\thanks{T.Jezo@liverpool.ac.uk}
\affiliation{Department of Physics, University of Durham, Durham DH1 3LE, UK\\
Department of Mathematical Sciences, University of Liverpool, Liverpool L69 3BX, UK}

\author{K.~Kova\v{r}\'{\i}k}
\thanks{kovarik@particle.uni-karlsruhe.de}
\affiliation{Institute for Theoretical Physics, Karlsruhe Institute of Technology,
Karlsruhe, D-76128, Germany}

\author{T.~Stavreva}
\thanks{stavreva@lpsc.in2p3.fr}
\affiliation{Laboratoire de Physique Subatomique et de Cosmologie, Universit\'e
Joseph Fourier/CNRS-IN2P3/INPG, \\
 53 Avenue des Martyrs, 38026 Grenoble, France}

\author{J.~Y.~Yu}
\thanks{yu@physics.smu.edu}
\affiliation{Southern Methodist University, Dallas, TX 75275, USA}


\begin{abstract}
We introduce a Hybrid Variable Flavor Number Scheme for heavy flavors,
denoted \hbox{H-VFNS}, which incorporates the advantages of both the traditional
Variable Flavor Number Scheme (VFNS) as well as the Fixed Flavor Number
Scheme (FFNS). By including an explicit $\nf$-dependence in both
the Parton Distribution Functions (PDFs) and the strong coupling constant
$\alpha_{S}$, we generate coexisting sets of PDFs and $\alpha_{S}$
for $\nf=\{3,4,5,6\}$ at any scale $\mu$ that are related analytically
by the $\msbar$ matching conditions. The \hbox{H-VFNS} resums the heavy
quark contributions and provides the freedom to choose the optimal
$\nf$ for each particular data set. Thus, we can fit selected HERA
data in a FFNS framework, while retaining the benefits of the VFNS
to analyze LHC data at high scales. We illustrate how such a fit can
be implemented for the case of both HERA and LHC data. 
\end{abstract}
\maketitle
\tableofcontents{}

\newpage{}

\section{Introduction\label{sec:Introduction}}

Parton distribution functions (PDFs) provide the essential link between
the theoretically calculated partonic cross-sections, and the experimentally
measured physical cross-sections involving hadrons and mesons. 
A good understanding of this link
is crucial if we are to make incisive tests of the standard model,
and search for subtle deviations which might signal new physics.

For precision analyses of PDFs, the heavy quarks (charm, bottom, and
top) must be properly taken into account; this is a non-trivial task
due to the different mass scales which enter the theory. There is
an extensive literature devoted to this question, and various heavy
flavor schemes have been devised which are used in modern 
global analyses of parton distribution functions.
The CTEQ global analyses of PDFs in nucleons \cite{Lai:2010vv,Gao:2013xoa} 
and nuclei \cite{Schienbein:2007fs,Schienbein:2009kk} employ as a default\footnote{In addition to the default scheme, many groups
also provide sets of PDFs obtained in other heavy flavor schemes.} 
the Aivazis-Collins-Olness-Tung (ACOT) scheme \cite{Aivazis:1993kh,Aivazis:1993pi} 
and refinements of it  \cite{Kramer:2000hn,Tung:2001mv}. 
Extensions of the ACOT scheme beyond NLO  \cite{Aivazis:1993pi,Kretzer:1998ju}
were recently presented in Refs.~\cite{Guzzi:2011ew,Stavreva:2012bs}.
The general ACOT scheme has also been applied to the case of 
DIS jet production~\cite{Kotko:2012ui,Kotko:2012kw} and $pp$ induced 
heavy quark production~\cite{Kniehl:2011bk}.
The default scheme of the MSTW PDFs \cite{Martin:2009iq} is the Thorne-Roberts (TR) factorization scheme 
\cite{Thorne:1998ga,Thorne:2006qt} and the NNPDF collaboration uses the FONLL method \cite{Cacciari:1998it} 
applied to deep inelastic scattering (DIS) \cite{Forte:2010ta} 
in its most recent PDF studies \cite{Ball:2011mu,Ball:2012cx}.
The ACOT, TR, and FONLL schemes are examples of (general mass)
variable flavor number schemes (VFNS).
Other groups like ABKM/ABM \cite{Alekhin:2009ni,Alekhin:2012ig}
and GJR/JR \cite{Gluck:2007ck,JimenezDelgado:2008hf} utilize the 
fixed flavor number scheme (FFNS) as their default option,
but include an option for other $N_F$ values~\cite{Alekhin:2009ni}.
The GJR/JR group also performs analyses in
VFNS~\cite{Gluck:2008gs,JimenezDelgado:2009tv}.
For recent reviews of the schemes see, e.g., \cite{Thorne:2008xf,Olness:2008px} and Sec.~22 in \cite{Binoth:2010ra}.

The ACOT scheme is based on the proof of factorization with massive quarks by Collins 
\cite{Collins:1998rz} which incorporates the flexibility of introducing separate matching and
switching scales (see Secs.~\ref{sec:review} and~\ref{sec:hfns}). This possibility has been discussed in the literature 
for some time \cite{Amundson:1998zk,Thorne:2008xf,Olness:2008px,Forte:2010ta}.
However, it is technically more complicated and
has never been implemented in a global analysis framework employing the ACOT scheme.
In this paper we study the VFNS in its most general formulation, with separate matching and switching scales,
and denote it as the Hybrid Variable Flavor Number Scheme (\hbox{H-VFNS}) in order to clearly distinguish it
from the {\em traditional} VFNS.

In the \hbox{H-VFNS} we generate {\em coexisting} sets of PDFs $f_{a}(x,\mu,\nf)$
and the strong coupling constant $\alpha_{s}(\mu,\nf)$ with $\nf=\{3,4,5,6\}$
which are related analytically by the precise $\msbar$ matching conditions.
This provides maximal flexibility, both in a global analysis and for
application of these PDFs, to choose the optimal subscheme (i.e.
the value of $\nf$) in which to compute a given observable. The freedom
of the \hbox{H-VFNS} allows an improved description of heavy flavor data
sets in a wide kinematic range.

The rest of this paper is organized as follows. In Sec.~\ref{sec:review}
we present a brief review of existing heavy flavor schemes before
we introduce our new \hbox{H-VFNS} in Sec.~\ref{sec:hfns}. In Sec.~\ref{sec:pdfs}
we investigate the $\nf$-dependence of the PDFs and $\alpha_{s}$,
followed by a discussion of the $\nf$-dependence of physical structure
functions in Sec.~\ref{sec:Physical-Structure-Functions}. In Sec.~\ref{sec:example}
we present an example of how the \hbox{H-VFNS} scheme could be employed for
a simultaneous study of low-scale data from HERA and high-scale data
from the LHC as they might enter a global analysis of PDFs. Finally,
in Sec.~\ref{sec:conclusions} we present our conclusions. Technical
details concerning the evolution of $\alpha_{s}$ and the PDFs as
well as the matching conditions between sets with different $\nf$
have been relegated to the appendix.

\section{Brief review of heavy flavor schemes\label{sec:review}}

There are several basic requirements that any complete theoretical
description of heavy quarks must satisfy in the context of perturbative
QCD (pQCD) to be valid in the full kinematic range from low to high
energies \cite{Collins:1998rz,Thorne:2008xf}. In particular, we focus
on the following three.
\begin{enumerate}
\item For energy scales $\mu\ll m$, the heavy quark of mass $m$ should
decouple from the theory. 
\item For energy scales $\mu\gg m$, physical observables must be infrared-safe
(IR-safe). 
\item Heavy quark mass effects should be properly taken into account.
\end{enumerate}
We now discuss/review some of the heavy flavor schemes used in the
literature in the light of the three basic requirements.

In the following, we denote a factorization (renormalization) scheme
with $\nf$ ($\nr$) active quark flavors in the initial state (in
quark loops) by $\FFS{\nf}{\nr}$. If not stated otherwise, we set
$\nf=\nr$ and write $\FFSb{\nf}$.

\subsection{Fixed Flavor Number Scheme}

A single scheme $\FFSb{\nf}$ with a fixed number of active quark
partons $\nf$ is called a fixed-flavor number scheme (FFNS). For
example, in the $N_{F}=3$ FFNS, $\FFSb3$, the gluon and the three
light quarks ($u,d,s$) are treated as active partons whereas the
heavy quarks are \emph{not} partons. They can only be produced in
loops and in the final state and their masses are fully retained in
the perturbative fixed order calculations. Similarly, it is possible
to define a $N_{F}=4$ FFNS, $\FFSb4$, and a $N_{F}=5$ FFNS, $\FFSb5$.

The FFNS satisfies the requirements 1 and 3. In particular, the final
state kinematics are exactly taken into account. Conversely, the FFNS
is not IR-safe because logarithms of the heavy quark mass $\alpha_{s}\ln(\mu/m)$
arise in each order of perturbation theory which will become large
for asymptotic energies $\mu\gg m$ so that they eventually spoil
the convergence of the perturbation series in $\alpha_{s}$. Therefore,
the FFNS cannot be reliably extended up to high energy scales such
as those required for analysis of the LHC data.

Despite the lack of IR-safety, the FFNS is widely used because it
is conceptually simple and a proper treatment of the final state kinematics
is crucial close to the heavy quark production threshold and for exclusive
studies of heavy quark production.

\subsection{Variable Flavor Number Scheme}

A variable-flavor number scheme (VFNS) is composed of a set of fixed
flavor number schemes $\FFSb{\nf}$ with different $\nf$ values.
The \textbf{matching scale} $\mu=\mum{\nf}$ specifies the scale at
which the PDFs and $\alpha_{s}$ in the scheme with $\nf+1$ flavors
are related to those with $\nf$ flavors. The matching scales are
of the order of the heavy quark mass $\mum{\nf}\simeq m_{\nf}$ ($\nf=4,5,6$)
in order to avoid large logarithms in the perturbatively calculable
matching conditions. The PDFs, $\alpha_{s}$, and observables are
computed in a sub-scheme $\FFSb{\nf}$, where $\nf=3,4,5,6$ is determined
by the energy scale $\mu$. We write this schematically as: 
\begin{equation}
\FFSb3\overset{\mum4}{\longrightarrow}\FFSb4\overset{\mum5}{\longrightarrow}\FFSb5\overset{\mum6}{\longrightarrow}\FFSb6.
\end{equation}

By construction, a VFNS satisfies heavy quark decoupling (requirement
1) since this is respected by the individual schemes $\FFSb{\nf}$
from which the VFNS is comprised. Furthermore, a VFNS is IR-safe (requirement
2) because it resums the $\alpha_{s}\,\ln(\mu/m_{N_{F}})$ terms to
all orders via the contribution from the heavy quark PDFs; hence,
it can be reliably extended to the region $\mu/m_{c,b}\to\infty$. 

The most delicate point to satisfy is the proper treatment of the heavy quark mass 
(requirement 3)~\cite{Collins:1998rz}:
\begin{itemize}
\item 
In the VFNS, since the UV counter-terms are the same as in the
$\msbar$ scheme, the evolution equations for the PDFs and $\alpha_{s}$
are \emph{exactly} those of a pure massless $\msbar$ scheme with $\nf$
active flavors; 
therefore, the information on the heavy quark masses enters
the PDF evolution only via the matching conditions between two sub-schemes.%
\footnote{For details about the $\msbar$ evolution of the heavy quarks see
Refs.~\cite{Collins:1986mp,Olness:1997yn}.}
\item The VFNS formalism allows all quark masses to be
retained in the calculation of the Wilson coefficients. 
While it is common to neglect the masses
of the lighter quarks for practical purposes, this simplification
is not necessary for the application of the VFNS; hence, the VFNS
fully retains all $\ord(m^{2}/\mu^{2})$ contributions. 
Furthermore,
multiple heavy quark masses can be treated precisely without loss
of accuracy, and this result is independent of whether the heavy quark
masses are large or small; hence, we have no difficulty addressing
contributions of $m_{c}$ and $m_{b}$ simultaneously in the VFNS. 
\item
Of course, the heavy quark masses can also be retained in the calculation
of the final state phase space of a given partonic subprocess. 
Here, the difficulty arises that 'collinear' heavy quarks in the evolution equations
do not appear in the partonic subprocesses and their effect on the phase space
is therefore not taken into account. For example, in DIS, the second heavy quark
produced by a gluon splitting is 'lost' in the leading order $\gamma^\star + c \to c$ subprocess
and theoretical calculations in a VFNS can overshoot the data close to the $c\bar{c}$ production
threshold, i.e., at low $Q^2$ and large $x$.
The problem can be overcome by incorporating the kinematical effect
of the second heavy quark via a slow rescaling variable resulting in the ACOT$_\chi$ 
scheme.\footnote{The details of the  ACOT$_\chi$ factorization scheme are in Ref.~\cite{Tung:2001mv}, and the 
factorization proof for S-ACOT$_\chi$ was demonstrated in Ref.~\cite{Guzzi:2011ew}.}
Subsequently, this procedure has also been adopted by the MSTW group~\cite{Thorne:2006qt}. 
 \item The ACOT$_\chi$ prescription provides a practical solution for the purpose of improving the quality
 of global analyses of PDFs in the VFNS since DIS structure function data ---forming the backbone of such analyses---
 are better described at low $Q^2$. 

However, there are some shortcomings of the $\chi$-prescription:
 \begin{inparaenum}[(i)]
 \item The convolution variable (at LO, the slow-rescaling variable) is not unique and different versions have been 
investigated in the literature
 \cite{Nadolsky:2009ge}.
\item As a matter of principle, production thresholds for more than one heavy quark pair (say 2 or 3 heavy quark pairs) cannot be
captured with a single slow-rescaling variable. Numerically, however, this will have negligible consequences (at NNLO precision).
\item Most importantly, a corresponding prescription has not yet been formulated for the hadroproduction of heavy quarks
\cite{Olness:1997yc,Kniehl:2004fy,Kniehl:2005mk,Kniehl:2008zza}
or other less inclusive observables.
Note, these shortcomings of the $\chi$-prescription 
do not apply to the general ACOT prescription.
 \end{inparaenum}
 \end{itemize}

The problems satisfying requirement 3 can be overcome/reduced by
switching to the $S^{(4)}$ scheme not at the charm mass but at a
larger scale. This possibility will be discussed in the next section.

\section{Hybrid Variable Flavor Number Scheme\label{sec:hfns}}

\begin{figure*}[t]
\centering{}
\includegraphics[clip,width=0.4\textwidth]{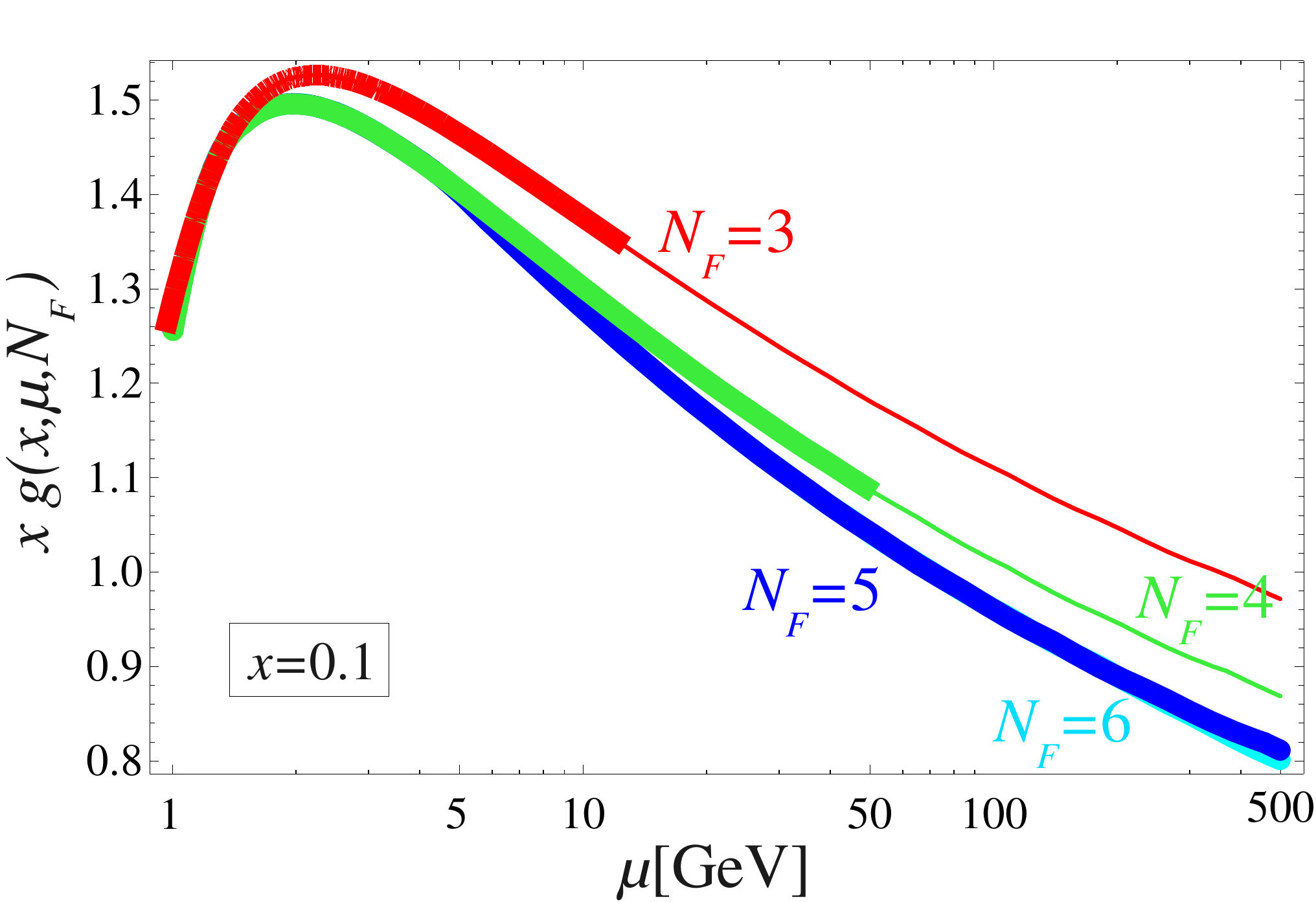}
\quad\quad
\includegraphics[width=0.38\textwidth]{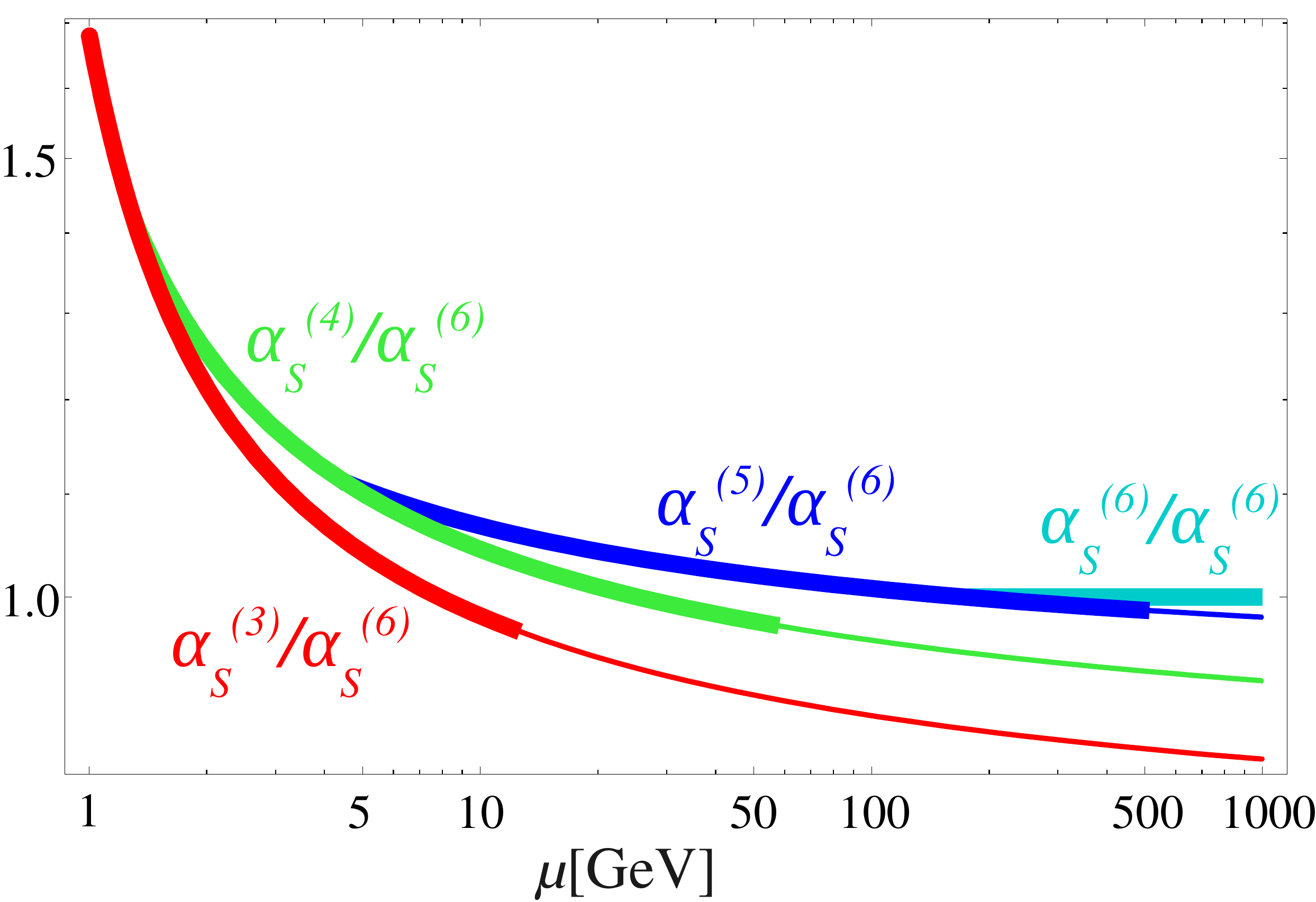}
\caption{Schematic of a \hbox{H-VFNS}:  PDF (left) and $\alpha_{S}$ (right) vs. $\mu$
for a selection of $\nf$ values. The preferred range of each $\nf$
branch is indicated by the thicker line. Thus, $f_{i}(x,\mu,\nf=3)$
can be used slightly above the $m_{c}$ transition, but for
very large $\mu$ scales the $\nf=4,5,6$ branches are preferred as
these resum the $m_{c}$ mass singularities. \label{fig:schematic}}
\end{figure*}

The traditional VFNS introduced in the previous section can be generalized
by introducing, in addition to the matching scales $\mum{\nf}$, separate
\textbf{switching scales} $\mut{\nf}$. The switching scale $\mut{\nf}$
prescribes where the transition from the scheme with $\nf$ flavors
to the one with $\nf+1$ flavors is performed. Below the switching
scale\emph{ }($\mu<\mut{\nf}$) \emph{physical observables} are calculated
in the $\FFSb{\nf}$ scheme , and above the switching scale ($\mut{\nf}<\mu$)
they are calculated in the $\FFSb{\nf+1}$ scheme. Thus, the
\hbox{H-VFNS} is a series of sub-schemes specified by: 
\begin{align}
{\mathrm{S}} & =\begin{cases}
\FFSb3; & \quad\mu\le\mut4\\
\FFSb4; & \quad\mut4<\mu\le\mut5\\
\FFSb5; & \quad\mut5<\mu\le\mut6\\
\FFSb6; & \quad\mut6<\mu
\end{cases}
\end{align}
We refer to this scheme as the hybrid variable flavor number scheme
(\hbox{H-VFNS}) in order to clearly distinguish it from the traditional VFNS
in which the matching and the switching (transition) scales are
equal. Indeed, in all practical applications to date these scales
have been identified with the heavy quark masses: $\mum{\nf}=\mut{\nf}=m_{\nf}$;
while this choice leads to considerable simplifications at the technical
level, it also brings some disadvantages which we will discuss below.%
\footnote{The choice
$\mum{\nf}=m_{\nf}$ eliminates terms of the
form $\ln(\mu/m_{\nf})$ in the matching conditions. Note, that we
generally prefer to choose $\mum{\nf}\leq\mut{\nf}$; technically,
we have the freedom to choose $\mum{\nf}>\mut{\nf}$, but this would
require a numerically unstable DGLAP ``backward-evolution'' from
the matching scale $\mum{\nf}$ down to the switching scale $\mut{\nf}$.%
}
The theoretical basis for the implementation of the presented
\hbox{H-VFNS} follows the general formulation of the ACOT
scheme given (and proven) in Ref.~\cite{Collins:1998rz}.

The essential technical step to implement the \hbox{H-VFNS} is to add an
explicit dependence on the number of active flavors, $\nf$, in both
the PDFs $f_{a}(x,\mu,\nf)$ and the strong coupling $\alpha_{S}(\mu,\nf)$.
This concept is illustrated notationally as: 
\begin{eqnarray*}
f_{i}(x,\mu) & \longrightarrow & f_{i}(x,\mu,\nf)\\
\alpha_{s}(\mu) & \longrightarrow & \alpha_{s}(\mu,\nf),
\end{eqnarray*}
 and we illustrate this schematically in Fig.~\ref{fig:schematic}
where we explicitly see the coexistence of PDFs and $\alpha_{S}$
for different $N_{F}$ values.%
    \footnote{
    The use of the 6-flavor $\alpha_S$ in the ratio in
    Fig.~\ref{fig:schematic} at low scales is just for illustration;
    for realistic calculations in the \hbox{H-VFNS} the 6-flavor $\alpha_S$ is
    used only above the $\mut6$ switching scale. Also the number of
    flavors used in $\alpha_S$ and PDFs is always matched.}

Instead of a single PDF, we will have a set of 4 coexisting PDFs,
$f_{a}(x,\mu,N_{F})$ with $N_{F}=\{3,4,5,6\}$, that are related
analytically by the $\msbar$ matching conditions (see Appendix~\ref{sub:App:pdf}).
Therefore, by knowing the PDFs for a specific $N_{F}$ branch, we
are able to compute the related PDFs for any other number of active
flavors.%
\footnote{This analytic relation is in contrast to, for example, the CTEQ5M
$N_{F}=\{3,4,5\}$ flavor fits~\cite{Lai:1999wy}, where each $N_{F}$
fit represents a separate phenomenological fit to the data set. Separately,
the MSTW $N_{F}=\{3,4,5\}$ flavor fits of Refs.~\cite{Martin:2006qz,Martin:2010db}
are related by $\msbar$ matching conditions.%
} Likewise, we have a set of 4 coexisting strong couplings, $\alpha_{S}(\mu,N_{F})$
for $N_{F}=\{3,4,5,6\}$, that are also related analytically by the
$\msbar$ matching conditions (see Appendix~\ref{sub:App:alphas}).

\subsection*{Generating the PDFs and $\alpha_{s}$ in the \hbox{H-VFNS}\label{sub:Generating-the-PDFs}}

These PDFs and $\alpha_{s}$ are computed using the following prescription. 
\begin{enumerate}
\item Parametrize the PDFs at a low initial scale $\mu_{0}=Q_{0}\sim1$~GeV;
as this is below the $m_{c,b,t}$ thresholds, this would correspond
to $N_{F}=3$. We also choose an initial value for $\alpha_{s}(\mu_{0},N_{F}=3)$
at the same scale.%
    \footnote{In practice, we obtain $\alpha_{s}(\mu_{0},N_{F}=3)$ by evolving
    the world average~\cite{Beringer:1900zz} $\alpha_{s}(M_{Z},N_{F}=5)=0.1184$
    down to $\mu_{0}$ using the renormalization group equation as described
    in Appendix~\ref{sub:App:alphas}.}
\item Starting at an initial scale $\mu_{0}$ with $N_{F}=3$, we evolve
the PDFs using the DGLAP evolution equations and $\alpha_{s}(\mu,N_{F})$
with the renormalization group equations up to $\mu_{max}$. We thus
obtain $f_{a}(x,\mu,N_{F}=3)$ and $\alpha_{s}(\mu,N_{F}=3)$ for
scales $\mu\in[\mu_{0},\mu_{max}]$. 
\item At $\mu=m_{c}$ we use the $\msbar$ matching conditions to compute
both the $N_{F}=4$ PDFs and $\alpha_{s}$ using the $N_{F}=3$ results.
We then use the $N_{F}=4$ evolution equations to obtain $f_{a}(x,\mu,N_{F}=4)$
and $\alpha_{s}(\mu,N_{F}=4)$ up to $\mu_{max}$. 
\item At $\mu=m_{b}$ we again use the $\msbar$ matching conditions to
compute both the $N_{F}=5$ PDFs and $\alpha_{s}$ using the $N_{F}=4$
results. We then use the $N_{F}=5$ evolution equations to obtain
$f_{a}(x,\mu,N_{F}=5)$ and $\alpha_{s}(\mu,N_{F}=5)$ up to $\mu_{max}$. 
\item At $\mu=m_{t}$ this procedure can be repeated again with $N_{F}=6$
for the top quark.%
\footnote{For maximum generality, we include the $N_{F}=6$ case of the top
quark; in practice, even for LHC processes there is little need to
resum these contributions.%
} 
\end{enumerate}
Because all the $N_{F}=\{3,4,5,6\}$ results for the
PDFs $f_{a}(x,\mu,N_{F})$ and the strong coupling $\alpha_{S}(\mu,N_{F})$
are retained,
the user has the freedom to choose which $N_{F}$ to use for a particular
calculation.%
\footnote{Note that there is a residual dependence on the involved matching
and switching scales (which is also present in traditional VFNS).
This is further discussed in Appendix~\ref{sub:App:pdf}.%
}
However, note that the number of active flavors used in $\alpha_{S}$
and in PDFs is always the same.

\subsection*{Properties of the \hbox{H-VFNS}}

Having generated a set of $N_{F}$-dependent PDFs and strong couplings,
we highlight two important properties. 
\begin{enumerate}
\item The PDFs and strong couplings with different $N_{F}$ flavors co-exist
simultaneously. 
\item The PDFs and strong couplings with one $N_{F}$ value have a precise
analytic relation to those with a different $N_{F}$ value which is
specified by the appropriate evolution equations and the $\msbar$
boundary conditions at $\mu=m_{c,b,t}$ (c.f., Appendix~\ref{sec:App:PDF_Alphas}). 
\end{enumerate}
Property 1) allows us to avoid dealing with an $N_{F}$ flavor transition
should it happen to lie right in the middle of a data set. For example,
if we analyze the HERA $F_{2}^{charm}$ data%
\footnote{Consider, for example, the data set of Ref.~\cite{Aktas:2005iw}.%
} which covers a typical range of $Q\sim[3,8]$~GeV, if we were to
use the traditional VFNS then the $N_{F}$ transition between 4 and
5 flavors would lie right in the middle of the analysis region; clearly
this is very inconvenient for the analysis. Because we can specify
the number of active flavors $N_{F}$ in the \hbox{H-VFNS}, we have the option
to \emph{not} activate the $b$-quark in the analysis even when $\mu>m_{b}$;
instead, we perform all our calculations of $F_{2}^{charm}$ using
$N_{F}=4$ flavors. This will avoid any potential discontinuities
in the PDFs and $\alpha_{s}$ in contrast to the traditional VFNS
which forces a transition to $\nf=5$ at the $b$-quark mass.

Property 2) allows us to use the $N_{F}=4$ PDF $f_{i}(x,\mu,N_{F}=4)$
extracted from the $F_{2}^{charm}$ data set and relate this to $N_{F}=5$
and $N_{F}=6$ PDFs that can be applied at high $\mu$ scales for
LHC processes. In this example note that all the HERA $F_{2}^{charm}$
data (both above and below $m_{b}$) influence the $N_{F}=5$ and
$N_{F}=6$ PDFs used for the LHC processes.%
\footnote{Conceptually, the HERA data above the bottom mass ($m_{b}<\mu)$ on
the $\nf=4$ branch is ``backward-evolved'' to the matching point
$\mum{5}=m_{b}$, and then ``forward-evolved'' for $N_{F}=5$ and
$N_{F}=6$. We outline this procedure in more detail in Sec.~\ref{sec:example}.
In particular, we show how such a fit can be performed using only
forward evolution, thus avoiding a (potentially unstable) numerical
backward evolution~\cite{Botje:2010ay}.
}

\subsection*{Challenges Resolved}

We can now see how this \hbox{H-VFNS} overcomes the challenges noted above.
While the traditional VFNS forced the user to transition from $N_{F}=4$
to $N_{F}=5$ at $\mu=m_{b}$ (for example), because the \hbox{H-VFNS} approach
retains the $N_{F}$ information we have the freedom to use the $N_{F}=4$
calculation for $\mu$ scales even above $m_{b}$.

The  \hbox{H-VFNS} also shares the benefits of the FFNS in that we can avoid
a $N_{F}$ transition which might lie in the middle of a data set.
Furthermore, while the FFNS cannot be extended to large scales due
to the uncanceled logs, the  \hbox{H-VFNS} can be used at high scales (such
as for LHC processes) because we retain the freedom to switch $N_{F}$
values and resum the additional logs where they are important.

Additionally, the  \hbox{H-VFNS} implementation gives the user maximum flexibility
in choosing where to switch between the $N_{F}$ and $N_{F}+1$ calculations.
Not only can one choose different switching points for different processes
(as sketched above), but we can make the switching point dependent
on the kinematic variables of the process. For example, the production
thresholds for charm/bottom quarks in DIS are given in terms of the
photon-proton center of mass energy $W^{2}\simeq Q^{2}(1-x)/x$; thus,
we could use this to define our switching scales.

An important operational question is: how far above the $\mu=m_{Q}$
can we reliably extend a particular $N_{F}$ framework. We know this
will have mass singular logs of the form $\alpha_{s}\ln(\mu/m_{Q})$,
so these will eventually spoil the
perturbative expansion of the coefficient functions.
We just
need to ensure that we transition to the $N_{F}+1$ result before
these logs obviate the perturbation theory. We will investigate this
question numerically in Sec.~\ref{sec:Physical-Structure-Functions}.

\subsection*{Relation to previous work}

In closing we want to note that many of the ideas
that we build upon here with the H-VFNS have been present in the literature
for some time. The proof of factorization paper by Collins~\cite{Collins:1998rz}
incorporates the flexibility of introducing separate matching and
switching scales, and applications to the ACOT scheme were outlined
in Ref.~\cite{Amundson:1998zk}, and Ref.~\cite{Thorne:2008xf}
provides a recent review of the situation. 
The separate $N_{F}$ sets
of the MSTW collaboration~\cite{Martin:2006qz} are precisely
defined by the $\msbar$ matching conditions~\cite{Buza:1996wv}
at ${\cal O}(\alpha_s)$.  
This is extended to higher order for MSTW~\cite{Martin:2010db}
and ABKM/ABM~\cite{Alekhin:2009ni}.
Additionally, the NNPDF group provides PDF sets with different numbers 
of active flavors in Refs.~\cite{Ball:2011mu,Ball:2012cx} for NLO and NNLO. 
The phenomenological implications of coexisting  $N_{F}$
PDF sets has been investigated in the MSTW and NNPDF
frameworks~\cite{Thorne:2012az,Ball:2013gsa}.
The extension of the ACOT scheme beyond NLO, where the PDF and $\alpha_{S}$
discontinuities appear, was presented in Ref.~\cite{Stavreva:2012bs}.
Putting these pieces together, and including the explicit $N_{F}$
dependence, allows us to construct a tractable implementation of the
H-VFNS with user-defined switching scales.

Operationally, we are able to provide maximum flexibility
with only a minimal extension of the PDF. A fully general framework
as described in Ref.~\cite{Amundson:1998zk} would require a separate
PDF grid (and associated evolution) for each data set with a distinct
matching or switching scale. With the implementation outlined in the
H-VFNS we are able to implement this economically with only three
PDF grids for $N_{F}=\{3,4,5\}$; this is possible for a number of
reasons as outlined below.

While we have imposed the choice $\mum{\nf}=m_{\nf}$,
we demonstrate in Appendix~\ref{sub:App:pdf} that when the matching
conditions are implemented correctly, particularly at higher orders,
the physical influence of this matching condition is minimal. On physical
grounds, the natural choice for the switching scale is at or above
the heavy quark mass scale $\mut{\nf}\ge m_{\nf}$. Additionally,
it is generally preferred to have the switching scale above the matching
scale $\mum{\nf}\le\mut{\nf}$ as this avoids the need for backward
evolution. Our implementation of the H-VFNS with $m_{\nf}=\mum{\nf}\le\mut{\nf}$
naturally accommodates these choices.

Therefore, our H-VFNS implementation economically
requires only three PDF grids (for $N_{F}=\{3,4,5\}$), yet provides
the user flexibility to use any switching scale,
and the choice of the fixed matching scale $\mum{\nf}=m_{\nf}$
has minimal impact on the physical results.

\section{$\nf$ Dependence of the PDFs and $\alpha_{s}$\label{sec:pdfs}}

In this study we are using an initial PDF parameterization based on
the {\tt nCTEQ} ``decut3'' set of Ref.~\cite{Kovarik:2010uv}. We use quark masses
of $m_{c}=1.3$~GeV, $m_{b}=4.5$~GeV, and $m_{t}=175$~GeV, with
a starting scale of $Q_{0}=1.2$ GeV which allows us to examine the
charm threshold. The full set of $N_{F}=\{3,4,5,6\}$ PDFs is generated
as described above using the $\msbar$ matching conditions applied
at the quark mass values.%
\footnote{These $N_{F}$-dependent PDFs are available on the nCTEQ web-page
at HEPForge.org.%
} The details of the matching are described in Appendix~\ref{sec:App:PDF_Alphas}.

\subsection{$\nf$ Dependence of the PDFs\label{sub:nf_PDF}}

We begin by illustrating the effect of the number of active flavors
$N_{F}$ on the PDFs, $f_{i}(x,\mu,N_{F})$. One of the simplest quantities
to examine is the momentum fraction $\left[\int_{0}^{1}x\, f_{i}(x)\, dx\right]$
carried by the PDF flavors as a function of the $\mu$-scale.

\begin{figure}[t]
\begin{centering}
\includegraphics[width=0.4\textwidth]{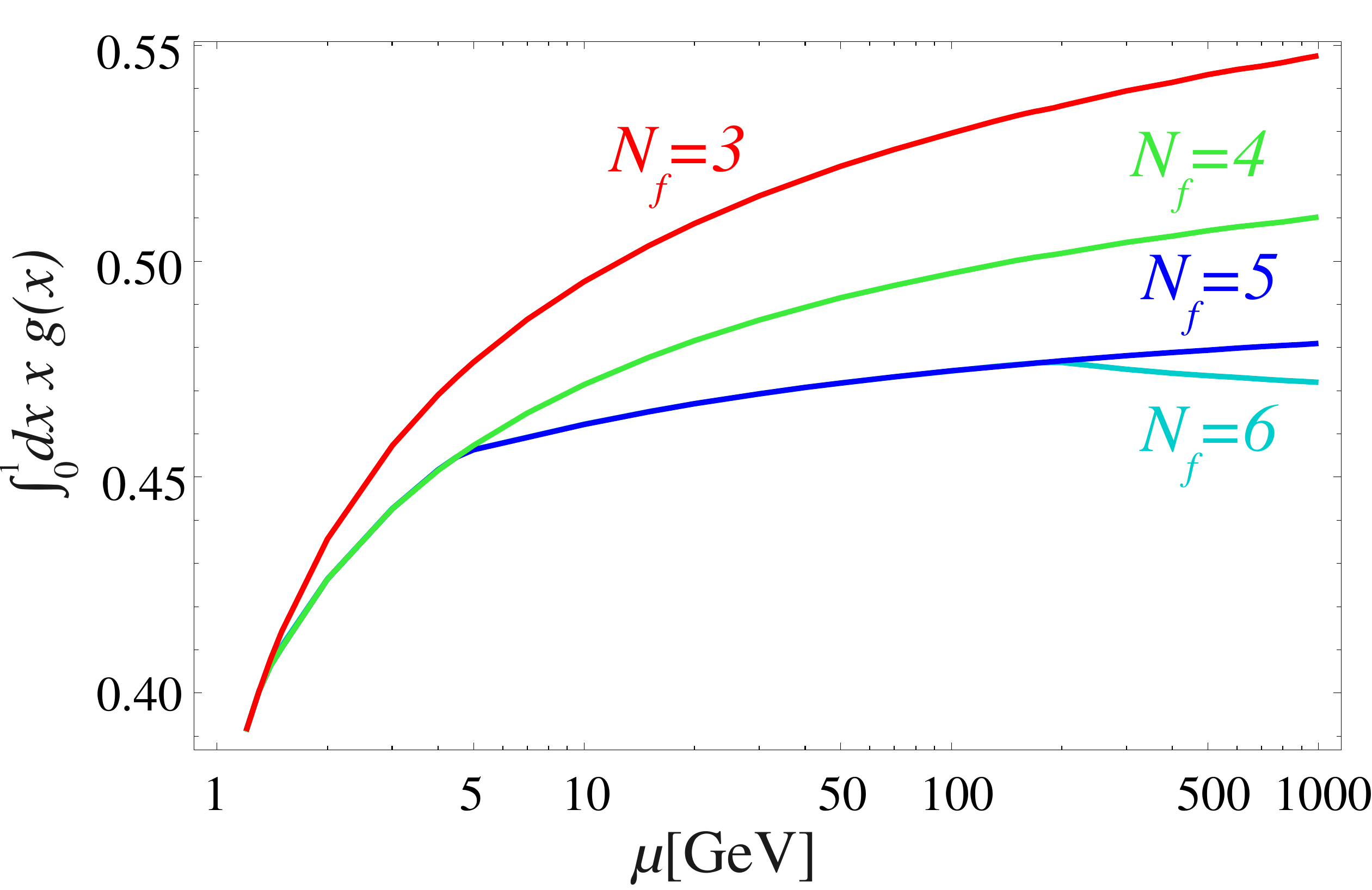}\quad{}\includegraphics[width=0.41\textwidth]{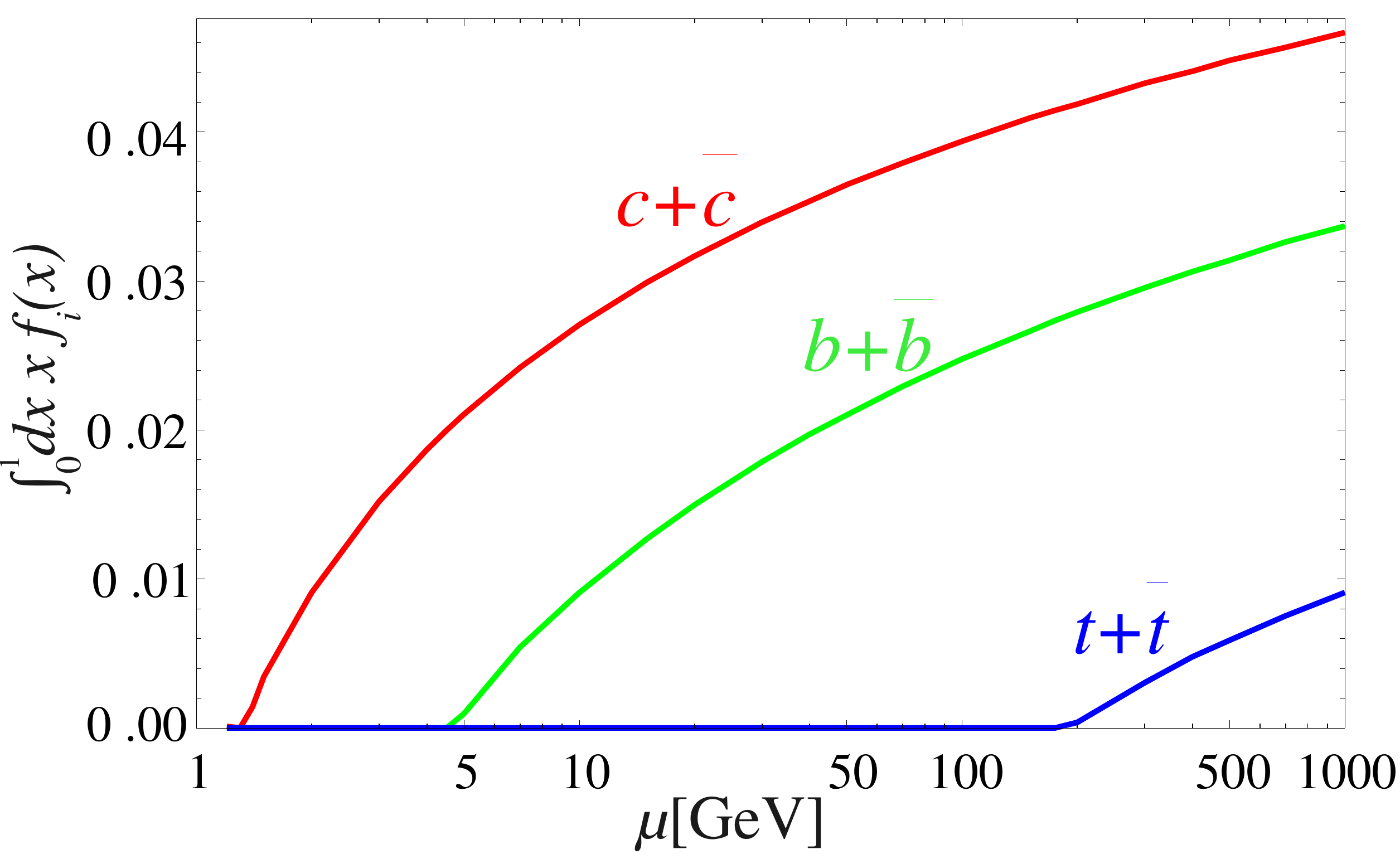} 
\par\end{centering}

\caption{(a) Gluon momentum fraction; (b) Momentum fraction for $c+\bar{{c}}$,
$b+\bar{{b}}$ and $t+\bar{{t}}$ quarks. The results have been obtained using
NLO PDFs ($\msbar$) with a 2-loop $\alpha_{S}$. \label{fig:momFrac} }
\end{figure}

Fig.~\ref{fig:momFrac} shows the gluon and heavy quark momentum
fractions as a function of the $\mu$ scale. For very low $\mu$ scales
all the curves coincide by construction; when $\mu<m_{c,b,t}$ the
charm, bottom, and top degrees of freedom will ``deactivate'' and
the $N_{F}=4,5,6$ results will reduce to the $N_{F}=3$ result.

As we increase the $\mu$ scale, we open up new channels. For example,
when $\mu>m_{c}$ the charm channel activates and the DGLAP evolution
will generate a charm PDF via the $g\to c\bar{c}$ process. Because
the overall momentum sum rule must be satisfied $\left[\sum_{i}\int_{0}^{1}x\, f_{i}(x)\, dx=1\right]$,
as we increase the momentum carried by the charm quarks, we must decrease
the momentum carried by the other partons. This interplay is evident
in Fig.~\ref{fig:momFrac}. In Fig.~\ref{fig:momFrac}-a, we see
that for $\mu=1000$~GeV, the momentum fraction of the $N_{F}=4$
gluon is decreased by $\sim4\%$ as compared to the $N_{F}=3$ gluon.
Correspondingly, in Fig.~\ref{fig:momFrac}-b we see that at $\mu=1000$~GeV,
the momentum fraction of the charm PDF is $\sim4\%$. Thus, when we
activate the charm in the DGLAP evolution, this depletes the gluon
and populates the charm PDF via $g\to c\bar{c}$ process.

In a similar manner, comparing the momentum fraction of the $N_{F}=5$
gluon to the $N_{F}=4$ gluon at $\mu=1000$~GeV we see the former
is decreased by $\sim3\%$; in Fig.~\ref{fig:momFrac}-b we see that
at $\mu=1000$~GeV the momentum fraction of the bottom PDF is $\sim3\%$.

\begin{figure}[h]
\begin{centering}
\includegraphics[width=0.41\textwidth]{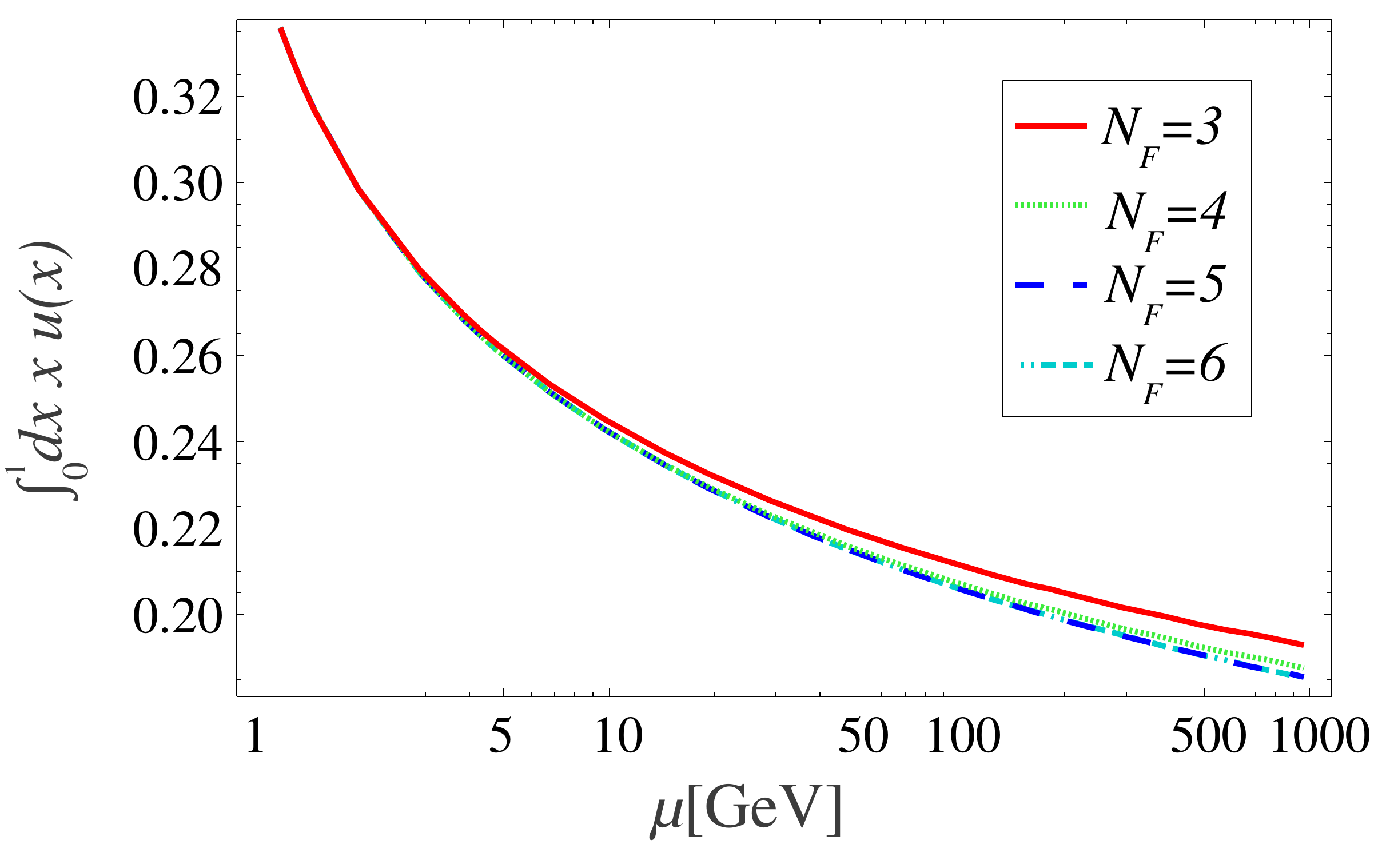}\quad{}\includegraphics[width=0.41\textwidth]{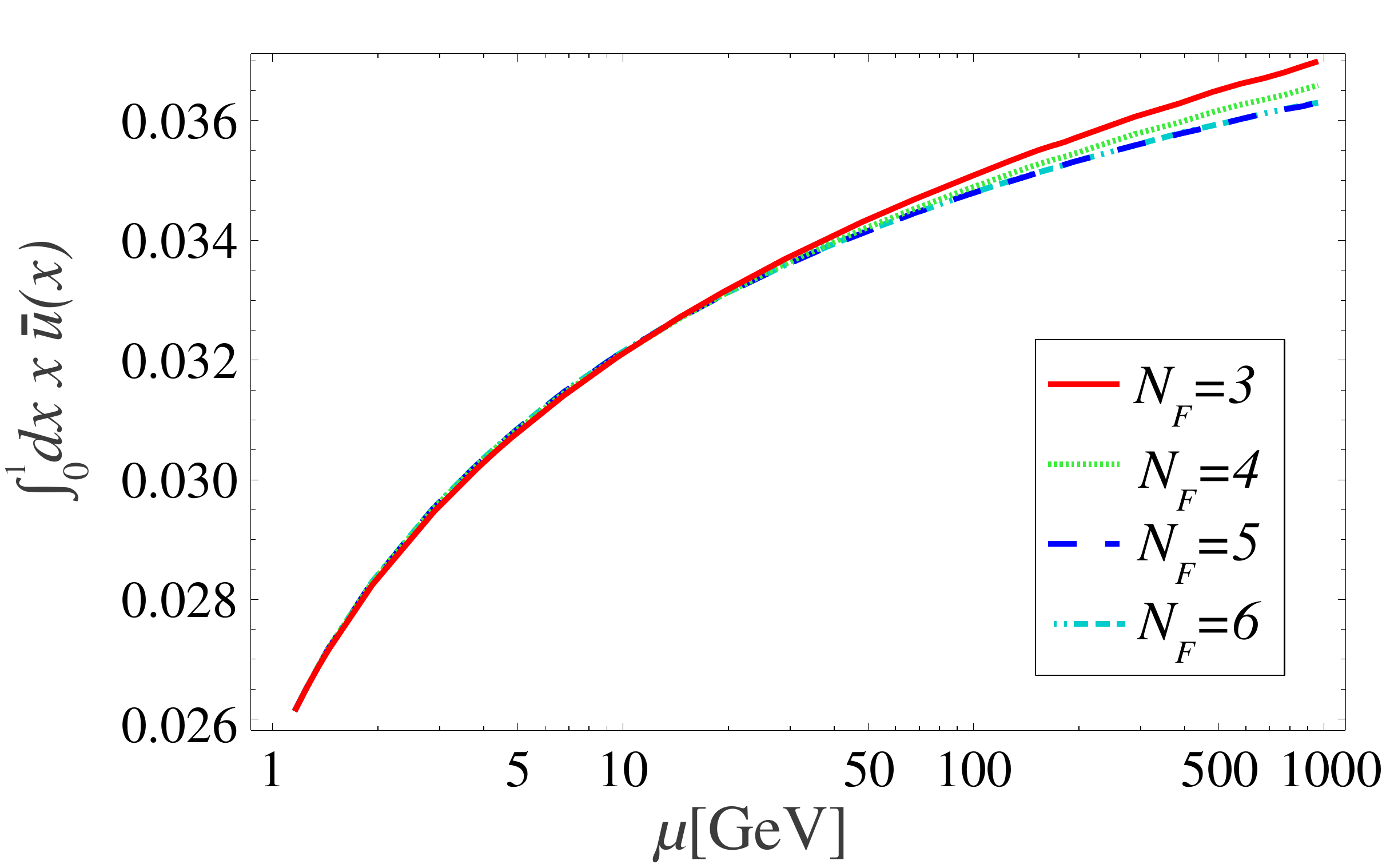} 
\par\end{centering}
\caption{Momentum fraction carried by the (a) $u$-quark and
(b) $\bar{u}$-quark in the 3, 4, 5, and 6 flavor schemes.
\label{fig:u-momFrac}}
\end{figure}

The gluon PDF is primarily affected by the heavy $N_{F}$ channels
as it couples via the $g\to c\bar{c},b\bar{b},t\bar{t}$ processes.
The effect on the light quarks $\{u,d,s\}$ is minimal as these only
couple to the heavy quarks via higher order processes ($u\bar{u}\to g\to c\bar{c}$).
This property is illustrated in Fig.~\ref{fig:u-momFrac} where we
display the $u$ and $\bar{u}$ quark momentum for different $N_{F}$ values.
While the $N_{F}$ variation yields a $\sim8\%$ momentum fraction
shift for the gluon, the total shift of the $u$ quark is only $\sim1\%$
of the momentum fraction.%
\footnote{For example, in Fig.~\ref{fig:u-momFrac}-a, we see the momentum
fraction change from $\sim20\%$ for $N_{F}=3$ to $\sim19\%$ for
$N_{F}=6$. %
}

\subsection{$\nf$ Dependence of $\alpha_{s}$}

The PDFs are only one piece of the full calculation; another essential
ingredient is the strong coupling constant $\alpha_{S}(\mu,N_{F})$.
The running coupling is sensitive to higher-order processes involving
virtual quark loops; hence, it depends on the number of active quarks,
and we make this dependence explicit with the $\alpha_{S}(\mu,N_{F})$
notation. More precisely, the strong coupling depends on the renormalization
scale $\mu_{R}$, in contrast to the factorization scale $\mu_{F}$.
However, for this work we have set $\mu_{R}=\mu_{F}=\mu$.

\begin{figure}
\begin{centering}
\includegraphics[width=0.35\textwidth]{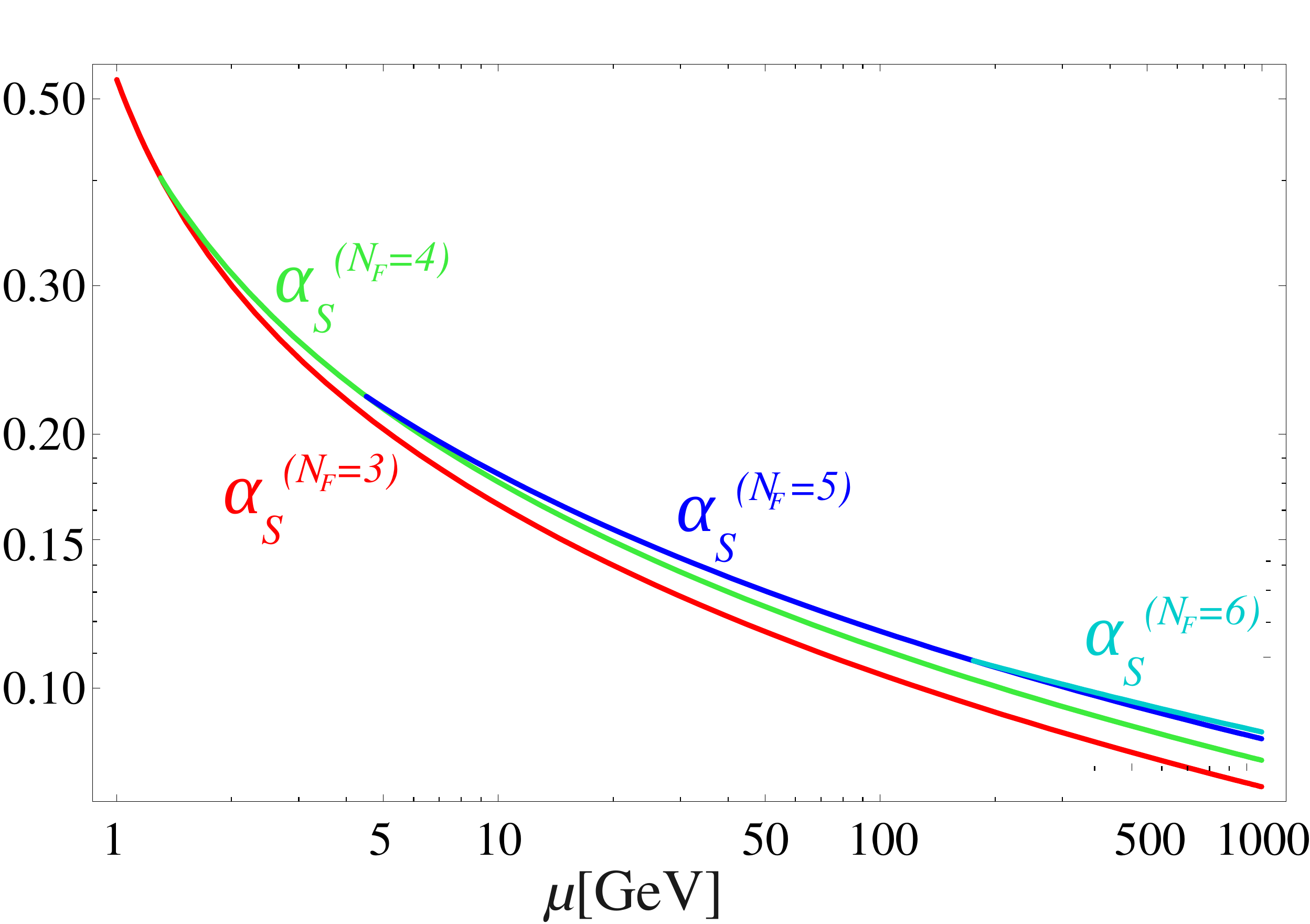}\quad{}\quad{}\quad{}\quad{}\includegraphics[width=0.35\textwidth]{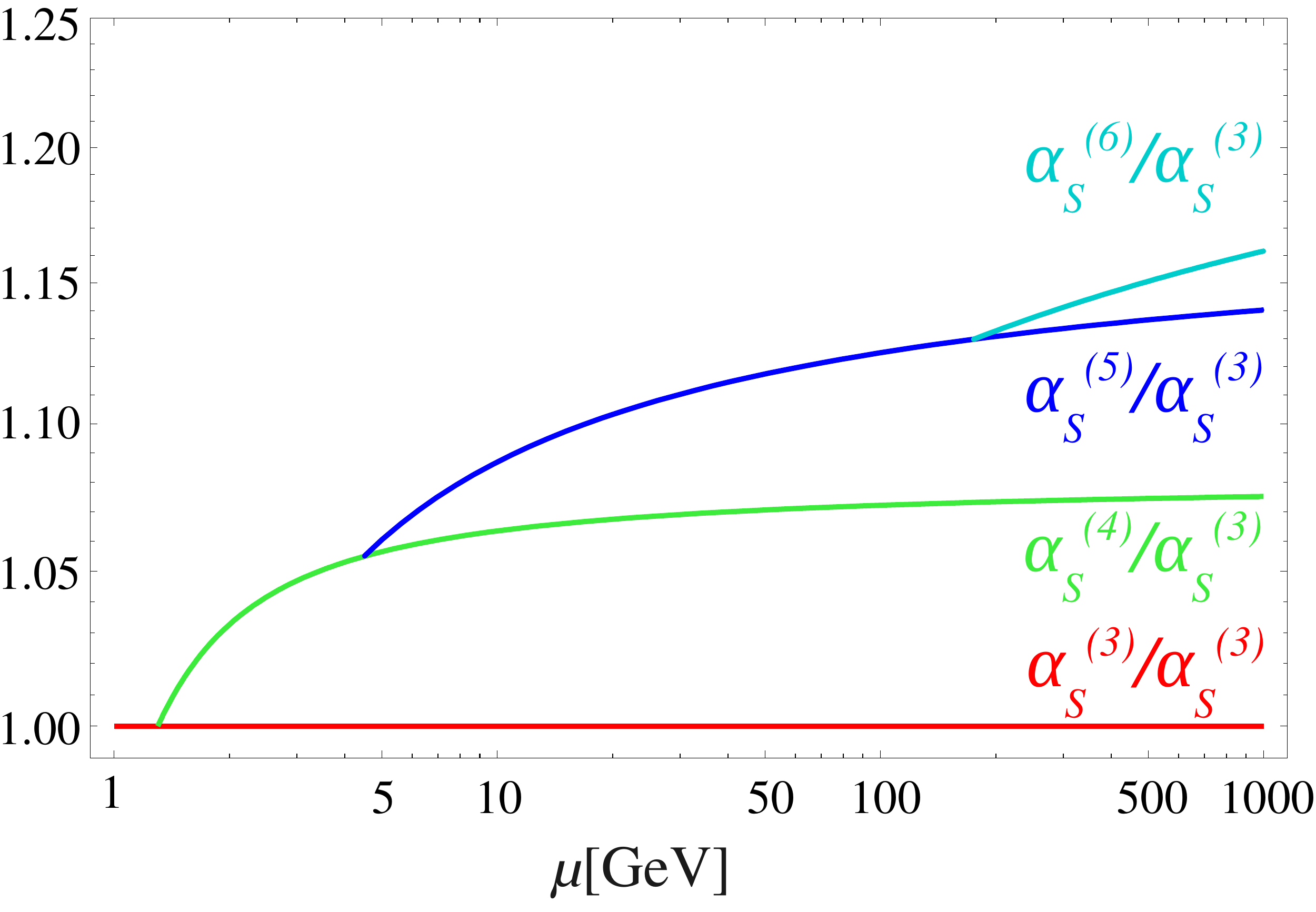} 
\par\end{centering}

\caption{(a) 2-loop $\alpha_{S}$ for different number of flavors;
(b) ratio of 3, 4, 5, and 6-flavor $\alpha_{S}$ to the 3-flavor one.
\label{fig:alphaS} }
\end{figure}

In Fig.~\ref{fig:alphaS} we display $\alpha_{S}(\mu,N_{F})$ vs.
$\mu$ for different $N_{F}$ values. We choose an initial $\alpha_{S}(\mu,N_{F})$
at a low $\mu=Q_{0}$ and $N_{F}=3,$ and evolve this to larger scales
using the NLO beta function. (See Appendix~\ref{sub:App:alphas}
for details.) As we saw in Fig.~\ref{fig:momFrac}, the $N_{F}$
transitions are evident.

There are strong constraints on $\alpha_{S}(\mu,N_{F})$ at low scales
($\mu\sim m_{\tau}$) from hadronic $\tau$ decays, and at high scales
($\mu\sim M_{Z}$) from LEP2 measurements~\cite{Beringer:1900zz};
thus, it is not trivial to satisfy both limits for a fixed value of $N_{F}$.

\subsection{Interplay between $\alpha_{S}(\mu,N_{F})$ and $g(x,\mu,N_{F})$}

If we could do an all-orders calculation for any physical observable,
this would be independent of $N_{F}$ and $\mu$; for finite-order
calculations, any residual $\mu$ and $N_{F}$ dependence is simply
an artifact of our truncated perturbation theory. Thus, the separate
contributions of the perturbative QCD result must conspire to compensate
the $\mu$ and $N_{F}$ dependence to the order of the calculation.

For example, when we activate the charm PDF, we find the gluon PDF
is decreased. Within the limits of the perturbation theory, we would
expect that the decreased contribution from the gluon initiated processes
would be (at least partially) compensated by the new charm initiated
processes. This compensation mechanism is clearly evident for the
calculation of $F_{2}^{charm}$; additionally, we find that because
the gluon initiated and charm initiated contribution generally have
opposite renormalization scale dependence, the resulting VFNS prediction
is more stable in $\mu$ as compared to the FFNS result~\cite{Aivazis:1993pi}.

Another compensating mechanism is evident when comparing Fig.~\ref{fig:momFrac}
and Fig.~\ref{fig:alphaS} where we note that the $N_{F}$ dependence
of $\alpha_{s}$ is generally opposite to that of the gluon PDF; this
observation is particularly interesting as many NLO contributions
are proportional to the combination $\alpha_{S}\times x\, g$. If
we consider the inclusive structure functions $F_{123L}$, for example,
the LO contributions are proportional to the electroweak couplings
and the quark PDFs -- both of which are relatively invariant under
changes in $N_{F}$. Thus, the primary effect of the $N_{F}$ dependence
will be to modify the NLO contributions which are dominantly proportional
to $\sim\alpha_{S}\times x\, g$. For these contributions, the $x\, g$
and $\alpha_{S}$ dependence will partially cancel each other out
so that the total result is relatively stable as a function of $N_{F}$~\cite{Aivazis:1993pi,Olness:1997yc}.

\begin{figure*}[b]
\subfloat[Different combinations of 3- and 5-flavor $\alphas^{(N_{R})}xg^{(N_{F})}(x)$
as a function of $x$, for different values of $\mu$: $5$~(left),
$10$~(middle) and $100$~(right) GeV. \label{fig:ASxg-N_x}]{\centering{}\includegraphics[width=0.32\textwidth]{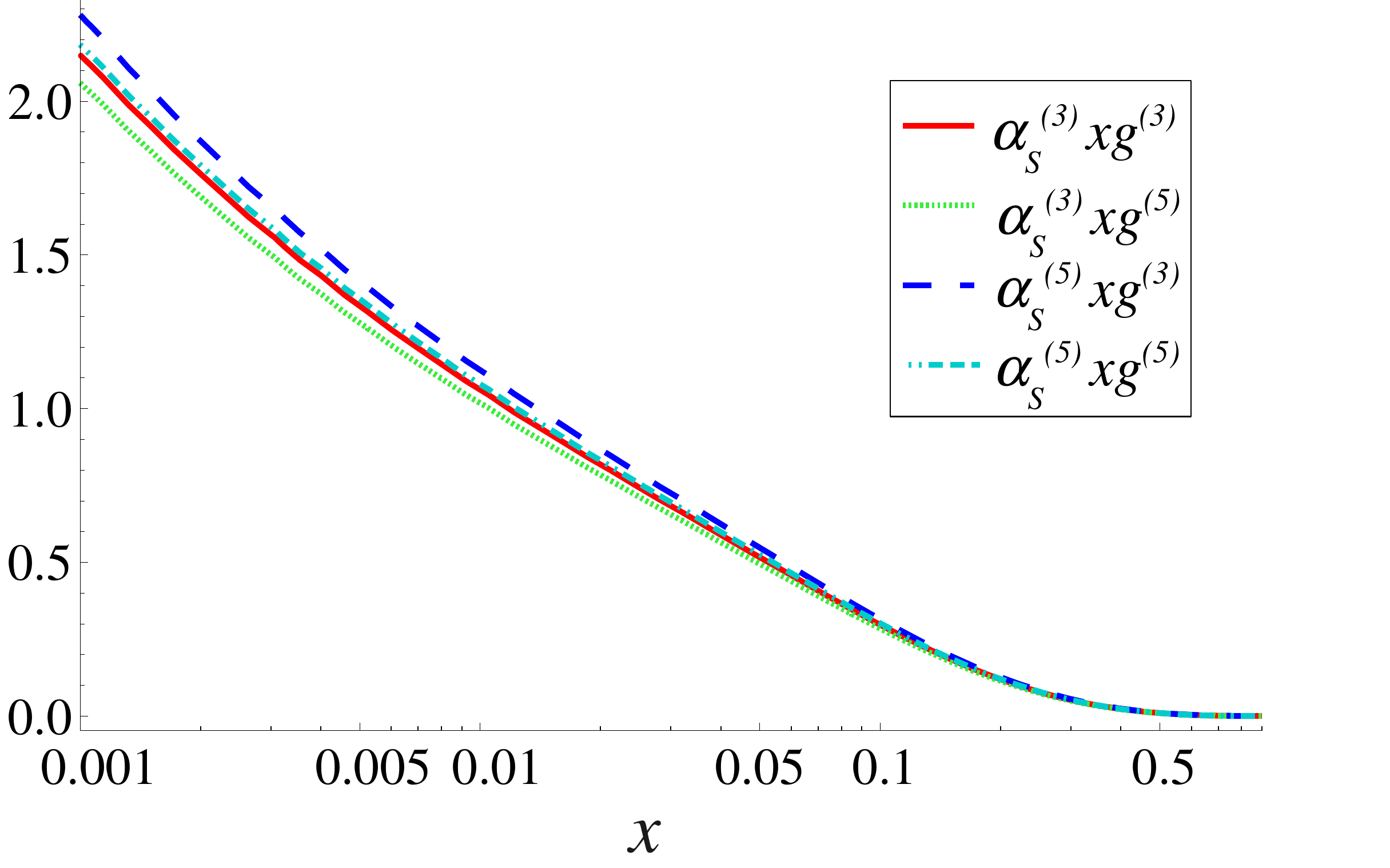}\quad{}\includegraphics[width=0.32\textwidth]{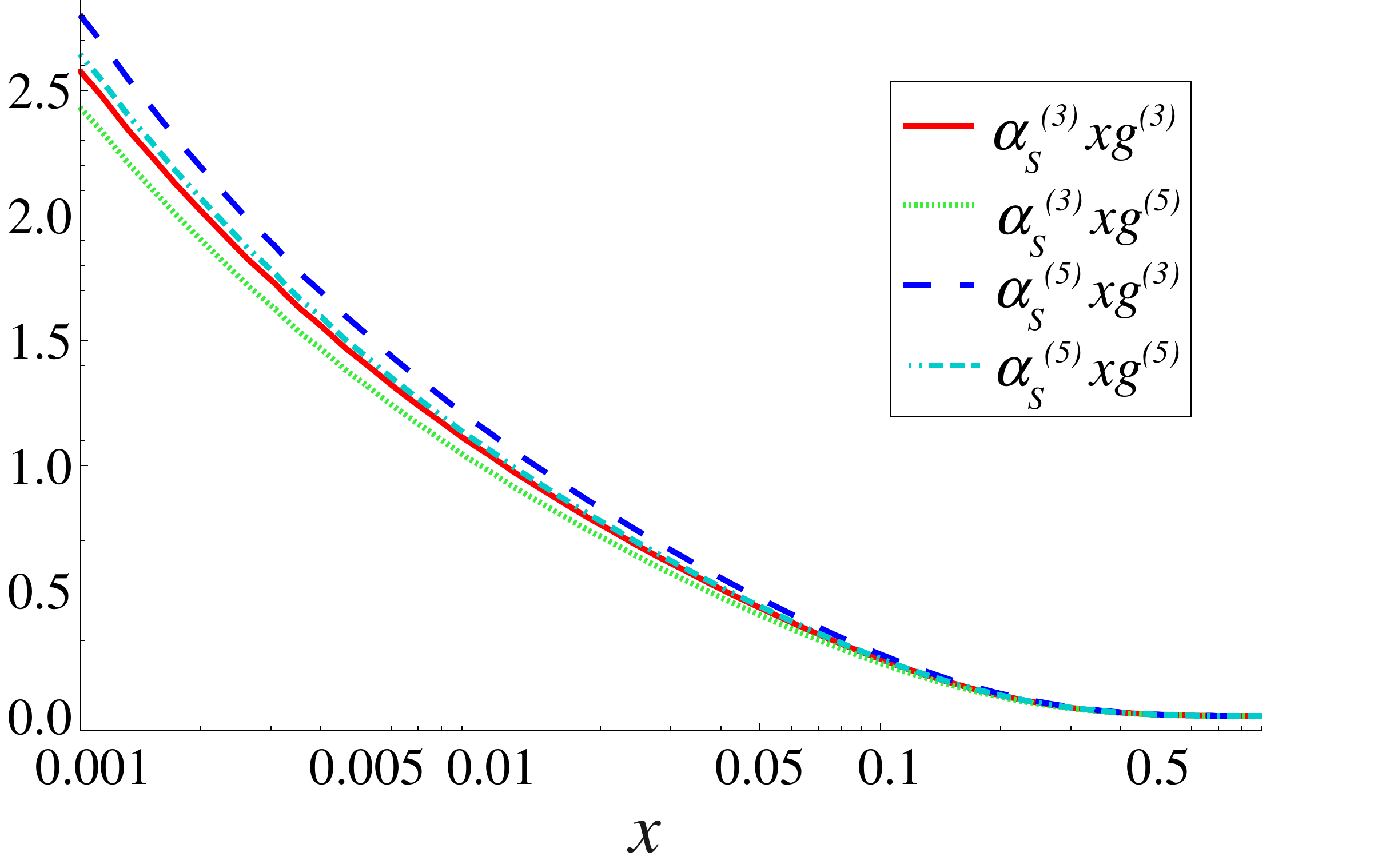}\quad{}\includegraphics[width=0.32\textwidth]{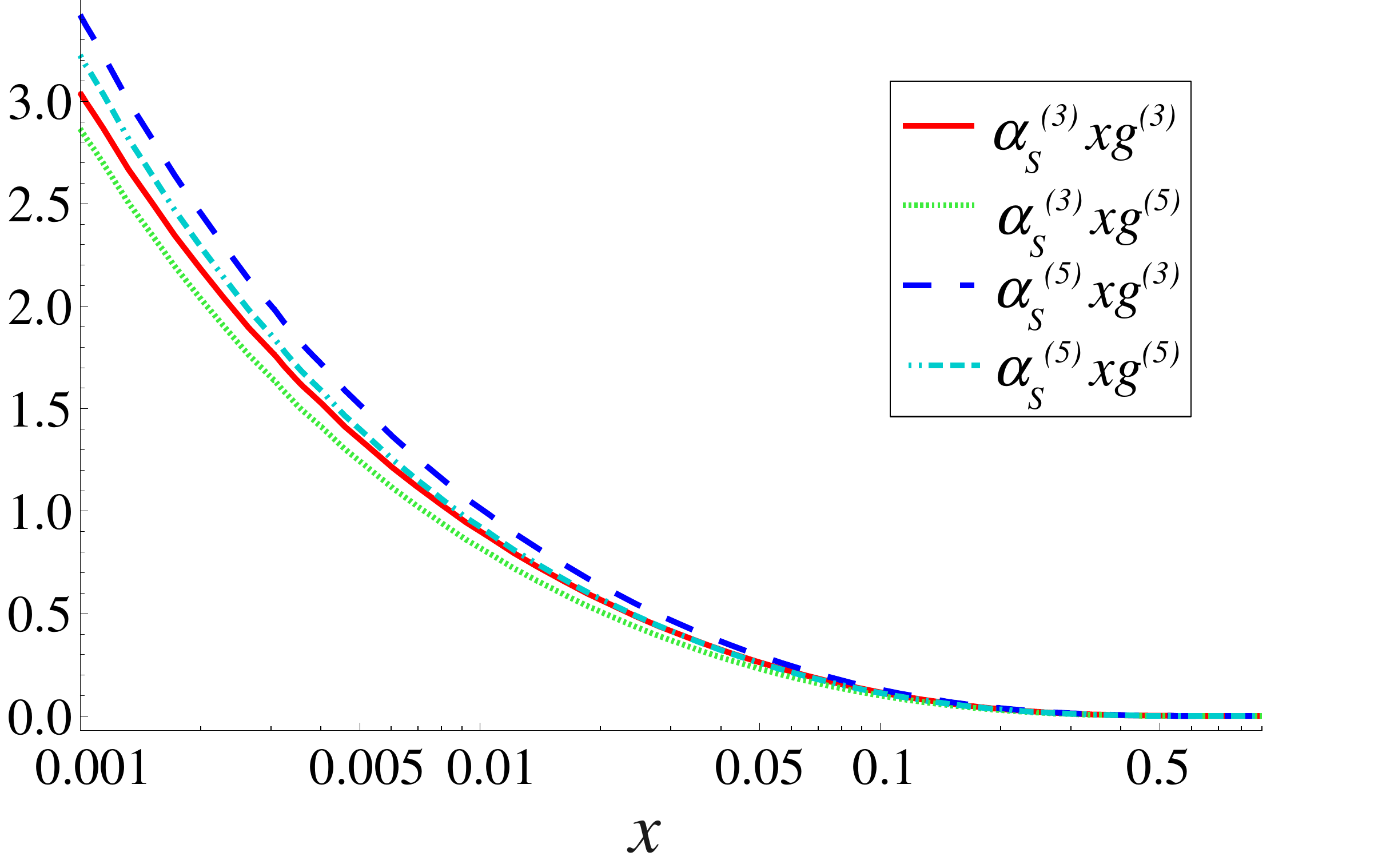}
}
\\
\subfloat[Ratio of different combination of 3- and 5-flavor $\alphas^{(N_{R})}xg^{(N_{F})}(x)$
and $\alphas^{(3)}xg^{(3)}(x)$ as a function of $x$, for different
values of $\mu$: $5$~(left), $10$~(middle) and $100$~(right)
GeV. \label{fig:ASxg-R_x} ]{\centering{}\includegraphics[width=0.32\textwidth]{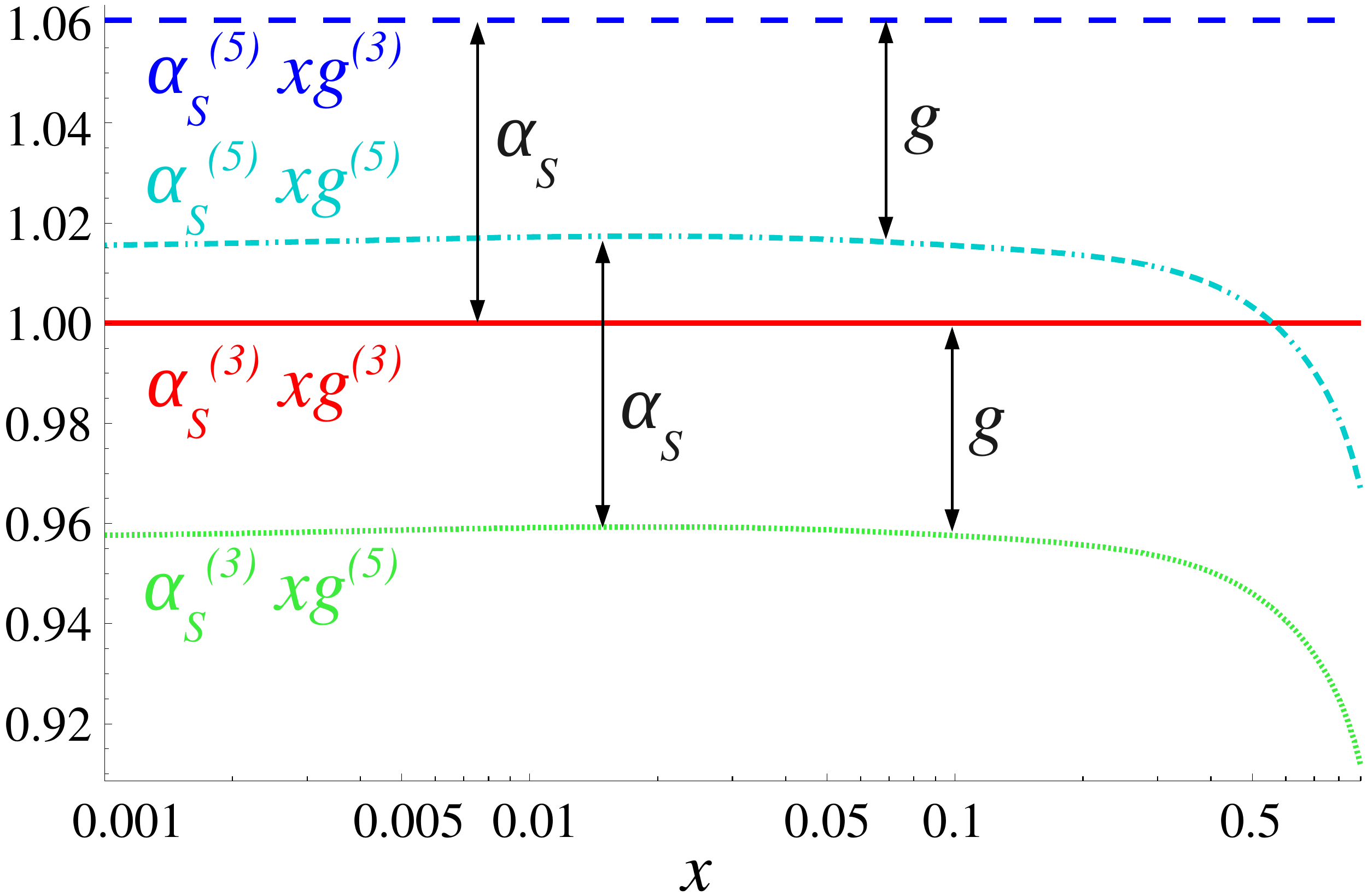}\quad{}\includegraphics[width=0.32\textwidth]{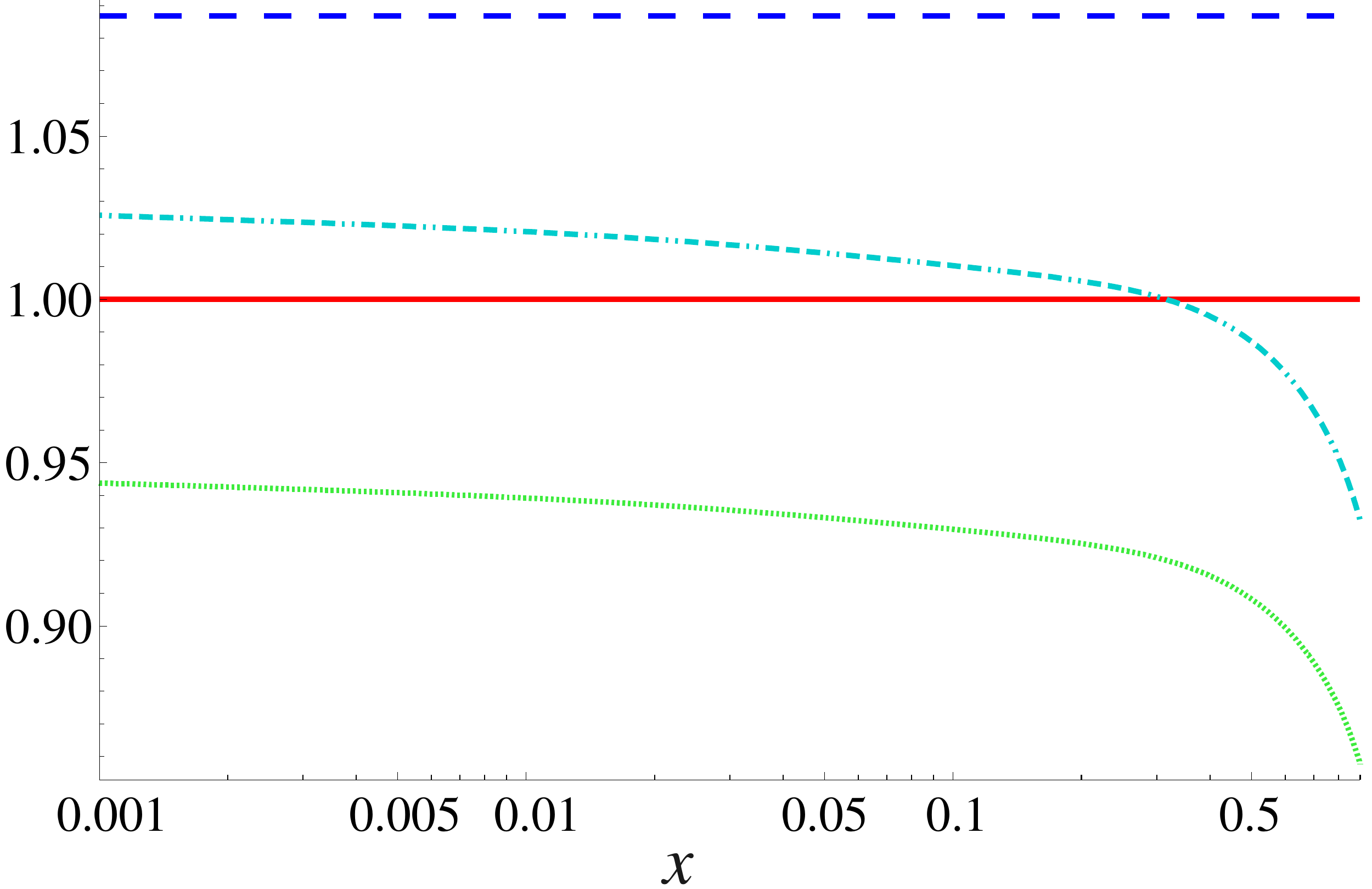}\quad{}\includegraphics[width=0.32\textwidth]{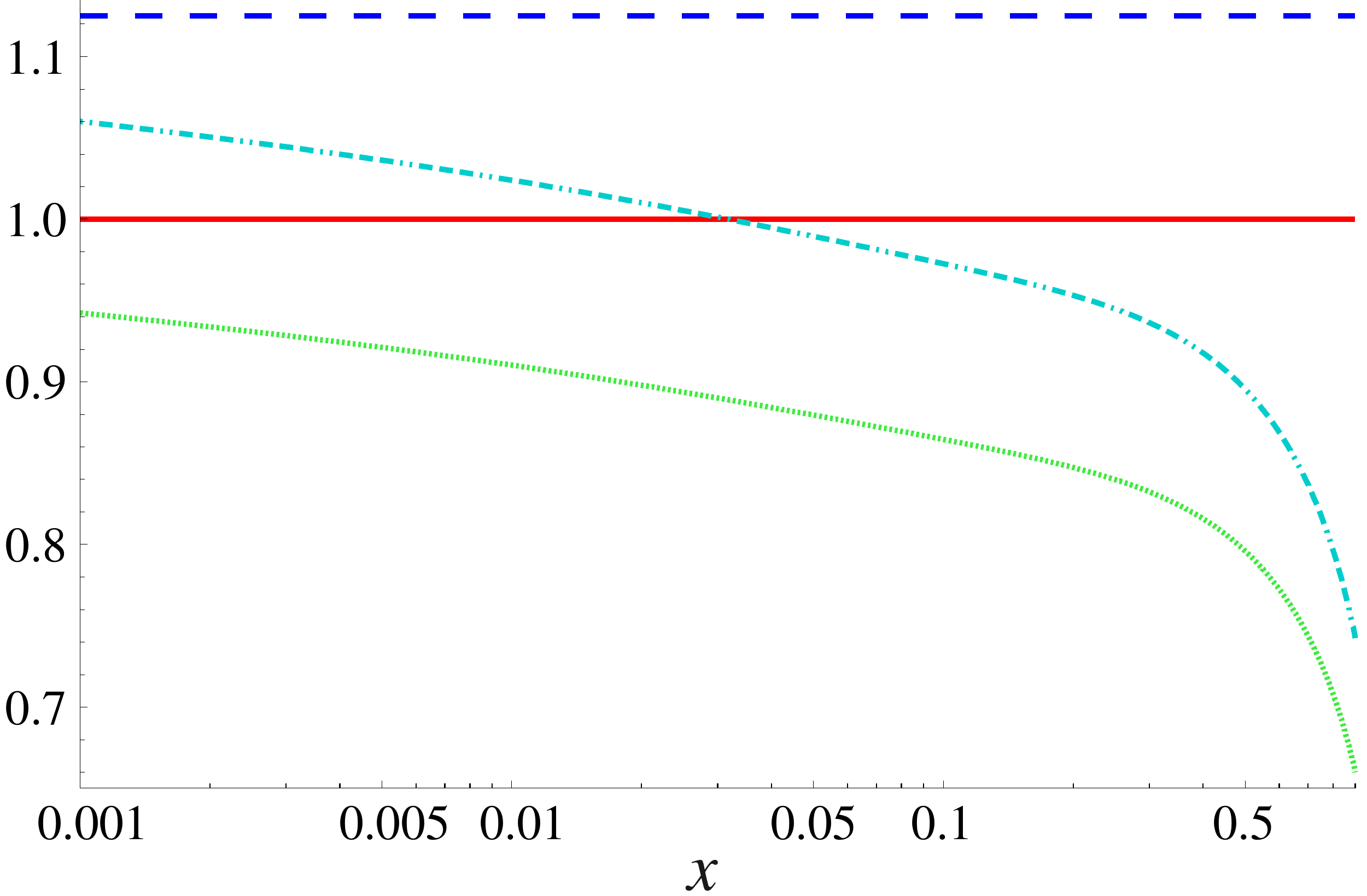}
}
\caption{$\alphas^{(N_{R})}xg^{(N_{F})}(x)$ as a function of $x$, for different
values of $\mu$. The curves are labeled $\alphas^{(N_{R})}xg^{(N_{F})}(x)=\{N_{R},N_{F}\}$
where the first term in braces indicates the $N_{R}$ for the $\alpha_{s}$
and the second indicates the $N_{F}$ for $g$.\label{fig:ASxg_x}}
\end{figure*}

\begin{figure*}[t]
\subfloat[Different combinations of 3- and 5-flavor $\alphas^{(N_{R})}xg^{(N_{F})}(x)$
as a function of $\mu$, for different values of $x$: $10^{-1}$~(left),
$10^{-3}$~(middle) and $10^{-5}$~(right).\label{fig:ASxg-N_Q}]{\centering{}\includegraphics[width=0.32\textwidth]{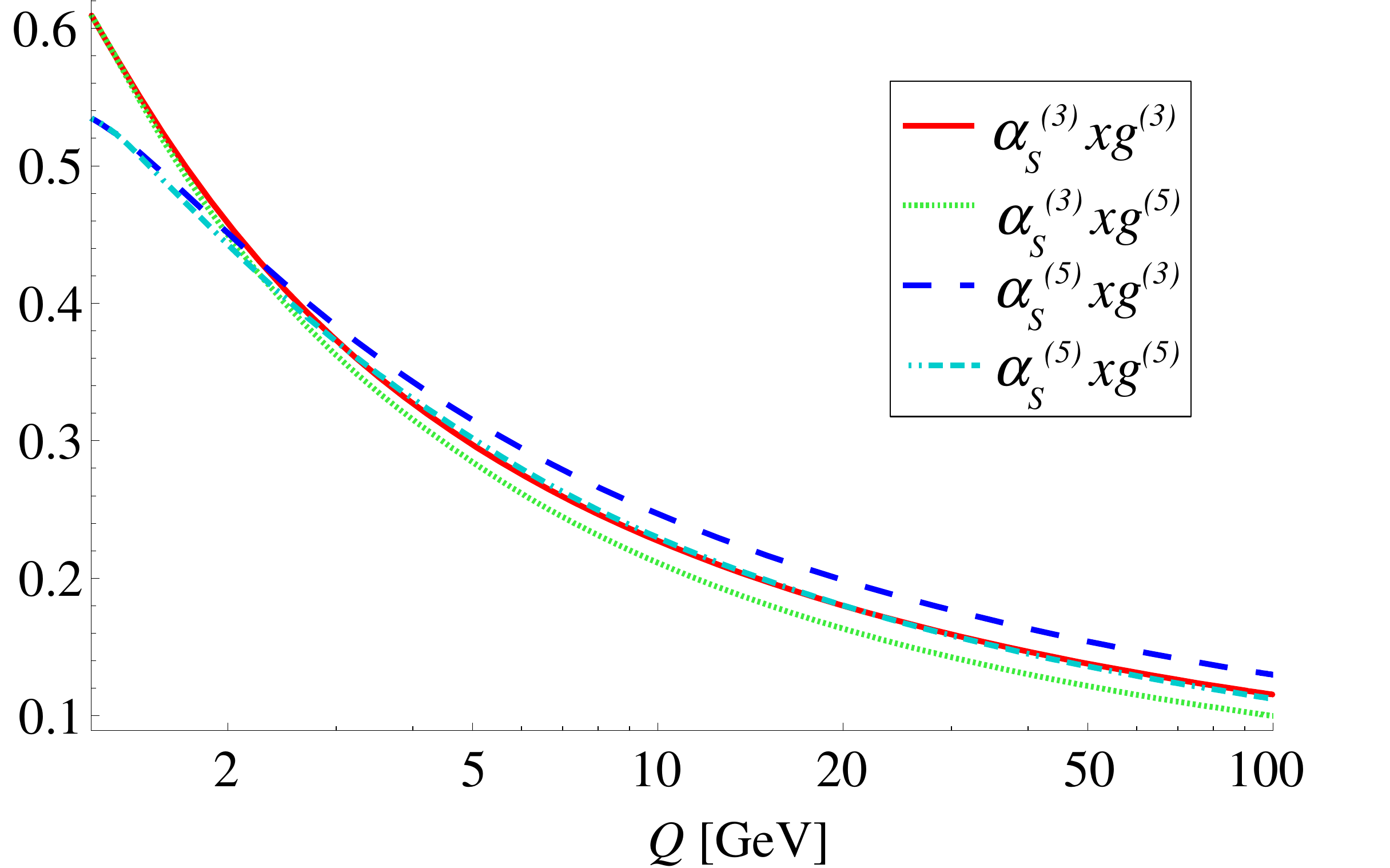}\quad{}\includegraphics[width=0.32\textwidth]{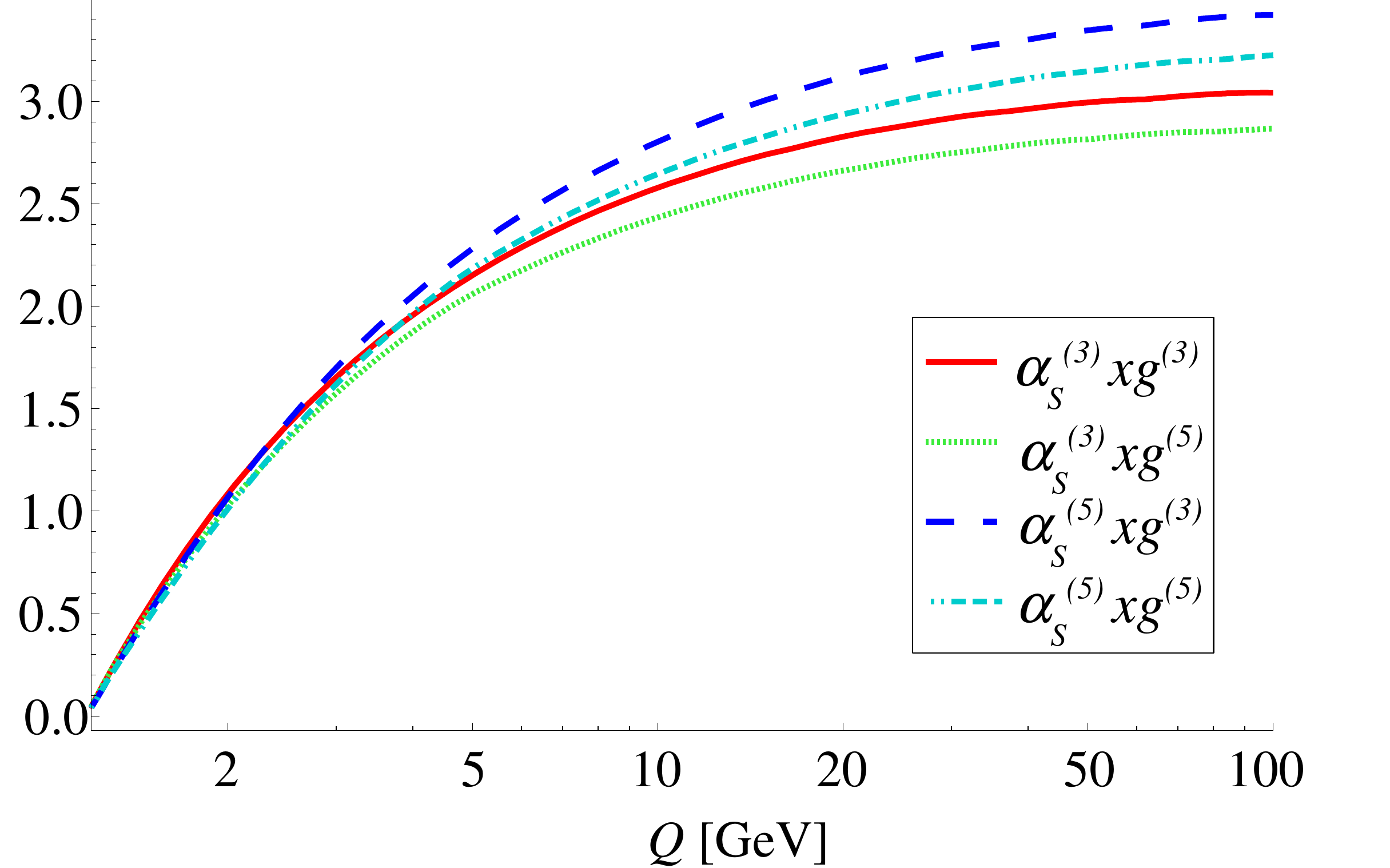}\quad{}\includegraphics[width=0.32\textwidth]{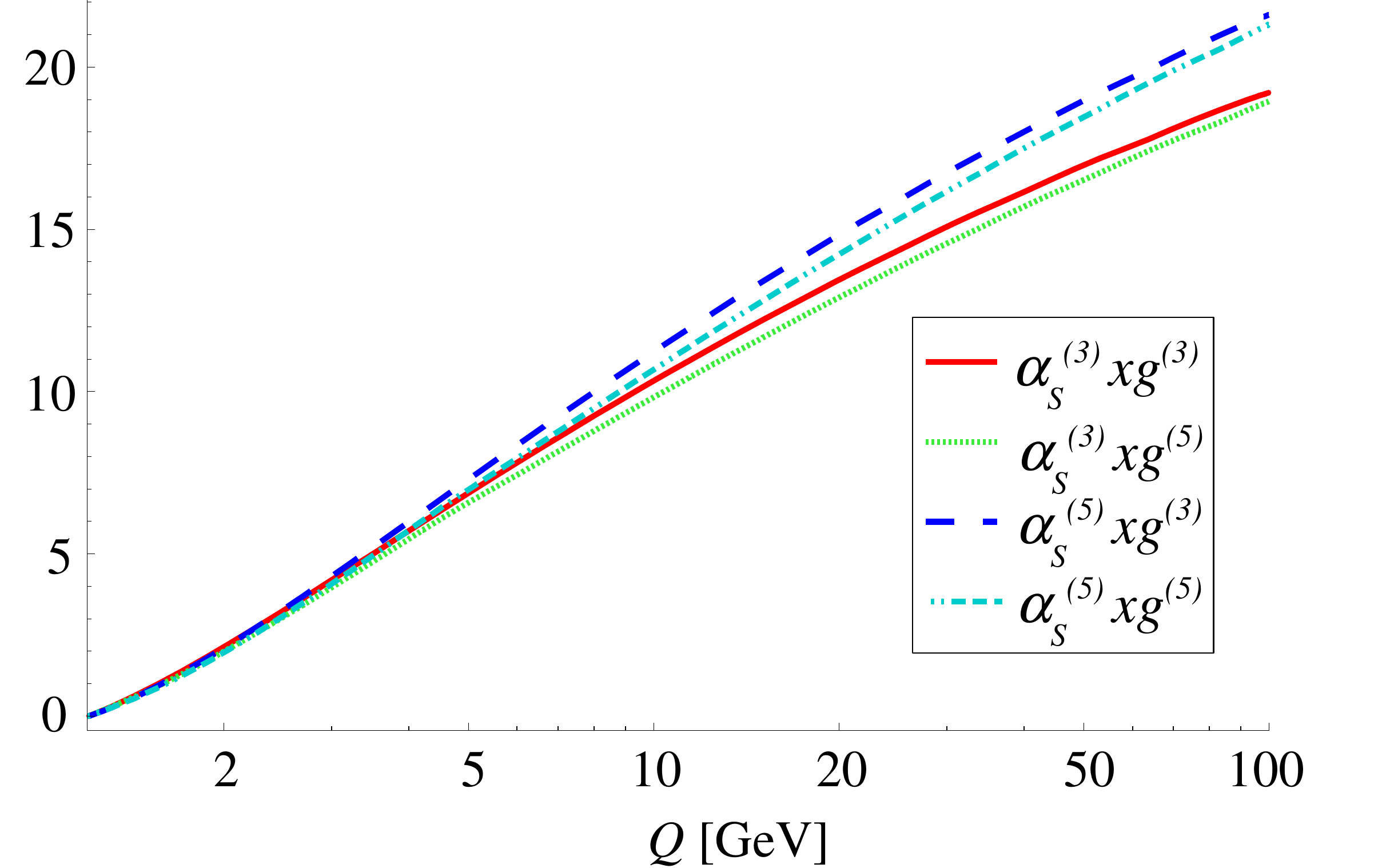}
}
\\
\subfloat[Ratio of different combination of 3- and 5-flavor $\alphas^{(N_{R})}xg^{(N_{F})}(x)$
vs. $\alphas^{(3)}xg^{(3)}(x)$ as a function of $\mu$, for different
values of $x$: $10^{-1}$~(left), $10^{-3}$~(middle) and $10^{-5}$~(right).
\label{fig:ASxg-R_Q} ]{\centering{}\includegraphics[width=0.32\textwidth]{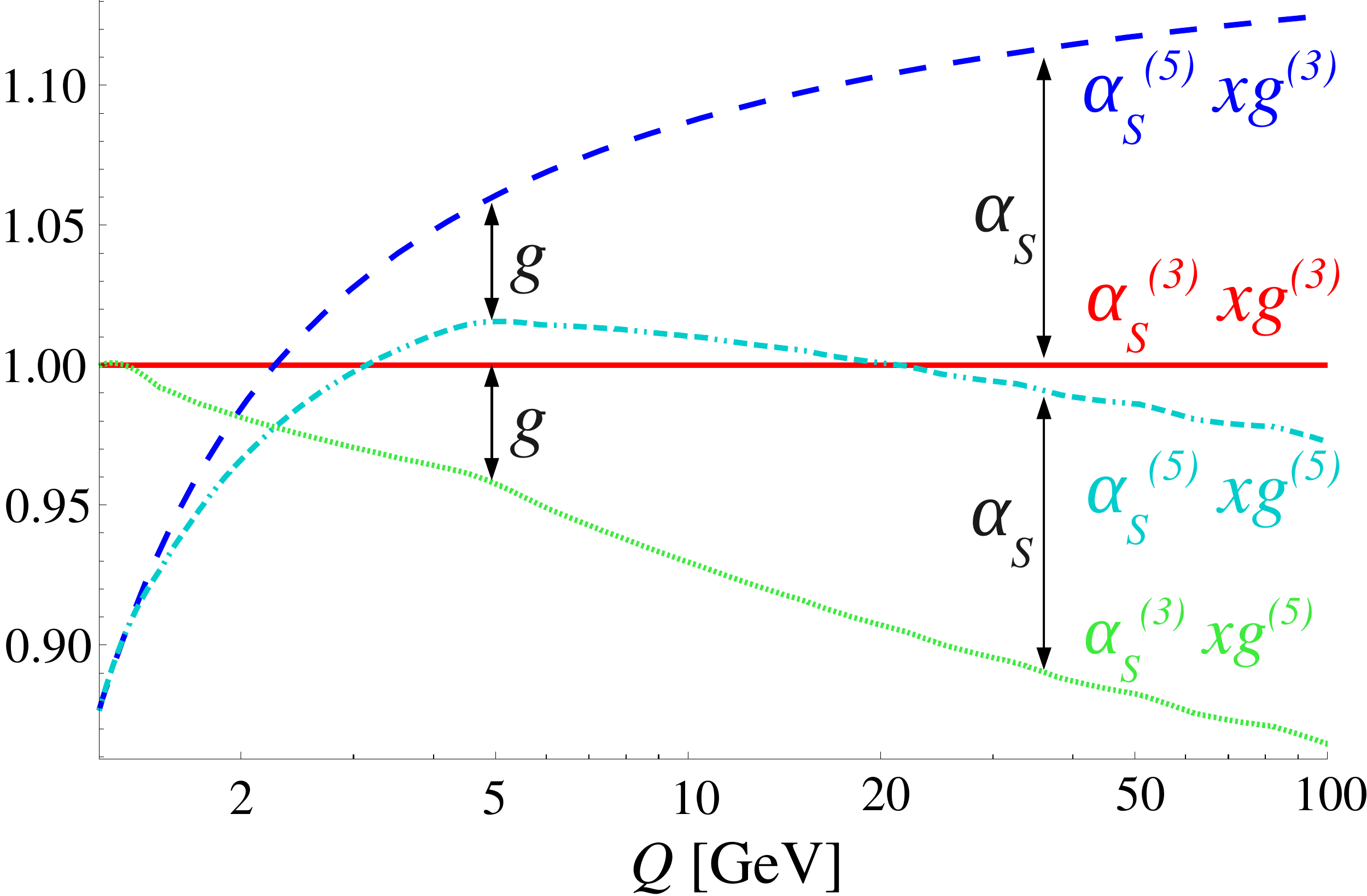}\quad{}\includegraphics[width=0.32\textwidth]{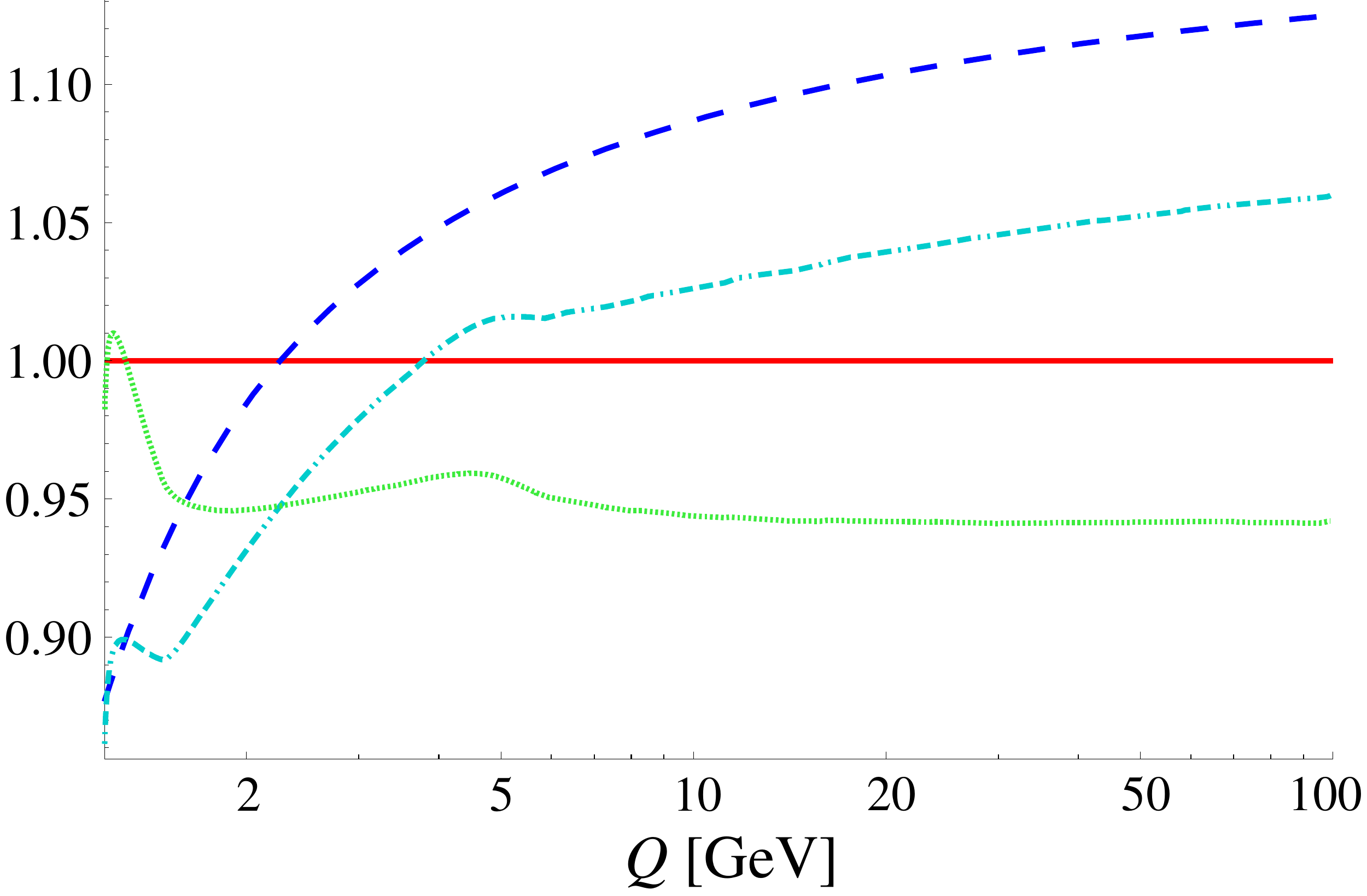}\quad{}\includegraphics[width=0.32\textwidth]{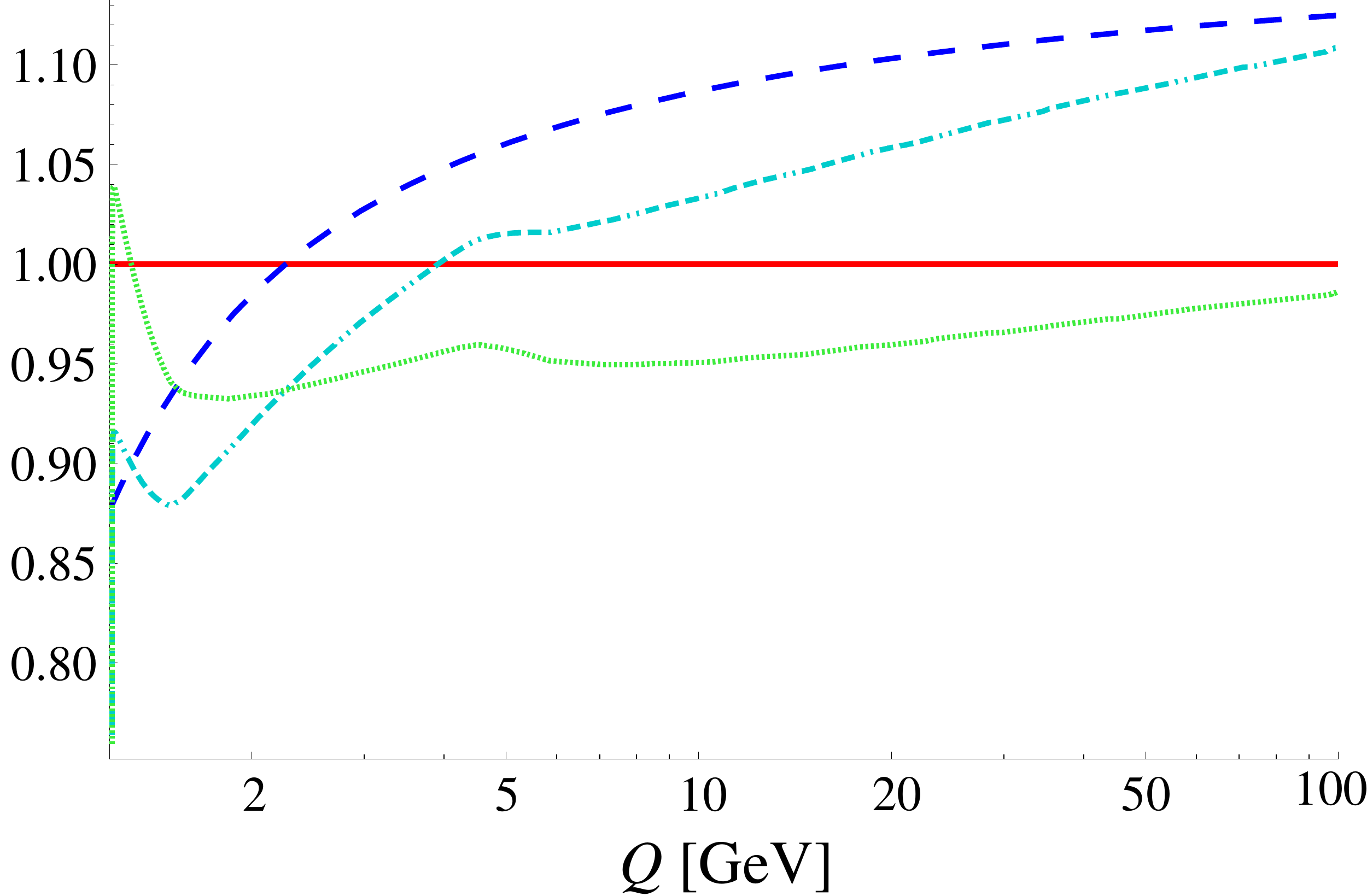}
}

\caption{$\alphas^{(N_{R})}xg^{(N_{F})}(x)$ as a function of $\mu$, for different
values of $x$. The curves are labeled $\alphas^{(N_{R})}g^{(N_{F})}(x)=\{N_{R},N_{F}\}$
where the first term in braces indicates the $N_{R}$ for the $\alpha_{s}$
and the second indicates the $N_{F}$ for $g$.\label{fig:ASxg_Q}}
\end{figure*}

To illustrate this mechanism, we show the combination $\alpha_{S}^{(N_R)}\times x\, g^{(N_F)}$
vs. $x$ (in Fig.~\ref{fig:ASxg_x}) and vs. $\mu$ (in Fig.~\ref{fig:ASxg_Q}).%
    \footnote{Note that we use here a 3(5)-flavor $\alpha_S$
    together with 5(3)-flavor PDFs only for illustrative purposes. In the
    actual implementation of the \hbox{H-VFNS} we always keep $N_R=N_F$.}
The compensating properties are best observed in the ratio plots (Figs.~\ref{fig:ASxg-R_x}
and \ref{fig:ASxg-R_Q}).

For example, in Fig.~\ref{fig:ASxg-R_x} for $\mu=5$~GeV we see
that if we start with $N_{F}=3$ for both $\alpha_{s}$ and $g$ (red
line), the effect of changing $N_{F}=5$ for $\alpha_{s}$ increases
$\alpha_{S}\times x\, g$ by 6\%; but, changing $N_{F}=5$ for the
gluon decreases $\alpha_{S}\times x\, g$ by roughly the same amount.
Hence, the combination $\alpha_{S}\times x\, g$ is relatively stable
under a change of $N_{F}$ as we see by comparing the curves labeled
$\{3,3\}$ (red) and $\{5,5\}$ (cyan). This is an example of how
the perturbation theory adjusts to yield a result that is (approximately)
independent of $N_{F}$ at a given order of perturbation theory.

In Fig.~\ref{fig:ASxg_Q} we show $\alpha_{S}\times x\, g$ vs. $\mu$
for a choice of $x$ values $\{10^{-1},10^{-3},10^{-5}\}$. While
$\{3,3\}$ (red) and $\{5,5\}$ (cyan) results are roughly comparable
for lower $\mu$ and higher $x$ values ($10^{-1}$), for smaller
$x$ values and larger $\mu$ the shift in the gluon is not sufficient
to compensate that of $\alpha_{s}$.

Reviewing Fig.~\ref{fig:ASxg_x} in more detail, we observe that
the $N_{F}$ compensation works well for lower $\mu$ values $\sim(5,10)$~GeV
across a broad range of $x$. For $\mu=5$~GeV, the curves labeled
$\{3,3\}$ (red) and $\{5,5\}$ (cyan) match within about $\sim2\%$
over much of the $x$ range. However, for larger $\mu=100$~GeV the
compensation between $\alphas$ and $g$ is diminished. We will see
this pattern again when we examine the physical structure functions,
and this difference is driven (in part) by uncanceled mass singularities
in the FFNS result.

\section{Physical Structure Functions vs. $\nf$ \label{sec:Physical-Structure-Functions}}

Having examined the unphysical (but useful) combination $\alphas\times xg$,
we now consider the physical observables $F_{2}$ and $F_{L}$ vs.
$N_{F}$. In Fig.~\ref{fig:F2} we display $F_{2}$ vs. $Q$ for
a choice of three $x$ values; the absolute values are shown in the
upper figures, and the ratios in the lower figures. Figure~\ref{fig:FL}
shows the corresponding plots for $F_{L}$. Both $F_{2}$ and $F_{L}$
were calculated at NLO and N3LO~\cite{Stavreva:2012bs} using 3 and
5 flavor  \hbox{H-VFNS} PDFs.%
   \footnote{As there is no complete N3LO massive calculation, we are using the
   approximation of Ref.~\cite{Stavreva:2012bs}; this is entirely sufficient
   for the purposes of this study.
   Note that in Ref.~\cite{Stavreva:2012bs}, the  PDF
   evolution is performed at NNLO by the QCDNUM~\cite{Botje:2010ay} code 
   which implements the $\msbar$ matching conditions~\cite{Buza:1996wv}
   which includes the resulting discontinuities.%
   }
We observe a number of patterns in these figures.

\subsubsection*{Low $Q$: $Q<m$}

At low $Q$ values, the $N_{F}=3$ and $N_{F}=5$ results coincide.
This is by design as once we go below the thresholds for $N_{F}=4,5$
the charm and bottom quarks are ``deactivated'' and all $N_{F}$
calculations reduce to the $N_{F}=3$ result.

At low $Q$ values, we also observe there is a significant difference
between the NLO and N3LO results; this difference arises from a number
of sources including the fact that at low $Q$ the value of $\alpha_{s}$
is large, hence the higher order corrections are typically larger
here.

\subsubsection*{High $Q$: $Q\gg m$}

As we move to larger $Q$ values, we notice two distinct features.

First, at large $Q$ we find the NLO and N3LO results tend to coincide.%
\footnote{The one exception is $F_{L}$ at large $x$ values; this suggests
that the higher order corrections in this kinematic region are large.
Recall that in the limit $m/Q\to0$ the LO contribution to $F_{L}$
vanishes, so it is not entirely surprising that this has large higher
order contributions.%
} Because $\alpha_{s}$ is decreasing at larger $Q$, the relative
importance of the higher order corrections is reduced.

Second, we see that the $N_{F}=3$ and $N_{F}=5$ results slowly diverge
from each other, both for the NLO and N3LO cases. This difference
can be traced to the uncanceled mass singularity in the $N_{F}=3$
calculations which is roughly proportional to $\alpha_{s}\ln[Q/m]$.
In the $N_{F}=5$ calculation, these logs are resummed into the heavy
quark PDFs; for the $N_{F}=3$ calculation, these logs are not resummed
and the calculation will be divergent in the limit $Q/m\to\infty$.

\begin{figure*}[t]
\subfloat[{{Inclusive $F_{2}$ as a function of $Q$ {[}GeV{]} for different
values of $x$: $10^{-1}$~(left), $10^{-3}$~(middle) and $10^{-5}$~(right).\label{fig:F2abs}
}}]{\centering{}\includegraphics[width=0.32\textwidth]{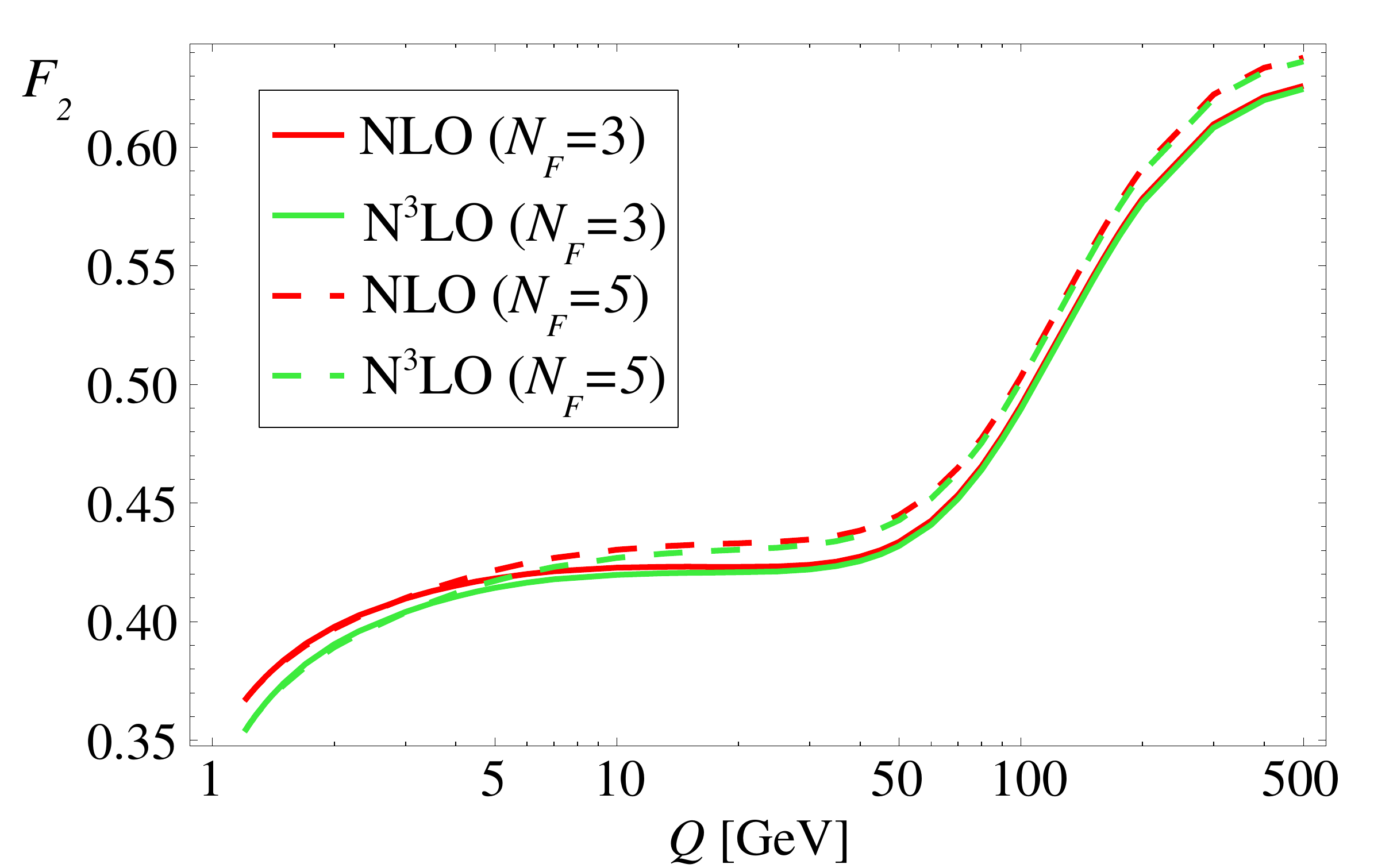}\quad{}\includegraphics[width=0.32\textwidth]{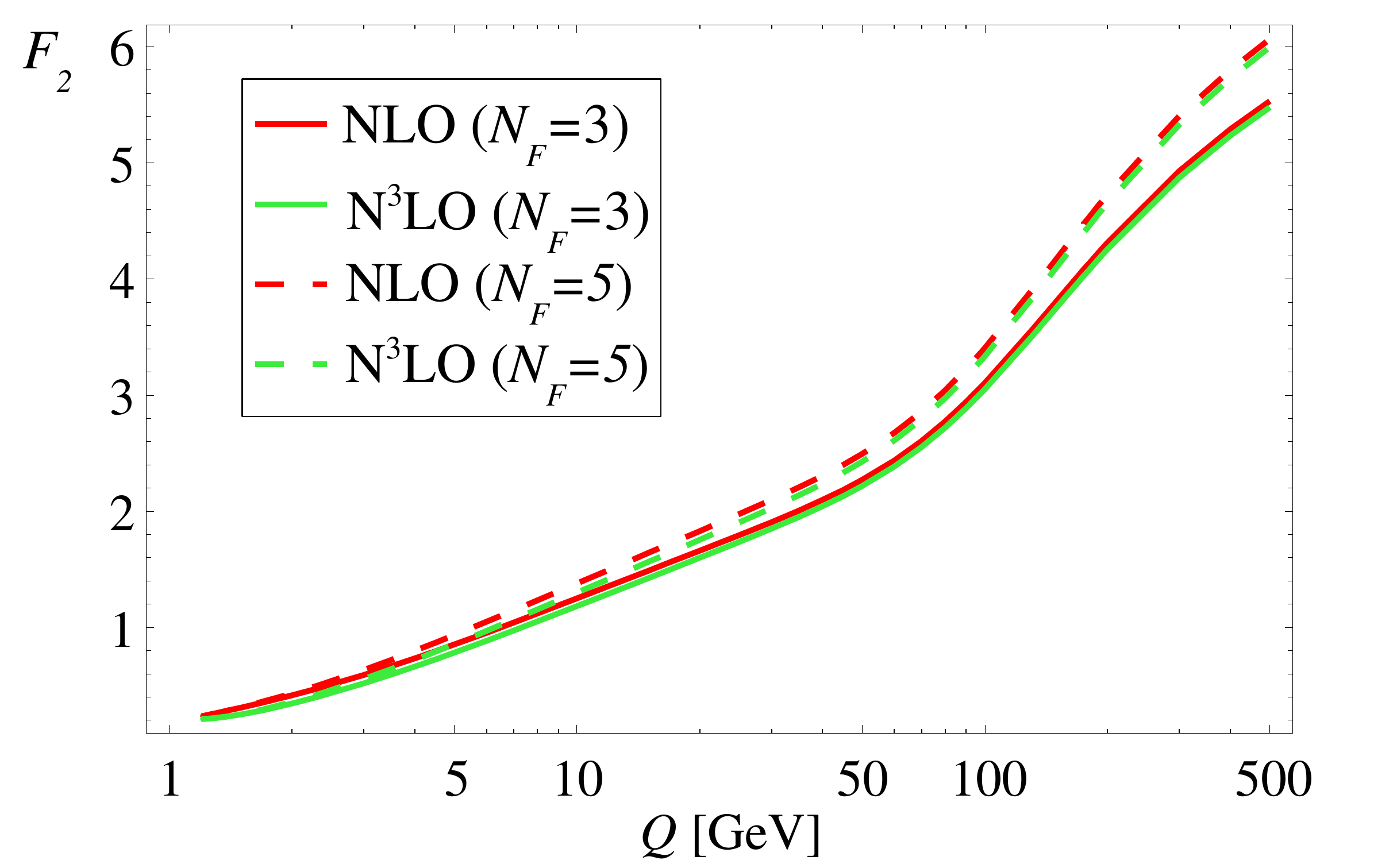}\quad{}
\includegraphics[width=0.32\textwidth]{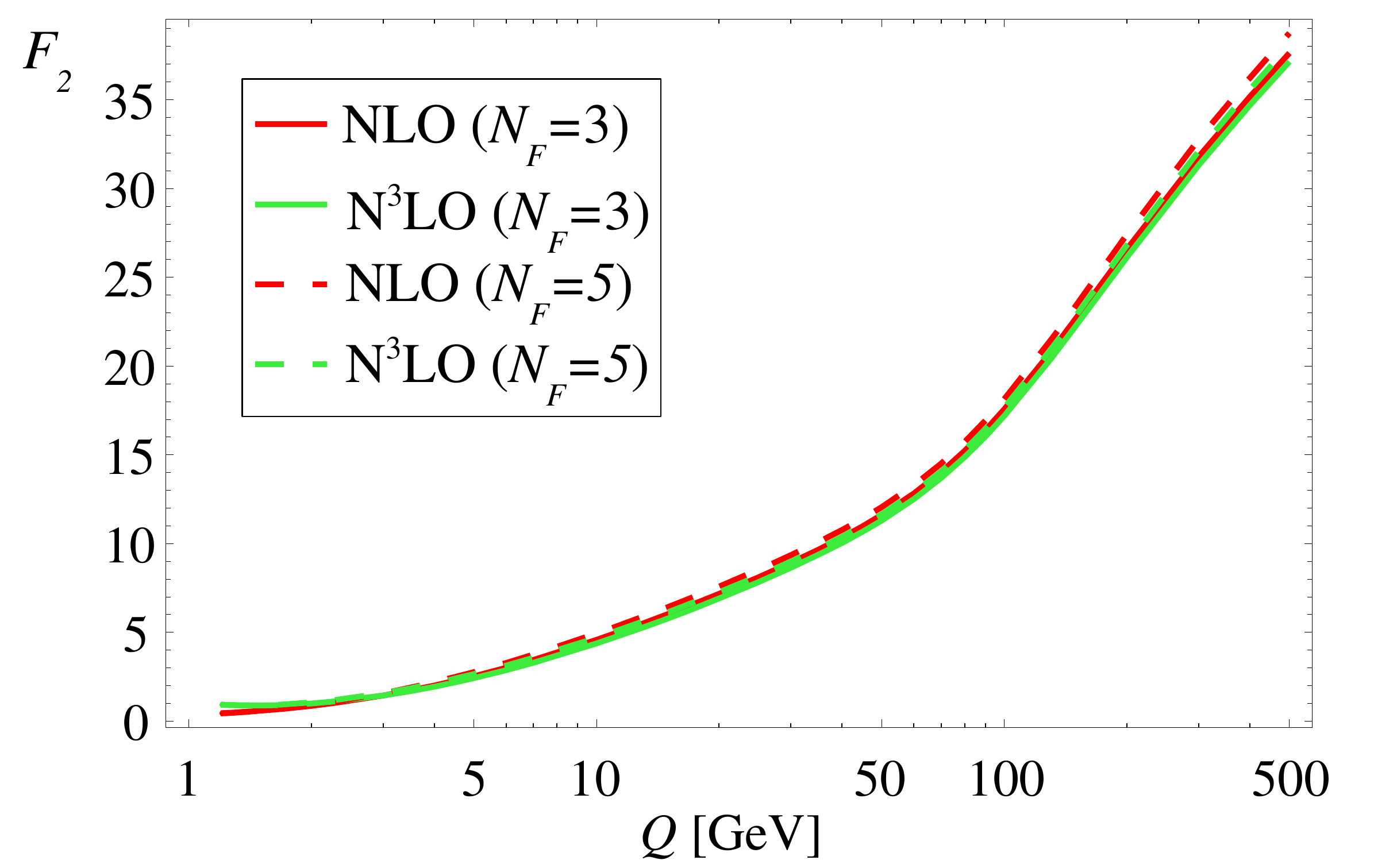}
}
\\
\subfloat[{{Ratio of inclusive $F_{2}$ as a function of $Q$ {[}GeV{]} for
different values of $x$: $10^{-1}$~(left), $10^{-3}$~(middle)
and $10^{-5}$~(right).\label{fig:F2r} }}]{\centering{}\includegraphics[width=0.32\textwidth]{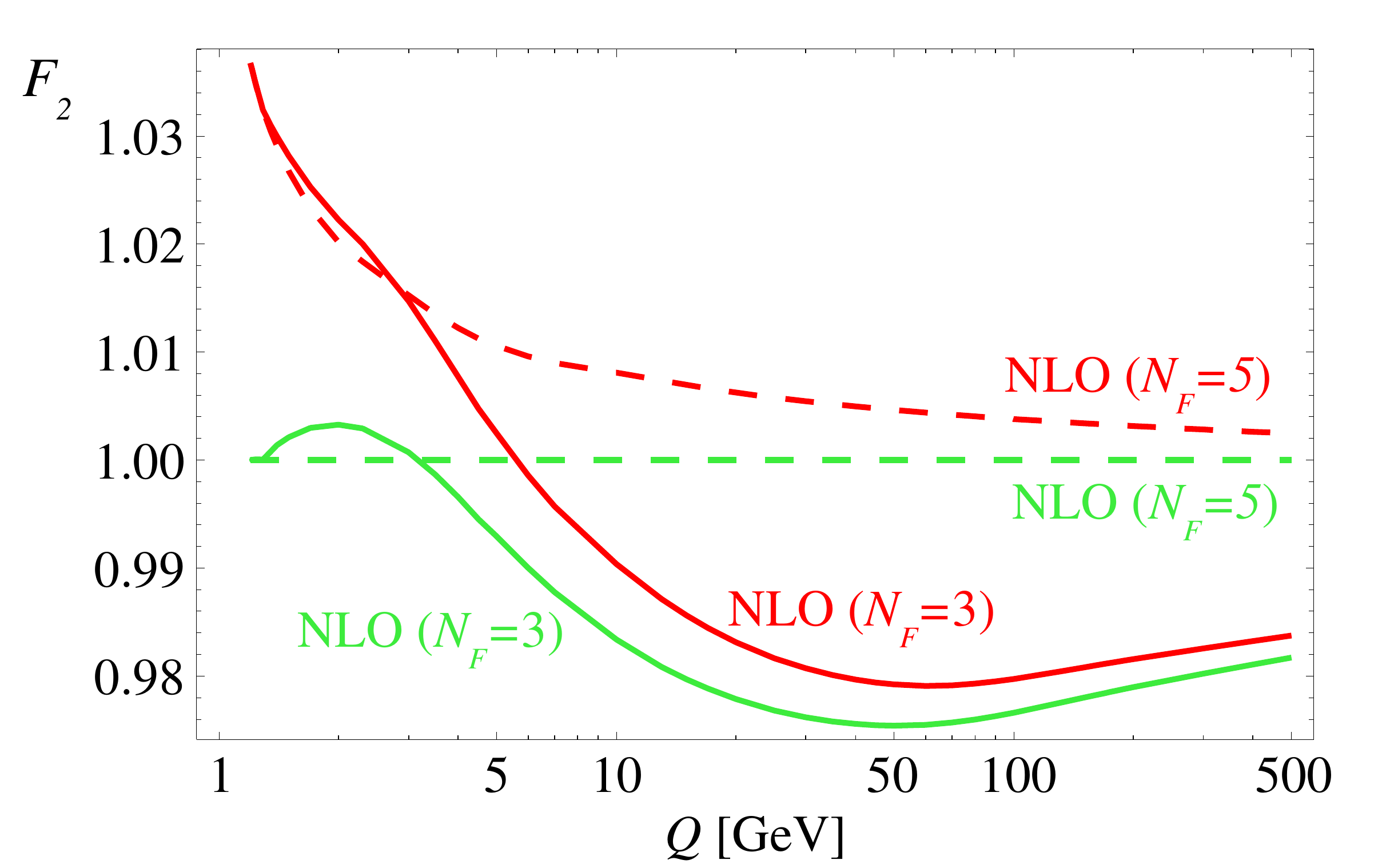}\quad{}\includegraphics[width=0.32\textwidth]{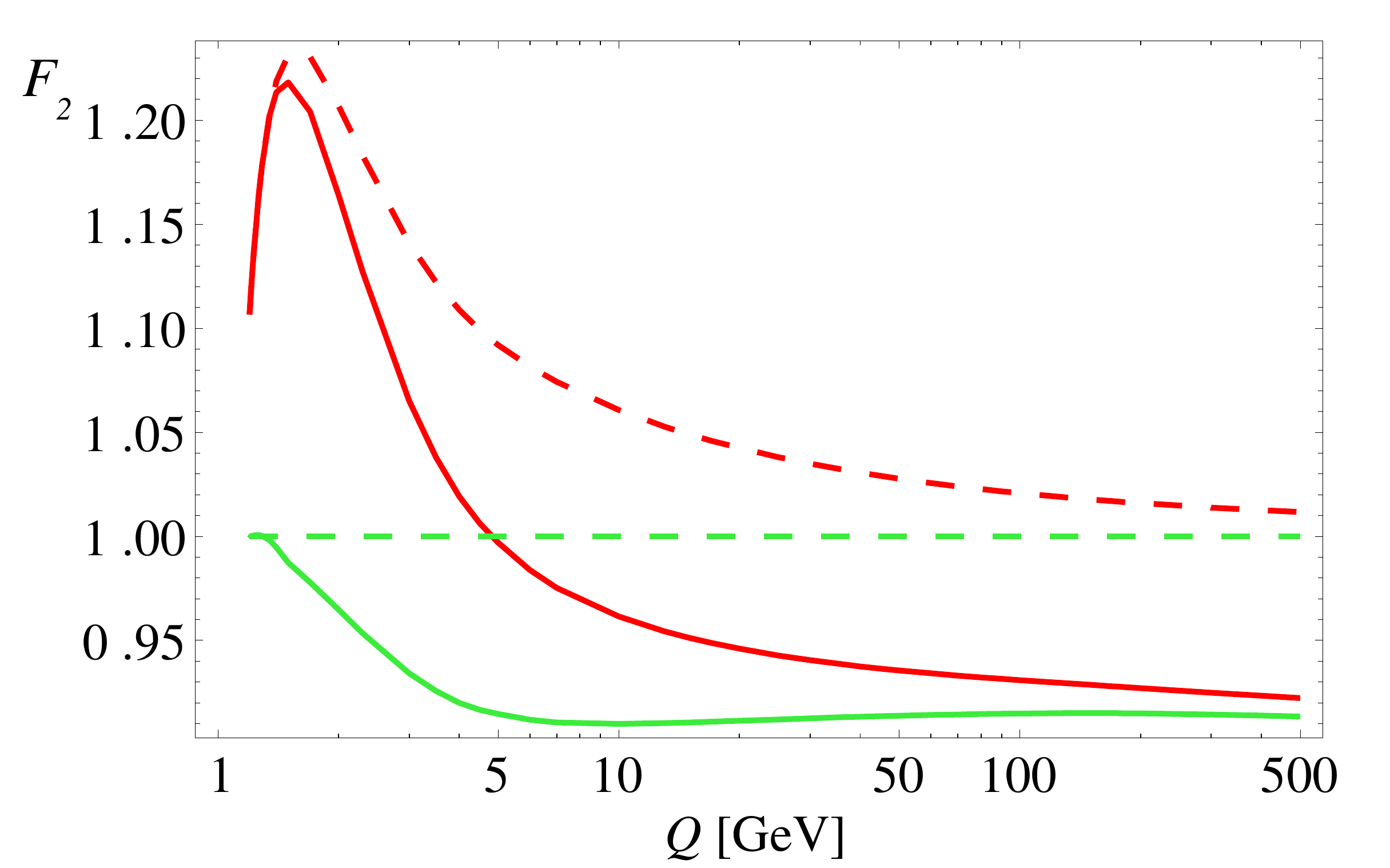}\quad{}\includegraphics[width=0.32\textwidth]{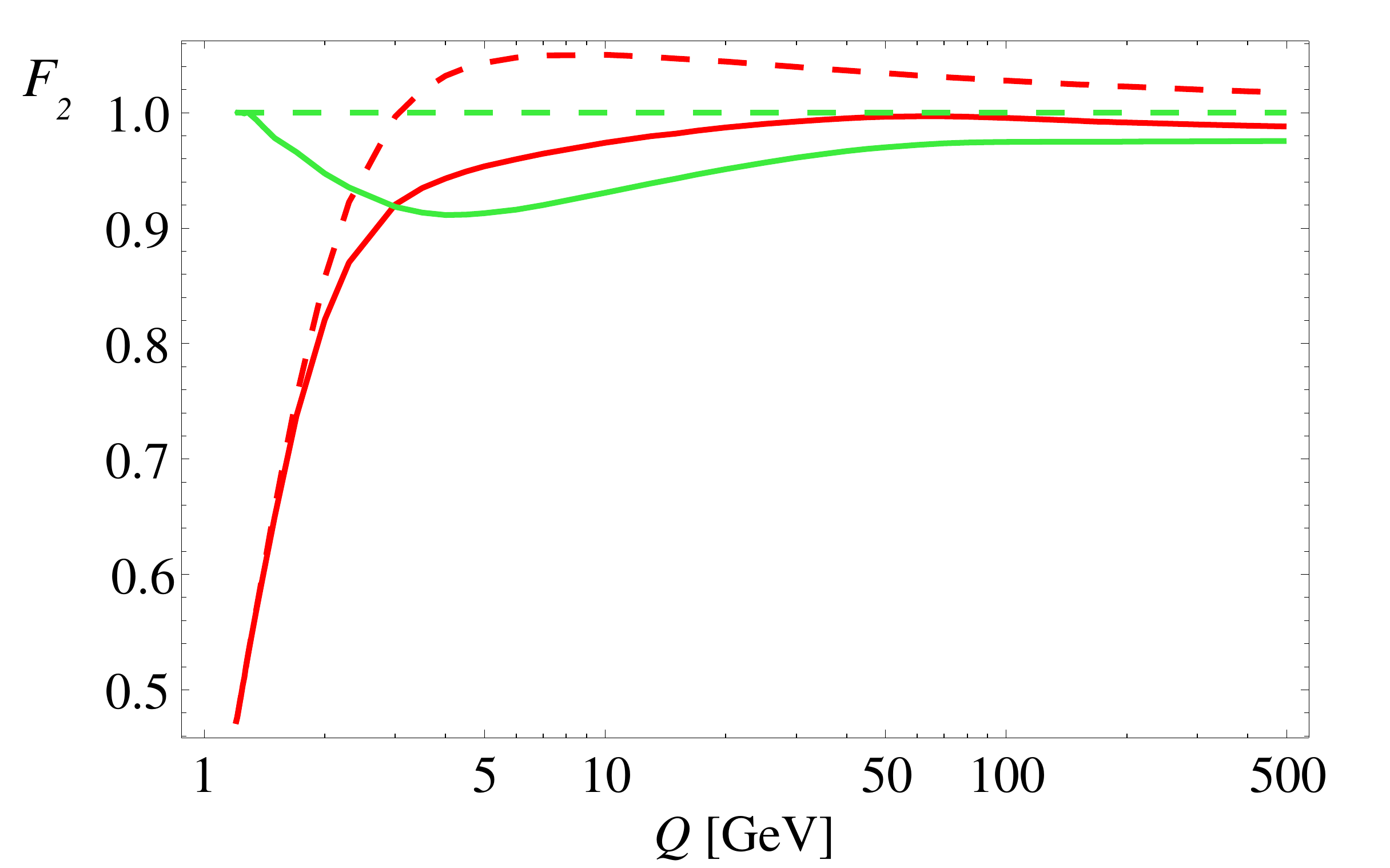}
}
\caption{Inclusive $F_{2}$ as a function of $Q$ {[}GeV{]} for different values
of $x$: $10^{-1}$~(left), $10^{-3}$~(middle) and $10^{-5}$~(right).
In these calculations we have chosen $\mu=Q$.
\label{fig:F2} }
\end{figure*}

\begin{figure*}[t]
\subfloat[{{Inclusive $F_{L}$ as a function of $Q$ {[}GeV{]} for different
values of $x$: $10^{-1}$~(left), $10^{-3}$~(middle) and $10^{-5}$~(right).\label{fig:FLabs}}}]{\centering{}\includegraphics[width=0.32\textwidth]{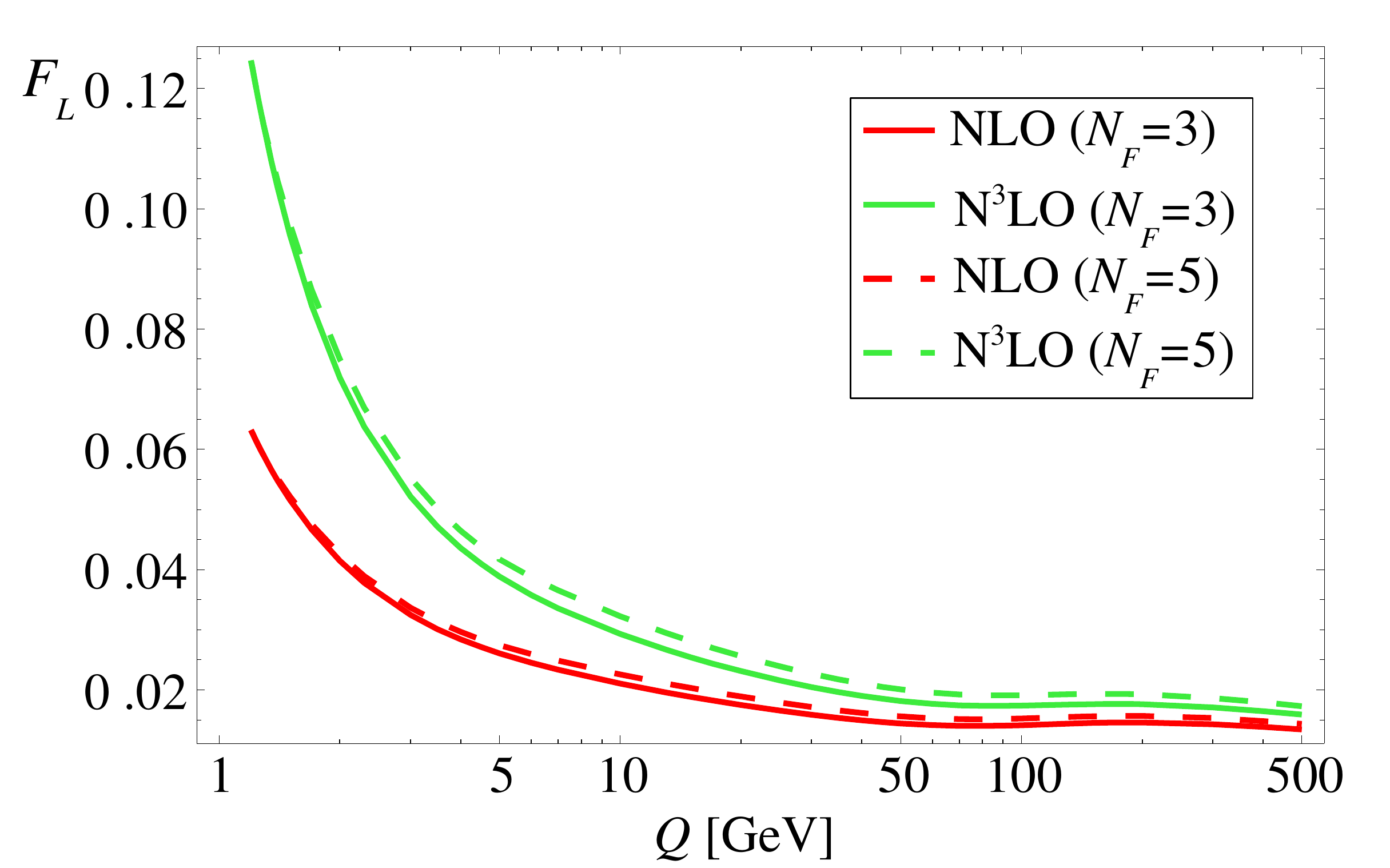}\quad{}\includegraphics[width=0.32\textwidth]{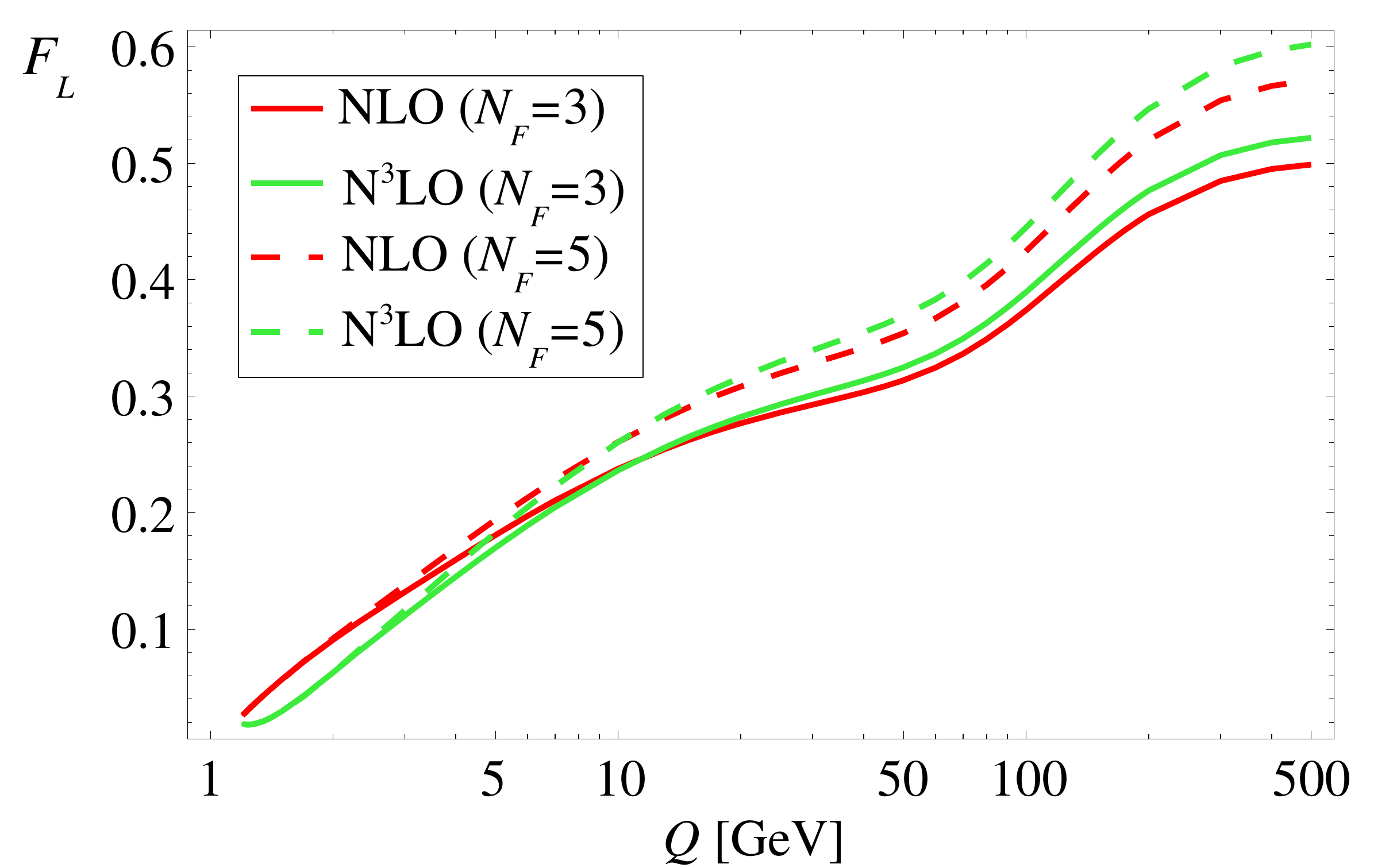}\quad{}\includegraphics[width=0.32\textwidth]{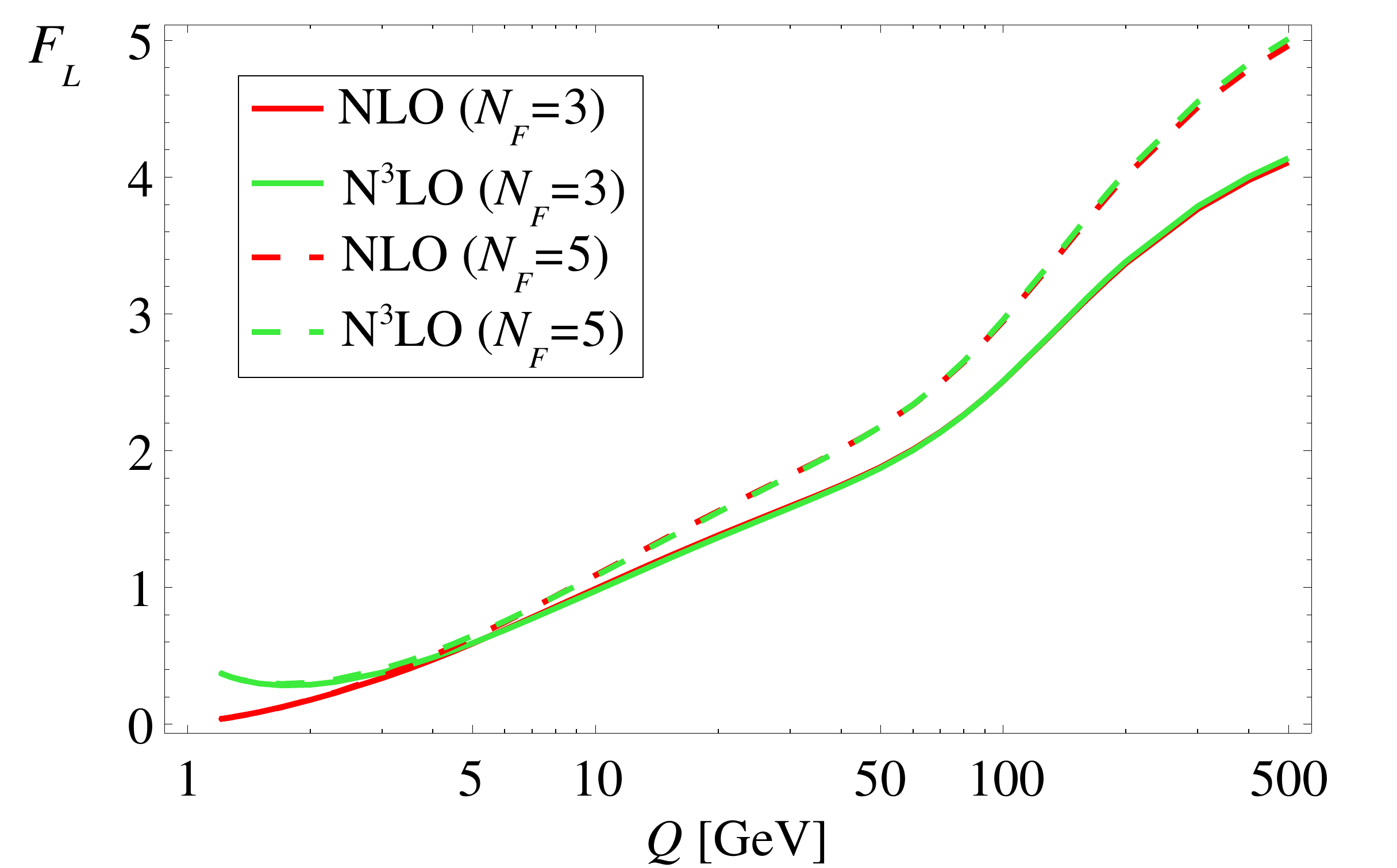}
}
\\
\subfloat[{{Ratio of inclusive $F_{L}$ as a function of $Q$ {[}GeV{]} for
different values of $x$: $10^{-1}$~(left), $10^{-3}$~(middle)
and $10^{-5}$~(right).\label{fig:FLr}}}]{\centering{}\includegraphics[width=0.32\textwidth]{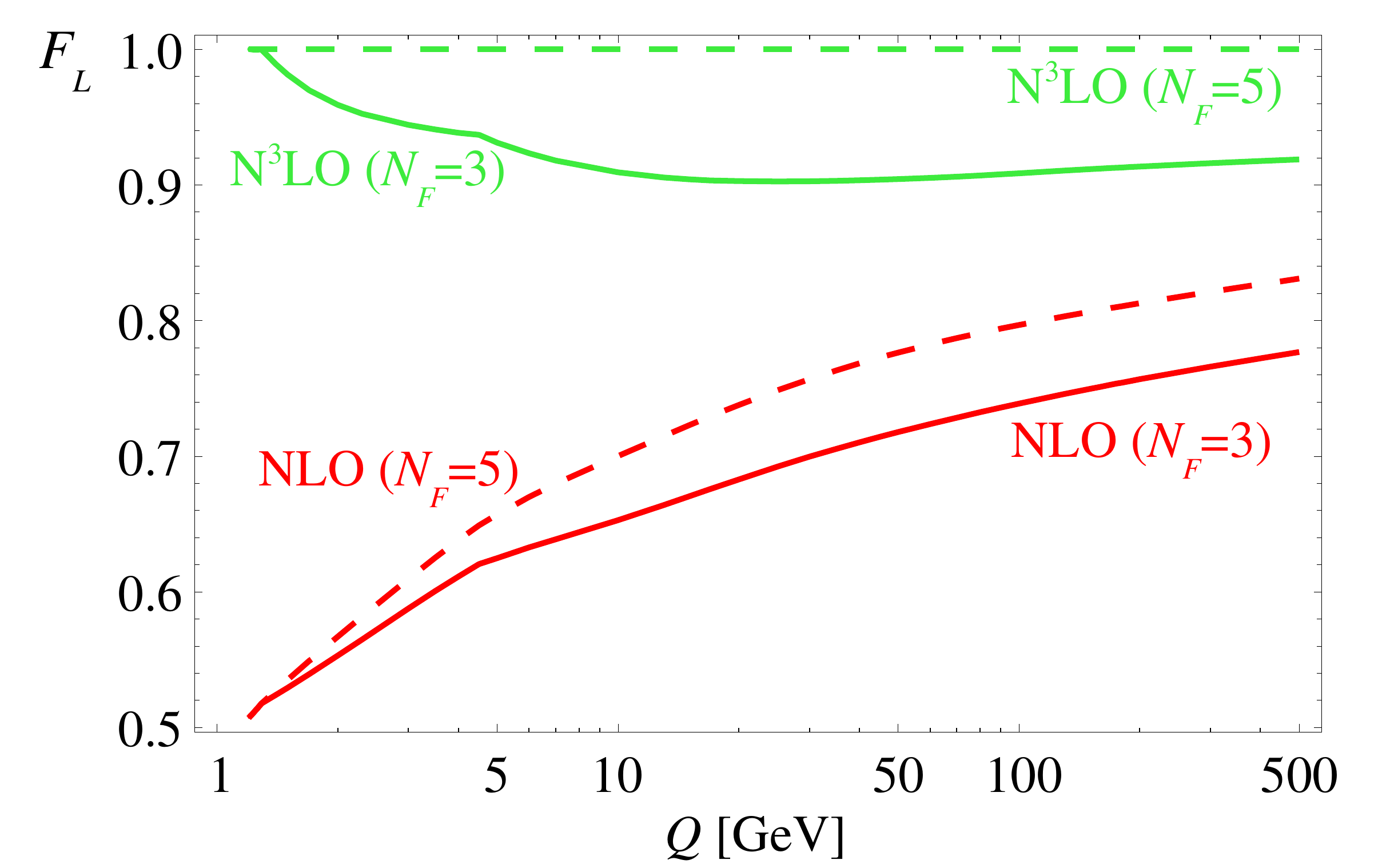}\quad{}\includegraphics[width=0.32\textwidth]{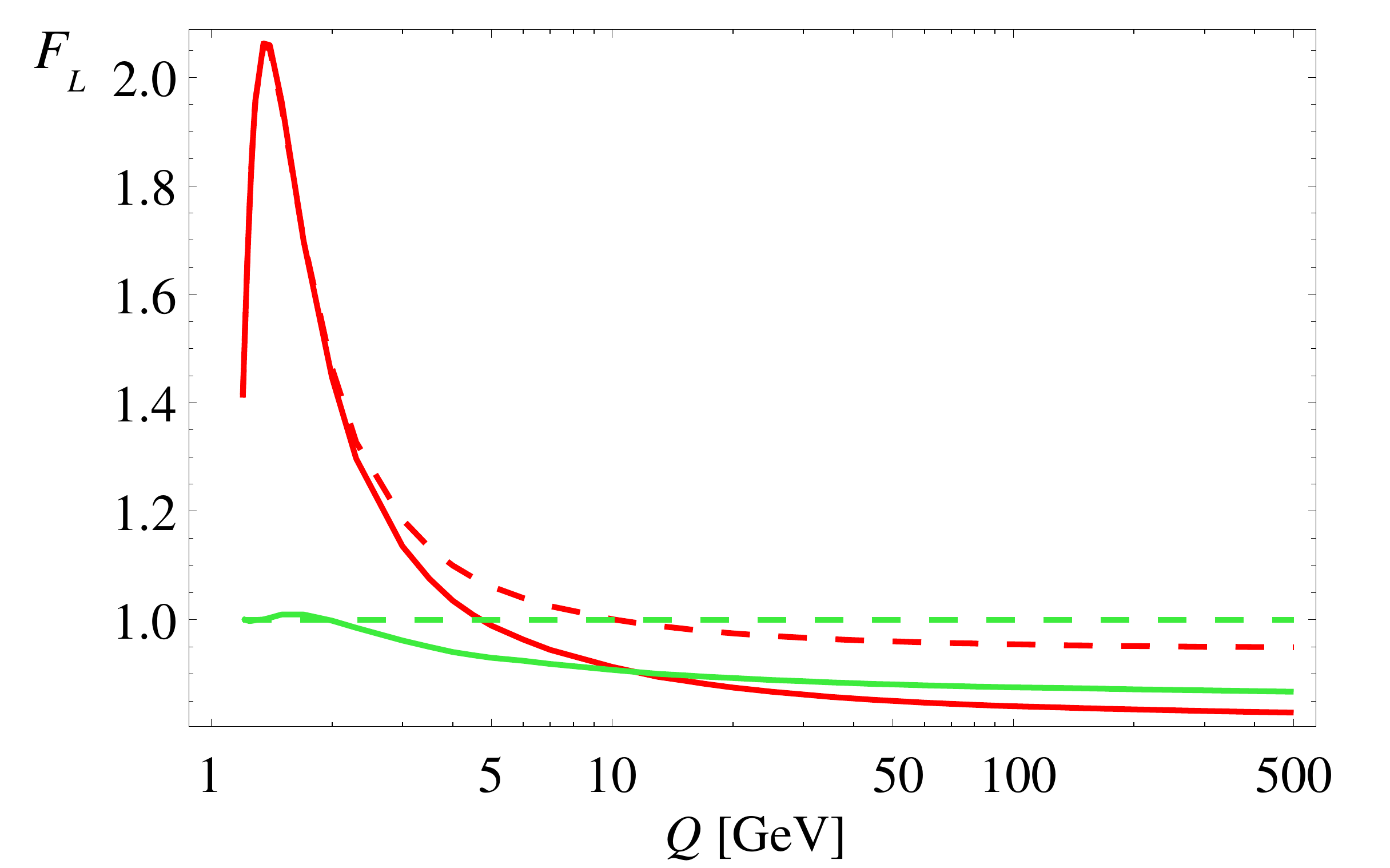}\quad{}\includegraphics[width=0.32\textwidth]{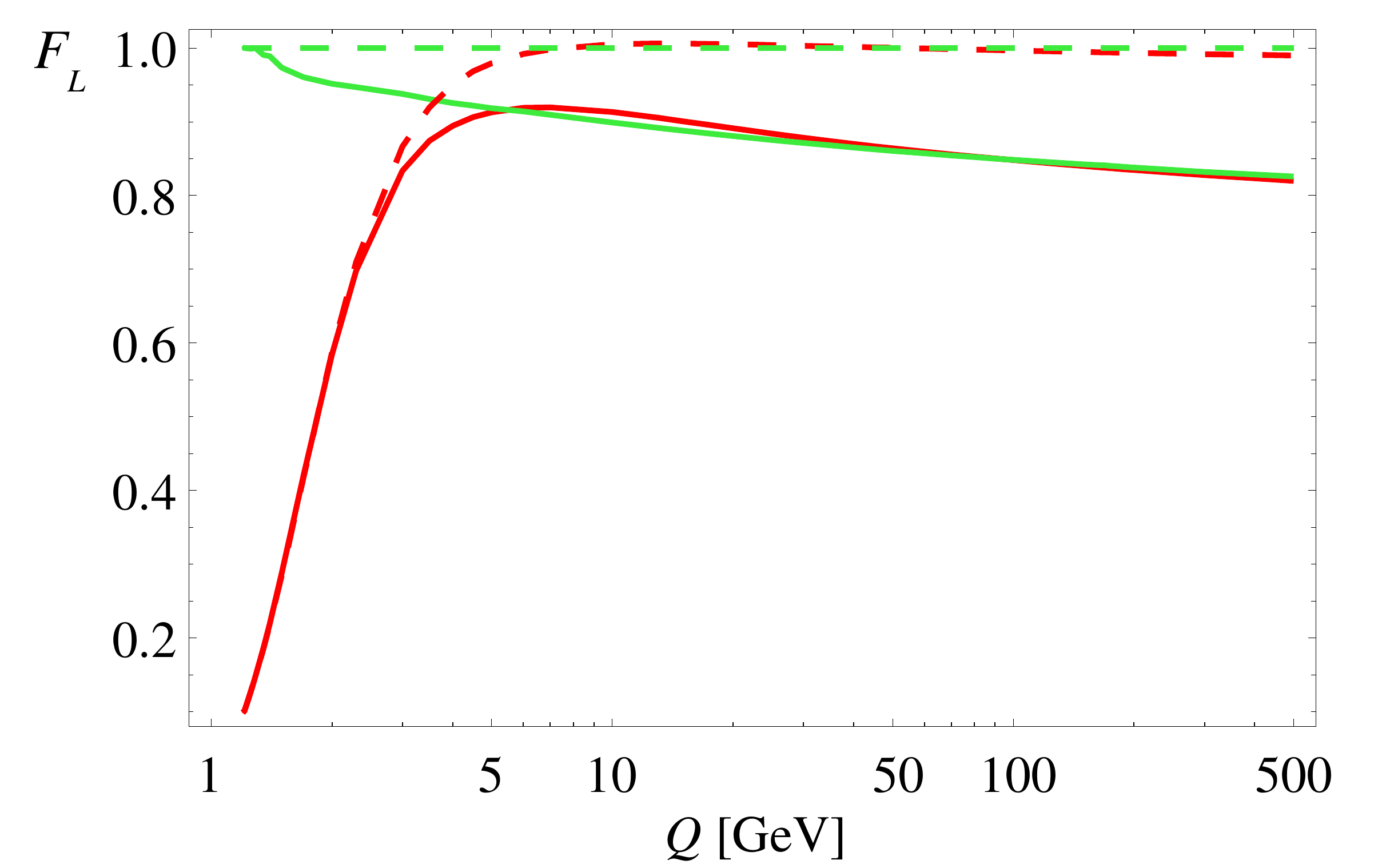}
}
\caption{Inclusive $F_{L}$ as a function of $Q$ {[}GeV{]} for different values
of $x$: $10^{-1}$~(left), $10^{-3}$~(middle) and $10^{-5}$~(right).
In these calculations we have chosen $\mu=Q$. \label{fig:FL} }
\end{figure*}

\subsubsection*{Intermediate $Q$: $Q\gtrsim m$}

We now come to the critical question: how far above the charm flavor
transition can we extend the $N_{F}=3$ FFNS calculation before the
uncanceled logs $\alpha_{s}\ln[Q/m]$ degrade the perturbation expansion.
By examining Figs.~\ref{fig:F2} and \ref{fig:FL} we can determine
the extent to which the $N_{F}=3$ and $N_{F}=5$ results diverge
due to these logs. For scales $\mu=Q$ a few times the quark mass
($m_{c}$) the difference is small; but for larger scales $\mu=Q\sim10m_{c}$
the difference can be in excess of 10\% depending on the specific
$x$ region. Also note, that while we are considering the inclusive
$F_{2,L}$, it is only the heavy quark components which are driving
the difference at large $\mu$ scales; for a less inclusive observable
(such as $F_{2}^{charm}$) this effect would be even more prominent.

\subsubsection{Recap}

To recap, in the three kinematic regions of interest we find the following. 
\begin{lyxlist}{00.00.0000}
\item [{$Q<m$}] In the low $Q$ region, we find the $N_{F}=3$ and $N_{F}=5$
results coincide; hence, in this region an $N_{F}=3$ FFNS result
will match with any VFNS result. Thus, we can use either the $N_{F}=3$
and $N_{F}=5$ calculation in this region. 
\item [{$Q\gg m$}] In the region of high $Q$, we find the $N_{F}=3$
and $N_{F}=5$ results diverge logarithmically due to the uncanceled
mass singularities, and in the limit $Q/m\to\infty$ the $N_{F}=3$
calculation contains divergent terms. Hence, in this region, we would
expect the VFNS $N_{F}=5$ result to be most reliable. 
\item [{$Q\gtrsim m$}] For $Q$ scales which are a few times the quark
mass or less, the $N_{F}=3$ and $N_{F}=5$ results are comparable;
for larger $Q$ scales, this difference will increase logarithmically
with the scale. Thus, we can use either the $N_{F}=3$ and $N_{F}=5$
calculation in this region, but as we move to larger scales we need
to transition to the $N_{F}=5$ in the VFNS. 
\end{lyxlist}
These conclusions are illustrated in Fig.~\ref{fig:schematic}, and
now we are able to make quantitative statements about the specific
regions of validity.

In summary, the $N_{F}$ dependent PDFs provide us the freedom to
choose the $N_{F}$ transitions where it is convenient for the analysis
of specific data sets; however, this freedom comes with the responsibility
that we must be aware of the mass singular logs and be sure not to
extend a particular $N_{F}$ FFNS calculation beyond its region of
reliability.

\section{An Example: From Low to High Scales\label{sec:example}}

We now finish with an example of how the \hbox{H-VFNS} scheme could be employed
for a simultaneous study of both a low-scale process ($\mu\sim m_{b}$)
at HERA%
\footnote{A relevant data set could be the recent analysis~\cite{Aaron:2011gp}
by the H1 experiment of $D^{*\pm}$ meson production and the extracted
$F_{2}^{charm}$ structure function.%
} and a high scale process ($\mu\gg m_{c,b}$) at the LHC.%
    \footnote{A relevant data set could be, for example, high-mass dilepton
    resonances~\cite{ATLAS-CONF-2013-017} or dijet mass
    spectrum~\cite{Chatrchyan:2013qha}, both analyses extend beyond one TeV.}

At HERA, a characteristic $Q$ range for the extraction of $F_{2}^{charm}$,
for example, is $\sim2<Q<10$~GeV and this spans the kinematic region
where the charm and bottom quarks become active in the PDF. These
analyses can be performed using a $N_{F}=3$ FFNS calculation as the
scales involved are not particularly large compared to the $m_{c,b,}$
scales. Additionally, the extraction of the $F_{2}^{charm}$ structure
function is often computed using the HVQDIS program~\cite{Harris:1997zq},
and this explicitly works in a $N_{F}=3$ FFNS. This approximation
is entirely adequate in this kinematic region as resummed logs are
not particularly large in the relevant $Q$ region. The $F_{2}^{charm}$
structure function 
extracted in~\cite{Aaron:2011gp}
is compared with predictions using $N_{F}=3$ FFNS
PDFs from CT10f3~\cite{Lai:2010vv} and MSTW2008f3~\cite{Martin:2010db}
and both yield good descriptions of the data.

Conversely, at the LHC the  $\mu$ range for new
particle searches via the Drell-Yan process can be in excess of a  TeV.
For this analysis, we would want to use 
$N_{F}=5$ so that the charm and bottom logs are 
resummed.\footnote{We could also use $N_{F}=6$, but the difference with the $N_{F}=5$
case is minimal.}

Because the  \hbox{H-VFNS} simultaneously provides $N_{F}=\{3,4,5,6\}$,
we can analyze the HERA data in a FFNS $N_{F}=3$ context while
also analyzing the LHC data in a $N_{F}=\{4,5,6\}$ VFNS context.

Operationally, we could perform a PDF fit to both a combination of
HERA and LHC data by implementing the following steps. 
\begin{enumerate}
\item Parametrize the PDFs at a low initial scale $\mu=Q_{0}\sim1$~GeV,
and generate a family of $N_{F}$ dependent PDFs as outlined in Sec.~\ref{sub:Generating-the-PDFs}. 
\item Fit the HERA $F_{2}^{charm}$ structure function data using $N_{F}=3$
``FFNS'' PDFs, $f_{i}(x,\mu,N_{F}=3)$ and $\alpha_{s}(\mu,N_{F}=3)$. 
\item Fit the high-scale LHC data using $N_{F}=4,5,6$ ``VFNS'' PDFs,
$f_{i}(x,\mu,N_{F}=4,5,6)$ and $\alpha_{s}(\mu,N_{F}=4,5,6)$.
\item Repeat steps 1) through 3) until we have a suitable minimum.
\end{enumerate}
Note, because we generate all the PDFs and $\alpha_{s}$ for all $N_{F}=\{3,4,5,6\}$
flavors in step 1), the separate $N_{F}$ branches are analytically
related. Furthermore, this is done using a ``forward'' DGLAP evolution;
no ``backward'' DGLAP evolution is required.

Also note that because we have access to all $N_{F}=\{3,4,5,6\}$
sets, there is no difficulty in performing the HERA analysis of step
2) and the LHC analysis of step 3) in different $N_{F}$ frameworks.

Finally, as we demonstrated in Sec\@.~\ref{sec:Physical-Structure-Functions},
the user is now responsible for ensuring each $N_{F}$ calculation
is not used beyond its range of validity. While it is now possible to
compute with $N_{F}=3$ at high $\mu$ scales,
this does not necessarily give a reliable result for the cross sections.

\subsubsection*{$N_{F}$ Conversion Factors}

Finally, we demonstrate how to use the family of $N_{F}$ dependent
PDFs to estimate the effect of changing from $N_{F}=3$ to $N_{F}=5$
in a calculation such as the extraction of $F_{2}^{charm}$ discussed
above. For example, the HVQDIS program~\cite{Harris:1995tu,Harris:1997zq}
works in a $N_{F}=3$ FFNS while many of the PDFs are only available
for $N_{F}=4,5$. If we have access to both
$N_{F}=3$ and $N_{F}=5$ PDFs, we can simply use the correct $N_{F}$
PDF set, and the conversion between the different $N_{F}$ sets is
simply given by the following identity: 
\[
f^{(N_{F}=5)}(x) = f^{(N_{F}=3)}(x)\left[\frac{f^{(N_{F}=5)}(x)}{f^{(N_{F}=3)}(x)}\right].
\]
The term in brackets above represents the ``correction factor''
in converting between $N_{F}=3$ and $N_{F}=5$ PDF sets. 

\begin{figure*}[t]
\begin{centering}
\includegraphics[width=0.4\textwidth]{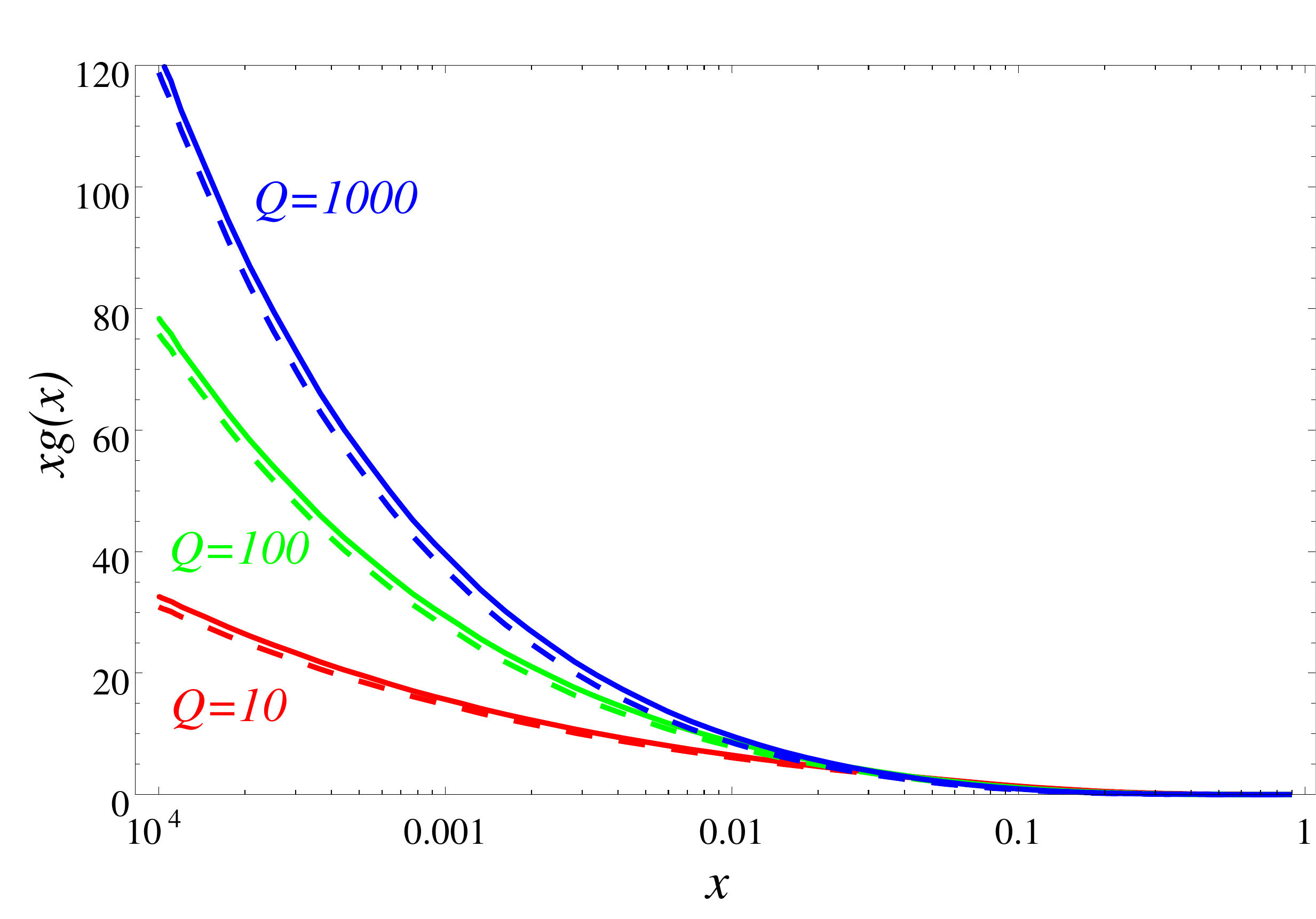}\quad{}\includegraphics[width=0.4\textwidth]{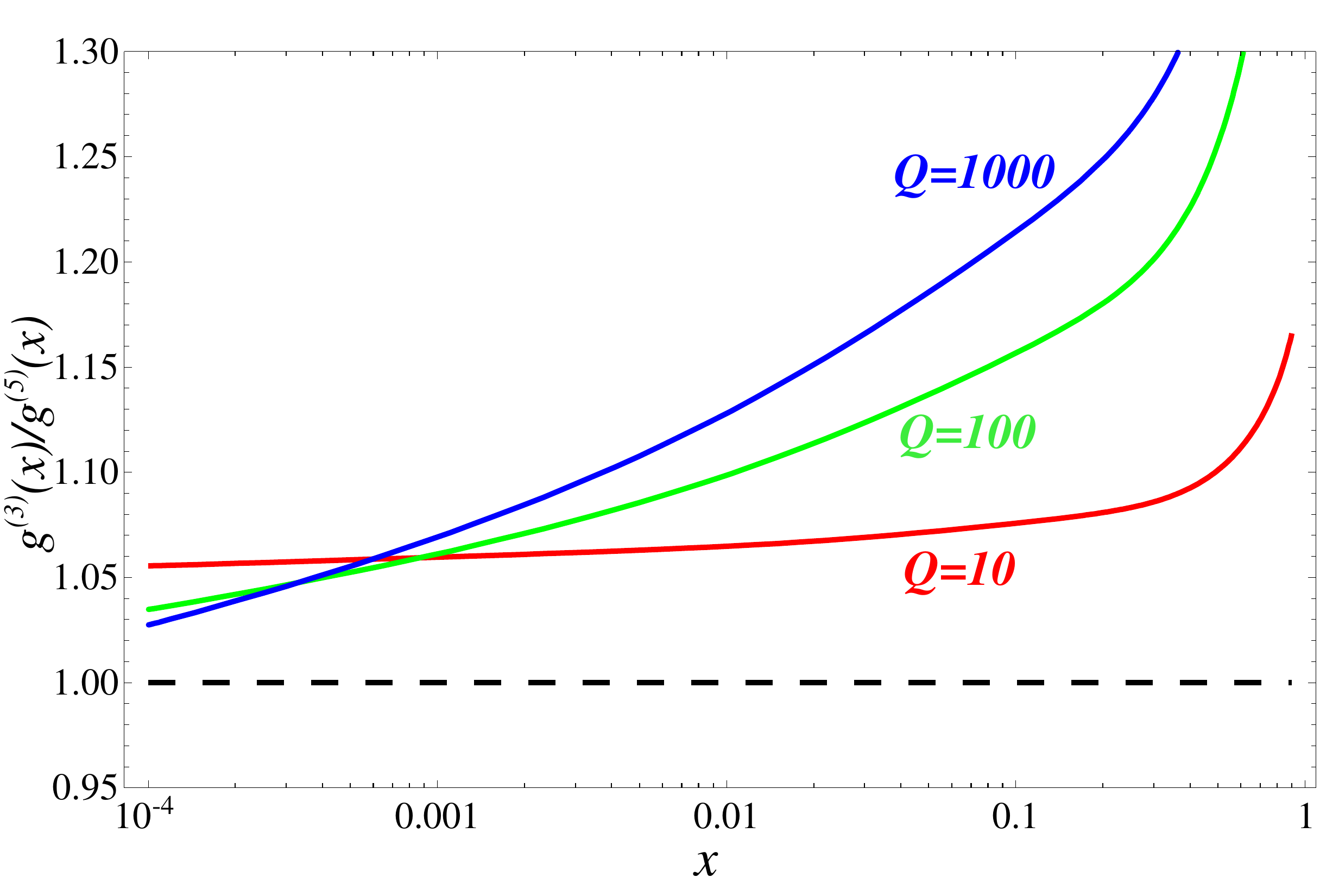} 
\par\end{centering}
\caption{(a) Comparison of 3-flavor (solid lines) and 5-flavor (dashed lines)
gluon for $Q=\{10,100,1000\}$ GeV; (b) ratio of 3- to 5-flavor gluon
for the corresponding $Q$ values.\label{fig:gluon-3vs5FLV}}
\end{figure*}

As we noted in Sec.~\ref{sub:nf_PDF}, the dominant effect of changing
from $N_{F}=3$ to $N_{F}=5$ was to deplete the gluon PDF which fed
the charm PDF via the $g\to c\bar{c}$ process. Therefore, we can
estimate this effect by comparing the shift of the gluon PDF for $N_{F}=3$
and $N_{F}=5$. This effect is shown in Fig.~\ref{fig:gluon-3vs5FLV}-a
where we plot the gluon PDF explicitly, and in Fig.~\ref{fig:gluon-3vs5FLV}-b
we plot the ratio. We see that even at the lowest $Q$ value displayed
(10~GeV) the shift in the gluon PDF is $\sim6\%$ and relatively
insensitive to $x$, except for the highest $x$ values. Because the
$x$ dependence is minimal, we can approximately extract this correction
factor from the convolution of the PDFs; thus, at scales $\mu\lesssim10$
GeV, we can estimate the effect of the $N_{F}=3$ to $N_{F}=5$ conversion
by simply rescaling the gluon PDF. 

For example, if we are looking at charm structure functions, this
is driven by the $\gamma g\to c\bar{c}$ process, plus higher order
corrections. Since this process is linear in the gluon PDF, the effect
would be approximately a constant overall shift; specifically, 6\%
for the case of $Q\sim10$~GeV.

Even if we do not have access to both the $N_{F}=3$ and $N_{F}=5$
PDF sets, the combination $[f^{(N_{F}=5)}/f^{(N_{F}=3)}]$ is driven
by the DGLAP evolution and only mildly sensitive to the detailed PDF;
hence, the above technique can still provide a rough approximation
as to the correction factor between the $N_{F}=3$ and $N_{F}=5$
PDFs.

\section{Conclusion\label{sec:conclusions}}

We have investigated the $N_{F}$ dependence of the
PDFs and proposed an extension of the traditional VFNS which we denote
the \hbox{H-VFNS}. In this scheme, we include an explicit $N_{F}$ dependence
in both the PDFs $f_{a}(x,\mu,N_{F})$ and strong coupling $\alpha_{S}(\mu,N_{F})$;
this provides the user the freedom, and responsibility, to choose
the appropriate $N_{F}$ values for each data set and kinematic region.

Our \hbox{H-VFNS} implementation economically requires only
four PDF grids (for $N_{F}=\{3,4,5,6\}$), yet provides the user flexibility
to use any switching scale. For a practical implementation of the
\hbox{H-VFNS}, we choose a fixed matching scale $\mum{\nf}=m_{\nf}$ and
demonstrate that this has minimal impact on the physical results.

The \hbox{H-VFNS} is able to simultaneously work with low
energy data (e.g., HERA data at low $Q^{2}$) in a $N_{F}=3$ FFNS
framework, while also incorporating high-scale LHC data in a $N_{F}=\{3,4,5,6\}$
framework. Additionally, this can be implemented without any backward
DGLAP evolution.

Although the PDFs and $\alpha_{S}(\mu,N_{F})$ are
discontinuous across flavor thresholds at higher orders, the \hbox{H-VFNS}
provides the user the flexibility to shift the $N_{F}$ transition
for individual data sets and kinematic regions to avoid complications.

Thus, the \hbox{H-VFNS} provides a valuable tool for fitting
data across a wide variety of processes and energy scales from low
to high.

\section*{Acknowledgments}

We thank
Sergey Alekhin,
Michiel Botje,
John Collins,
Kateria Lipka,
Pavel Nadolsky,
Voica Radescu,
Randall Scalise, 
and
the members of the HERA-Fitter group
for valuable discussions.
F.I.O., I.S., and J.Y.Y. acknowledge the hospitality of CERN, DESY,
Fermilab, and Les Houches where a portion of this work was performed.
This work was partially supported by the U.S. Department of Energy
under grant {DE-FG02-13ER41996}, and the Lighter Sams
Foundation. The research of T.S. is supported by a fellowship from
the Théorie LHC France initiative funded by the CNRS/IN2P3. This work
has been supported by \textit{Projet international de cooperation
scientifique} PICS05854 between France and the USA. T.J. was supported
by the Research Executive Agency (REA) of the European Union under
the Grant Agreement number PITN-GA-2010-264564 (LHCPhenoNet).

\appendix

\section{Evolution and Matching Conditions \label{sec:App:PDF_Alphas}}

\subsection{$\alphas$ Evolution \& Matching Conditions\label{sub:App:alphas}}

The running of the $\alphas(\mu,N_{F})$ is given by the renormalization
group equation: 
\[
\mu^{2}\frac{d\alpha_{s}}{d\mu^{2}}=\beta(\alpha_{s})=-\left(b_{0}\alpha_{s}^{2}+b_{1}\alpha_{s}^{3}+b_{2}\alpha_{s}^{4}+...\right).
\]
 At the NLO (2-loop) level, which we use in this work, we obtain

\begin{eqnarray}
\alphas(\mu^{2},N_{F}) & = & \frac{1}{b_{0}\ln\frac{\mu^{2}}{\Lambda^{2}}}\left(1-\frac{b_{1}}{b_{0}^{2}}\frac{\ln\left(\ln\mbox{\ensuremath{\frac{\mu^{2}}{\Lambda^{2}}}}\right)}{\ln\mbox{\ensuremath{\frac{\mu^{2}}{\Lambda^{2}}}}}\right),\nonumber \\
\end{eqnarray}
where $b_{0}=(33-2N_{F})/12\pi$ and $b_{1}=(153-19N_{F})/24\pi^{2}$.
The $N_{F}$ dependence arises from the virtual quark loops which
enter at higher orders.

The relation of $\alpha_{s}$ across flavor thresholds for $N_{F}$
and $N_{F}+1$ flavors is computed to be~\cite{Beringer:1900zz}:

\begin{eqnarray*}
\alpha_{S}(\mu^{2},N_{F+1}) & = & \alpha_{S}(\mu^{2},N_{F})\,\biggl[1+\\
 &  & \left.\sum_{k=1}^{\infty}\sum_{\ell=0}^{k}\, c_{k\ell}\,\left[\alpha_{S}(\mu^{2},N_{F})\right]^{k}\,\ln^{l}\left(\frac{\mu^{2}}{m^{2}}\right)\right],
\end{eqnarray*}
 where $c_{10}=0$ and $c_{20}=-11/72\pi^{2}$. Thus, even if we perform
the matching at $\mu=m$ we find 
\[
\alpha_{S}(m^{2},N_{F}+1)=\alpha_{S}(m^{2},N_{F})\,+c_{20}\alpha_{S}^{3}(m^{2},N_{F})
\]
 such that there is a ${\cal O}\left(\alpha_{S}^{3}\right)$ discontinuity
in $\alpha_{S}$.

In the above, the $\mu$ scale appearing in the argument
of $\alpha_{S}$ is more precisely the renormalization scale $\mu_{R}$;
this is distinguished from the factorization scale $\mu_{F}$ appearing
in the argument of PDF. However, in this work, we choose to set $\mu_{R}=\mu_{F}=\mu$.

Note also that there are in fact two versions of
the FFNS scheme, which are characterized by different treatment of
the number of active flavors entering $\alphas$ (denoted here as
$N_{R}$, to be distinguished from the number of flavors entering
PDF evolution $N_{F}$). In the ``classical'' FFNS, $N_{R}=N_{F}$.
In the modified version, $N_{R}$ is incremented across flavor thresholds
as in the VFNS while $N_{F}$ remains fixed. Discussion of advantages
and disadvantages of these two formulations of FFNS can be found in~\cite{Martin:2006qz,Gluck:2006ju}.
In particular, allowing $N_{R}$ to vary can help the running $\alphas$
accommodate experimental constraints from both high ($\sim M_{Z}$)
and low ($\sim m_{\tau}$) scales~\cite{Beringer:1900zz,Berger:2010rj}.

\subsection{PDF Evolution \& Matching Conditions\label{sub:App:pdf}}

\begin{figure*}[t]
\includegraphics[clip,width=0.3\textwidth]{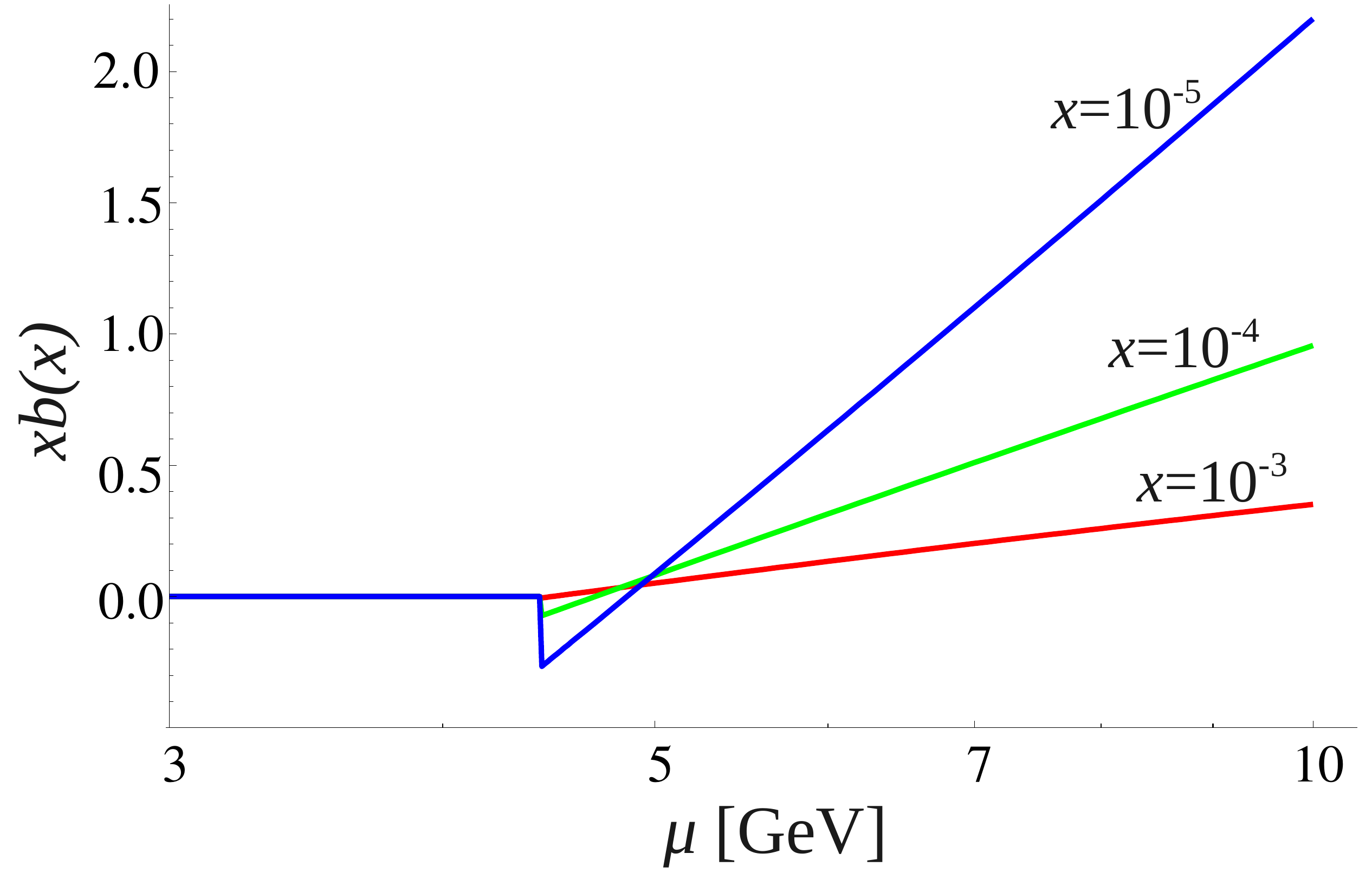}\quad{}\includegraphics[width=0.3\textwidth]{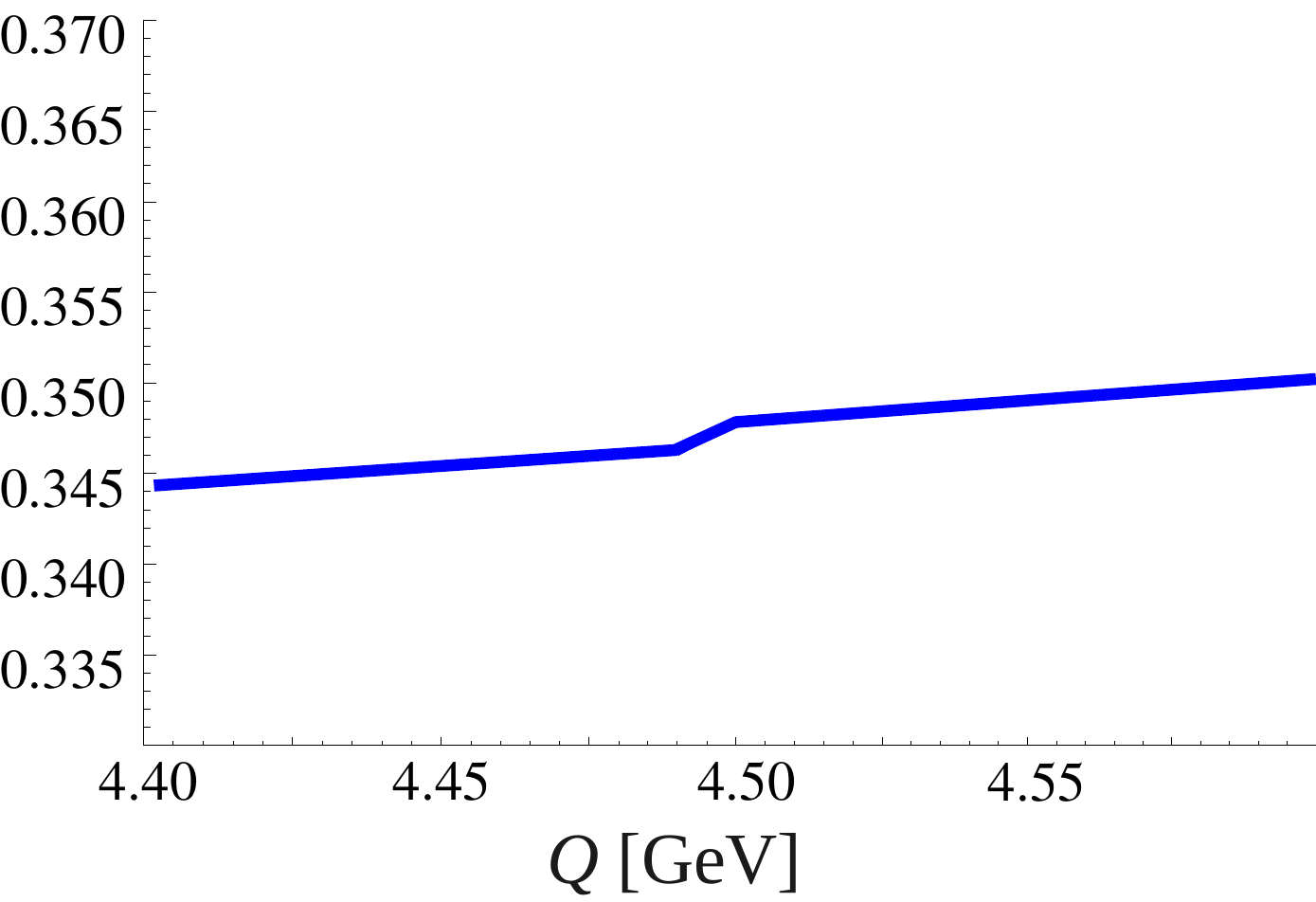}\quad{}\includegraphics[width=0.293\textwidth]{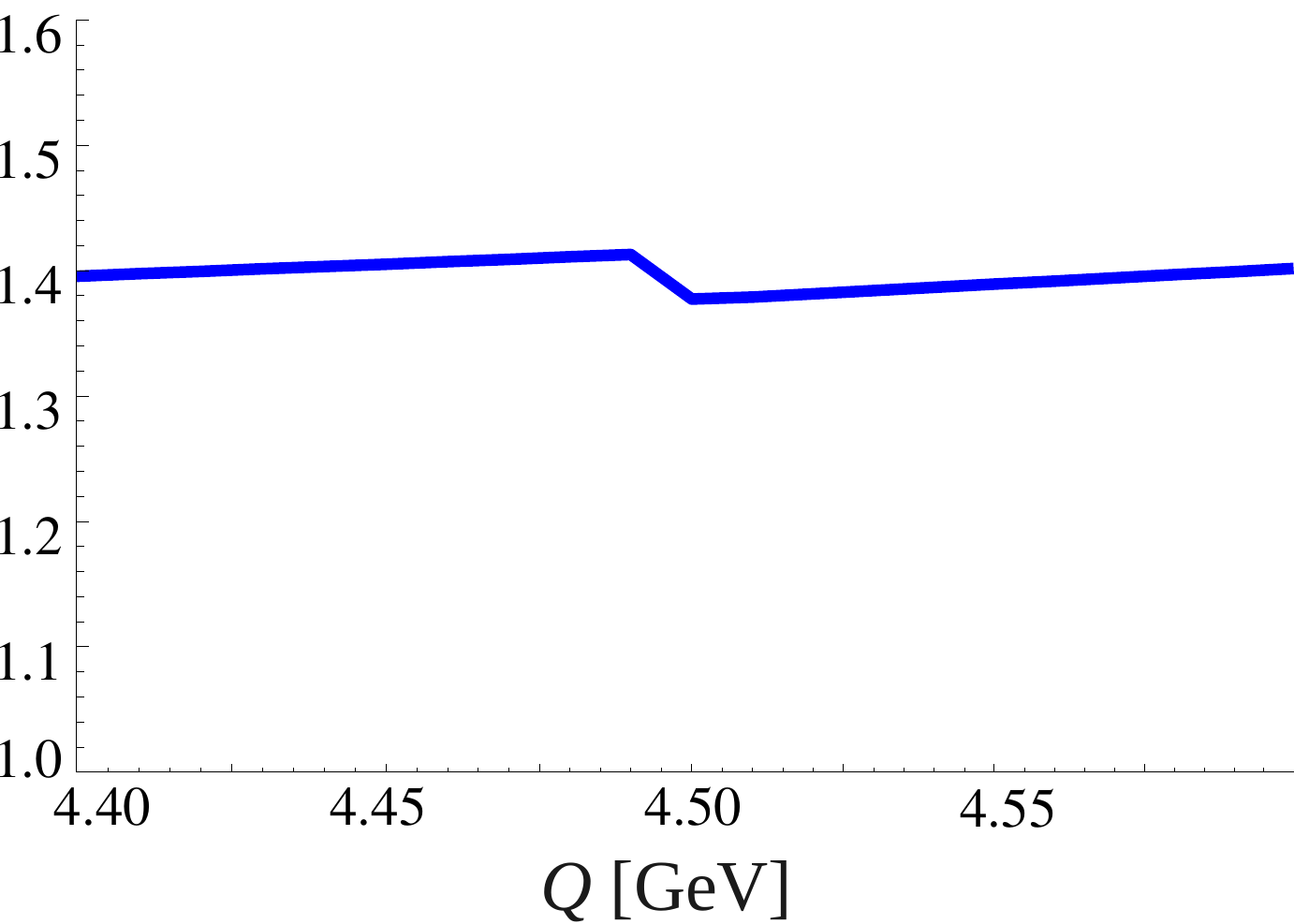}
\caption{a) Discontinuity in the $b$-quark PDF $f_{b}(x,\mu)$ at NNLO vs.
$\mu$ for $x$=\{$10^{-3},10^{-4},10^{-5}\}$. b) \& c) Discontinuity
for $F_{L}$ vs. $Q$ for $x$=\{$10^{-3},10^{-5}\}$ (left to right)
at NNLO in the region of the bottom mass, $m_{b}=4.5$~GeV as computed
in Ref.~\cite{Stavreva:2012bs}. \label{fig:F2Ldisc}}
\end{figure*}

\begin{figure*}[t]
\subfloat[Comparison of charm NLO matching and DGLAP PDFs for two choices of
$x$.]{\centering{}\includegraphics[width=0.2\textwidth]{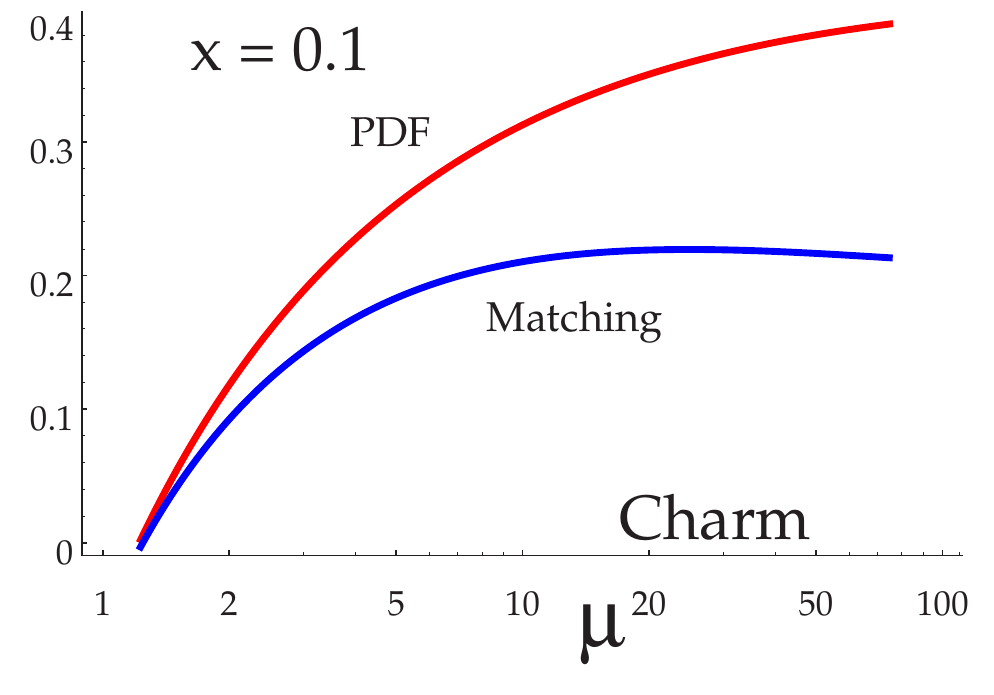}\quad{}\includegraphics[bb=0bp 0bp 286bp 200bp,width=0.2\textwidth]{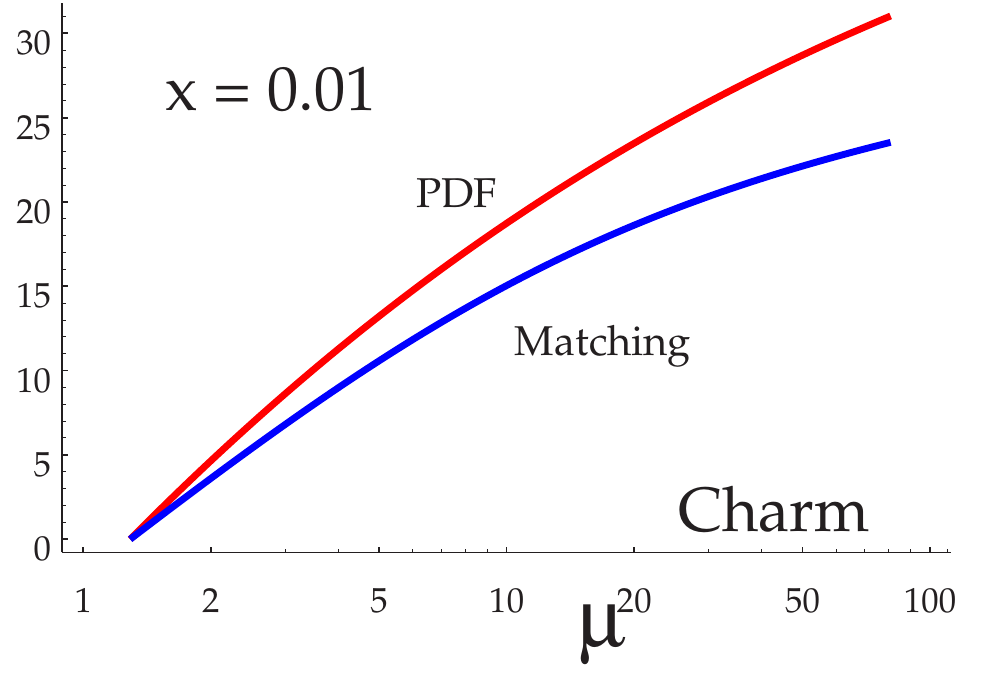}

}\subfloat[Comparison of bottom NLO matching and DGLAP PDFs for two choices of
$x$.]{\centering{}\includegraphics[width=0.2\textwidth]{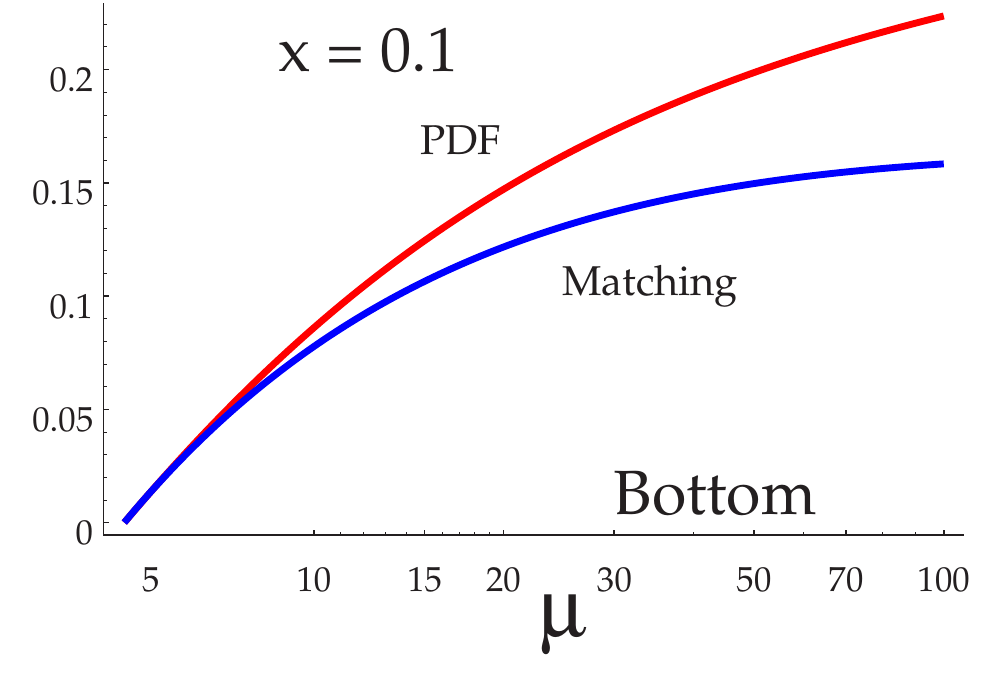}\quad{}\includegraphics[bb=0bp 0bp 286bp 200bp,width=0.2\textwidth]{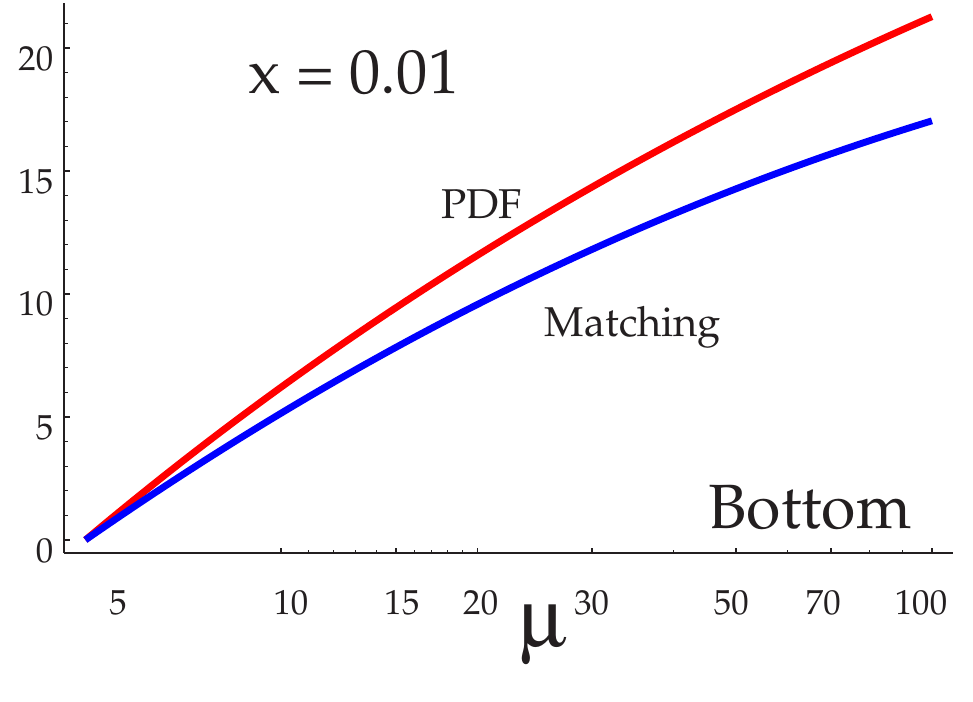}

}

\caption{Comparison of charm and bottom matching conditions at NLO with the
DGLAP evolved PDFs for $x=\{0.1,0.01\}$.
\label{fig:pertVSpdf}}
\end{figure*}

The relation of the PDFs with $N_{F}+1$ flavors to that of $N_{F}$
can be computed perturbatively~\cite{Buza:1995ie,Buza:1996wv}.
The explicit form of these matching conditions can be found
e.g. in eqs.~(2.37)-(2.41) and Appendix~B of Ref.~\cite{Buza:1996wv}.
For the purpose of further discussion we show here
only a symbolic form of the matching conditions
\begin{equation}
\label{eq:fmatching}
\tilde{f}_{i}(x,\mu,N_{F}+1) = A^{ij}\otimes\, \tilde{f}_{j}(x,\mu,N_{F}),\\
\end{equation}
where
\begin{equation}
\label{eq:aij}
\begin{split}
A^{ij} &= \delta^{ij} + \frac{\alphas}{2\pi}\left(a_{1}^{ij}+b_1^{ij}\,\ln\left[\frac{\mu^{2}}{m^{2}}\right]\right)
\\&\quad
+ \left(\frac{\alphas}{2\pi}\right)^{2}\,\left(a_{2}^{ij}
      + b_2^{ij}\,\ln\left[\frac{\mu^{2}}{m^{2}}\right] + c_2^{ij}\,\ln^{2}\left[\frac{\mu^{2}}{m^{2}}\right]\right)+...
\end{split}
\end{equation}
In the above equation $\tilde{f}_i$ can be a combination of
light parton densities ($f_i+f_{\bar{i}}$),
heavy parton densities ($f_H+f_{\bar{H}}$),
the singlet combination of parton densities $\Sigma$,
or the gluon.
Note that there is an implicit summation over the above combinations.
Coefficients $a^{ij}$, $b^{ij}$, $\cdots$ can be computed perturbatively.
While we have not indicated it explicitly, all quantities 
on the RHS of Eq.~\eqref{eq:fmatching} (including $\alpha_S(N_F)$)
are evaluated with $N_F$ flavors, and those 
on the LHS are evaluated with $N_F +1$ flavors.

Note that the QCDNUM~\cite{Botje:2010ay} program includes the NNLO
evolution with the discontinous and NNLO matching conditions.

In the $\msbar$ scheme the $a_{1}^{ij}$ term is computed to be zero, while the $a_{2}^{ij}$
term is non-zero. Because $a_{1}^{ij}=0$, if we perform the matching
between $N_{F}$ and $N_{F}+1$ flavors at $\mu=m$, the $\ln(\mu/m)$
terms vanish and we find at NLO {[}${\cal O}(\alphas^{1})${]} that
$f_{i}(x,\mu=m,N_{F}+1)=f_{i}(x,\mu=m,N_{F})$; that is, the PDFs
are continuous. This is why, at NLO, the VFNS implemented the matching
automatically at $\mu=m$. Because $a_{2}^{ij}\not=0$, at NNLO and
beyond the PDFs will acquire discontinuities of ${\cal O}(\alphas^{2})$;
therefore, there is no longer any special benefit obtained by forcing
the $N_{F}$ transition at $\mu=m$.

For example, the discontinuity of the $b$-quark PDF is shown in Fig.~\ref{fig:F2Ldisc}-a,
and curiously this yields a slightly negative value just above the
transition point for $f_{b}(x,\mu\gtrsim m,N_{F}=5)$. There is a
corresponding discontinuity in the gluon PDF (not shown) which has
a positive shift, as it must to ensure the PDF sum rules are satisfied.

These discontinuities exhibit themselves in the physical observables
such as the structure functions as shown in Fig.~\ref{fig:F2Ldisc}-b
and Fig.~\ref{fig:F2Ldisc}-c. These discontinuities are formally
higher order, and will be reduced order by order as we extend the
perturbation theory. It is interesting to note that $F_{L}$ for the
larger $x$ value ($10^{-3}$) has a slightly positive discontinuity
while at the smaller $x$ value ($10^{-5}$) the discontinuity is
negative. This reflects the shift between the (positive) gluon and
the (negative) quark contributions in the different $x$ regions.
It is this mixture of the gluon and the quark terms which will ensure
the physical observable is continuous up to the specified order of
perturbation theory, while the PDF will always remain discontinuous
at ${\cal O}(\alphas^{2})$.

In the presented \hbox{H-VFNS}, we choose to compute the matching between
$N_{F}$ and $N_{F}+1$ flavors at $\mu=m$ (because the logs vanish);
however, since we retain both the $N_{F}$ and $N_{F}+1$ PDFs for
$\mu\geq m$, the user has the choice to compute in either the $N_{F}$
or $N_{F}+1$ framework, whichever is more suitable. Because the
traditional VFNS did not provide PDFs for $N_{F}$ flavors at $\mu\geq m$,
this was previously not an option.

The matching conditions of Eqs.~\eqref{eq:fmatching} and \eqref{eq:aij}
essentially represent a perturbative expansion of the DGLAP evolution
equations, up to an additional constant term $a_{k}^{ij}$.

We observe that if we choose to perform the matching not at $\mu=m$
but instead at a higher scale such as $\mu=2m$, the PDF boundary
condition for the heavy quark is not $f_{c}(x,\mu,N_{F}=4)=0$. Instead,
the correct condition at NLO is: 
\begin{eqnarray}
f_{c}(x,\mu,N_{F} & = & 4)\simeq0+\frac{\alphas(\mu,N_{F}=3)}{2\pi}\,\ln\left[\frac{\mu^{2}}{m_{c}^{2}}\right]\,\times\nonumber \\
 &  & \times\quad P_{qg}\otimes g(x,\mu,N_{F}=3)+...\label{eq:nloMatching}
\end{eqnarray}
 Note, the LHS uses $N_{F}=4$ PDFs and the RHS uses $N_{F}=3$ PDFs.

These matching conditions are displayed in Fig.~\ref{fig:pertVSpdf}
where we compare these to the DGLAP evolved PDF distribution at NLO.
We see for scales near the matching point $\mu\sim m$, the differences
are small. However, if the matching is performed away from the $\mu\sim m$
region, then the differences are larger. This is because the matching
of Eq.~\eqref{eq:nloMatching} is only computed to NLO, so it only
includes a single partonic splitting, while the DGLAP evolution resums
an infinite tower of partonic emissions. The difference comes from
the missing second-order splittings which are proportional to
$\alphas^{2}\,\ln(\mu/m)$. If we repeat this exercise and compute
the matching to NNLO,%
\footnote{ An example of NNLO matching is provided in Ref.~\cite{Alekhin:2009ni}.%
} then we will include the $\alphas^{2}\,\ln(\mu/m)$ contributions,
but miss the $\alphas^{3}\,\ln(\mu/m)$. Thus the curves in Fig.~\ref{fig:pertVSpdf}
will remain comparable for a larger range of $\mu\gtrsim m$.

In this analysis, our matching scale is always taken to be the quark
mass, $\mum{\nf} = m_{\nf}$. This provides us the benefit that the
PDF with $\nf$ active flavors is defined for all values above
$\mu=m_{\nf}$ without invoking backward-evolution.

In the traditional VFNS, the switching scale $\mut{\nf}$ was forced
to be equal to the matching scale, which was set to the quark masses:
$\mut{\nf}=\mum{\nf}=m_{\nf}$. For the \hbox{H-VFNS}, the switching scale
$\mut{\nf}$ is not predefined by the PDF set but can freely be
chosen by the user.

The resulting PDFs will, to some extent, depend on the matching scale
$\mum{\nf}$, but as Fig.~\ref{fig:pertVSpdf} demonstrates this
effect will be insignificant so long as $\mum{\nf}\sim m_{\nf}$.
Likewise, resulting observables will, to some extent, depend on the
switching scale $\mut{\nf}$, but as Figs.~\ref{fig:F2} and~\ref{fig:FL}
demonstrate this effect will be insignificant so long as we do stay
within the region of validity.

\bibliographystyle{apsrev}
\bibliography{bibNF}

\begin{thebibliography}{57}
\expandafter\ifx\csname natexlab\endcsname\relax\def\natexlab#1{#1}\fi
\expandafter\ifx\csname bibnamefont\endcsname\relax
  \def\bibnamefont#1{#1}\fi
\expandafter\ifx\csname bibfnamefont\endcsname\relax
  \def\bibfnamefont#1{#1}\fi
\expandafter\ifx\csname citenamefont\endcsname\relax
  \def\citenamefont#1{#1}\fi
\expandafter\ifx\csname url\endcsname\relax
  \def\url#1{\texttt{#1}}\fi
\expandafter\ifx\csname urlprefix\endcsname\relax\def\urlprefix{URL }\fi
\providecommand{\bibinfo}[2]{#2}
\providecommand{\eprint}[2][]{\url{#2}}

\bibitem[{\citenamefont{Lai et~al.}(2010)\citenamefont{Lai, Guzzi, Huston, Li,
  Nadolsky et~al.}}]{Lai:2010vv}
\bibinfo{author}{\bibfnamefont{H.-L.} \bibnamefont{Lai}},
  \bibinfo{author}{\bibfnamefont{M.}~\bibnamefont{Guzzi}},
  \bibinfo{author}{\bibfnamefont{J.}~\bibnamefont{Huston}},
  \bibinfo{author}{\bibfnamefont{Z.}~\bibnamefont{Li}},
  \bibinfo{author}{\bibfnamefont{P.~M.} \bibnamefont{Nadolsky}},
  \bibnamefont{et~al.}, \bibinfo{journal}{Phys.Rev.}
  \textbf{\bibinfo{volume}{D82}}, \bibinfo{pages}{074024}
  (\bibinfo{year}{2010}), \eprint{1007.2241}.

\bibitem[{\citenamefont{Gao et~al.}(2013)\citenamefont{Gao, Guzzi, Huston, Lai,
  Li et~al.}}]{Gao:2013xoa}
\bibinfo{author}{\bibfnamefont{J.}~\bibnamefont{Gao}},
  \bibinfo{author}{\bibfnamefont{M.}~\bibnamefont{Guzzi}},
  \bibinfo{author}{\bibfnamefont{J.}~\bibnamefont{Huston}},
  \bibinfo{author}{\bibfnamefont{H.-L.} \bibnamefont{Lai}},
  \bibinfo{author}{\bibfnamefont{Z.}~\bibnamefont{Li}}, \bibnamefont{et~al.}
  (\bibinfo{year}{2013}), \eprint{1302.6246}.

\bibitem[{\citenamefont{Schienbein et~al.}(2008)}]{Schienbein:2007fs}
\bibinfo{author}{\bibfnamefont{I.}~\bibnamefont{Schienbein}}
  \bibnamefont{et~al.}, \bibinfo{journal}{Phys. Rev.}
  \textbf{\bibinfo{volume}{D77}}, \bibinfo{pages}{054013}
  (\bibinfo{year}{2008}), \eprint{0710.4897}.

\bibitem[{\citenamefont{Schienbein et~al.}(2009)\citenamefont{Schienbein, Yu,
  Kovarik, Keppel, Morfin et~al.}}]{Schienbein:2009kk}
\bibinfo{author}{\bibfnamefont{I.}~\bibnamefont{Schienbein}},
  \bibinfo{author}{\bibfnamefont{J.}~\bibnamefont{Yu}},
  \bibinfo{author}{\bibfnamefont{K.}~\bibnamefont{Kovarik}},
  \bibinfo{author}{\bibfnamefont{C.}~\bibnamefont{Keppel}},
  \bibinfo{author}{\bibfnamefont{J.}~\bibnamefont{Morfin}},
  \bibnamefont{et~al.}, \bibinfo{journal}{Phys.Rev.}
  \textbf{\bibinfo{volume}{D80}}, \bibinfo{pages}{094004}
  (\bibinfo{year}{2009}), \eprint{0907.2357}.

\bibitem[{\citenamefont{Aivazis
  et~al.}(1994{\natexlab{a}})\citenamefont{Aivazis, Olness, and
  Tung}}]{Aivazis:1993kh}
\bibinfo{author}{\bibfnamefont{M.~A.~G.} \bibnamefont{Aivazis}},
  \bibinfo{author}{\bibfnamefont{F.~I.} \bibnamefont{Olness}},
  \bibnamefont{and} \bibinfo{author}{\bibfnamefont{W.-K.} \bibnamefont{Tung}},
  \bibinfo{journal}{Phys. Rev.} \textbf{\bibinfo{volume}{D50}},
  \bibinfo{pages}{3085} (\bibinfo{year}{1994}{\natexlab{a}}),
  \eprint{hep-ph/9312318}.

\bibitem[{\citenamefont{Aivazis
  et~al.}(1994{\natexlab{b}})\citenamefont{Aivazis, Collins, Olness, and
  Tung}}]{Aivazis:1993pi}
\bibinfo{author}{\bibfnamefont{M.}~\bibnamefont{Aivazis}},
  \bibinfo{author}{\bibfnamefont{J.~C.} \bibnamefont{Collins}},
  \bibinfo{author}{\bibfnamefont{F.~I.} \bibnamefont{Olness}},
  \bibnamefont{and} \bibinfo{author}{\bibfnamefont{W.-K.} \bibnamefont{Tung}},
  \bibinfo{journal}{Phys.Rev.} \textbf{\bibinfo{volume}{D50}},
  \bibinfo{pages}{3102} (\bibinfo{year}{1994}{\natexlab{b}}),
  \eprint{hep-ph/9312319}.

\bibitem[{\citenamefont{Kramer et~al.}(2000)\citenamefont{Kramer, Olness, and
  Soper}}]{Kramer:2000hn}
\bibinfo{author}{\bibfnamefont{M.}~\bibnamefont{Kramer}},
  \bibinfo{author}{\bibfnamefont{F.~I.} \bibnamefont{Olness}},
  \bibnamefont{and} \bibinfo{author}{\bibfnamefont{D.~E.} \bibnamefont{Soper}},
  \bibinfo{journal}{Phys.Rev.} \textbf{\bibinfo{volume}{D62}},
  \bibinfo{pages}{096007} (\bibinfo{year}{2000}), \eprint{hep-ph/0003035}.

\bibitem[{\citenamefont{Tung et~al.}(2002)\citenamefont{Tung, Kretzer, and
  Schmidt}}]{Tung:2001mv}
\bibinfo{author}{\bibfnamefont{W.-K.} \bibnamefont{Tung}},
  \bibinfo{author}{\bibfnamefont{S.}~\bibnamefont{Kretzer}}, \bibnamefont{and}
  \bibinfo{author}{\bibfnamefont{C.}~\bibnamefont{Schmidt}},
  \bibinfo{journal}{J.Phys.} \textbf{\bibinfo{volume}{G28}},
  \bibinfo{pages}{983} (\bibinfo{year}{2002}), \eprint{hep-ph/0110247}.

\bibitem[{\citenamefont{Kretzer and Schienbein}(1998)}]{Kretzer:1998ju}
\bibinfo{author}{\bibfnamefont{S.}~\bibnamefont{Kretzer}} \bibnamefont{and}
  \bibinfo{author}{\bibfnamefont{I.}~\bibnamefont{Schienbein}},
  \bibinfo{journal}{Phys.Rev.} \textbf{\bibinfo{volume}{D58}},
  \bibinfo{pages}{094035} (\bibinfo{year}{1998}), \eprint{hep-ph/9805233}.

\bibitem[{\citenamefont{Guzzi et~al.}(2012)\citenamefont{Guzzi, Nadolsky, Lai,
  and Yuan}}]{Guzzi:2011ew}
\bibinfo{author}{\bibfnamefont{M.}~\bibnamefont{Guzzi}},
  \bibinfo{author}{\bibfnamefont{P.~M.} \bibnamefont{Nadolsky}},
  \bibinfo{author}{\bibfnamefont{H.-L.} \bibnamefont{Lai}}, \bibnamefont{and}
  \bibinfo{author}{\bibfnamefont{C.-P.} \bibnamefont{Yuan}},
  \bibinfo{journal}{Phys.Rev.} \textbf{\bibinfo{volume}{D86}},
  \bibinfo{pages}{053005} (\bibinfo{year}{2012}), \eprint{1108.5112}.

\bibitem[{\citenamefont{Stavreva et~al.}(2012)\citenamefont{Stavreva, Olness,
  Schienbein, Jezo, Kusina et~al.}}]{Stavreva:2012bs}
\bibinfo{author}{\bibfnamefont{T.}~\bibnamefont{Stavreva}},
  \bibinfo{author}{\bibfnamefont{F.}~\bibnamefont{Olness}},
  \bibinfo{author}{\bibfnamefont{I.}~\bibnamefont{Schienbein}},
  \bibinfo{author}{\bibfnamefont{T.}~\bibnamefont{Jezo}},
  \bibinfo{author}{\bibfnamefont{A.}~\bibnamefont{Kusina}},
  \bibnamefont{et~al.}, \bibinfo{journal}{Phys.Rev.}
  \textbf{\bibinfo{volume}{D85}}, \bibinfo{pages}{114014}
  (\bibinfo{year}{2012}), \eprint{1203.0282}.

\bibitem[{\citenamefont{Kotko and
  Slominski}(2012{\natexlab{a}})}]{Kotko:2012ui}
\bibinfo{author}{\bibfnamefont{P.}~\bibnamefont{Kotko}} \bibnamefont{and}
  \bibinfo{author}{\bibfnamefont{W.}~\bibnamefont{Slominski}},
  \bibinfo{journal}{Phys.Rev.} \textbf{\bibinfo{volume}{D86}},
  \bibinfo{pages}{094008} (\bibinfo{year}{2012}{\natexlab{a}}),
  \eprint{1206.4024}.

\bibitem[{\citenamefont{Kotko and
  Slominski}(2012{\natexlab{b}})}]{Kotko:2012kw}
\bibinfo{author}{\bibfnamefont{P.}~\bibnamefont{Kotko}} \bibnamefont{and}
  \bibinfo{author}{\bibfnamefont{W.}~\bibnamefont{Slominski}}, pp.
  \bibinfo{pages}{819--822} (\bibinfo{year}{2012}{\natexlab{b}}),
  \eprint{1206.3517}.

\bibitem[{\citenamefont{Kniehl et~al.}(2011)\citenamefont{Kniehl, Kramer,
  Schienbein, and Spiesberger}}]{Kniehl:2011bk}
\bibinfo{author}{\bibfnamefont{B.}~\bibnamefont{Kniehl}},
  \bibinfo{author}{\bibfnamefont{G.}~\bibnamefont{Kramer}},
  \bibinfo{author}{\bibfnamefont{I.}~\bibnamefont{Schienbein}},
  \bibnamefont{and}
  \bibinfo{author}{\bibfnamefont{H.}~\bibnamefont{Spiesberger}},
  \bibinfo{journal}{Phys.Rev.} \textbf{\bibinfo{volume}{D84}},
  \bibinfo{pages}{094026} (\bibinfo{year}{2011}), \eprint{1109.2472}.

\bibitem[{\citenamefont{Martin et~al.}(2009)\citenamefont{Martin, Stirling,
  Thorne, and Watt}}]{Martin:2009iq}
\bibinfo{author}{\bibfnamefont{A.~D.} \bibnamefont{Martin}},
  \bibinfo{author}{\bibfnamefont{W.~J.} \bibnamefont{Stirling}},
  \bibinfo{author}{\bibfnamefont{R.~S.} \bibnamefont{Thorne}},
  \bibnamefont{and} \bibinfo{author}{\bibfnamefont{G.}~\bibnamefont{Watt}},
  \bibinfo{journal}{Eur. Phys. J.} \textbf{\bibinfo{volume}{C63}},
  \bibinfo{pages}{189} (\bibinfo{year}{2009}), \eprint{0901.0002}.

\bibitem[{\citenamefont{Thorne and Roberts}(1998)}]{Thorne:1998ga}
\bibinfo{author}{\bibfnamefont{R.~S.} \bibnamefont{Thorne}} \bibnamefont{and}
  \bibinfo{author}{\bibfnamefont{R.~G.} \bibnamefont{Roberts}},
  \bibinfo{journal}{Phys. Rev.} \textbf{\bibinfo{volume}{D57}},
  \bibinfo{pages}{6871} (\bibinfo{year}{1998}), \eprint{hep-ph/9709442}.

\bibitem[{\citenamefont{Thorne}(2006)}]{Thorne:2006qt}
\bibinfo{author}{\bibfnamefont{R.}~\bibnamefont{Thorne}},
  \bibinfo{journal}{Phys.Rev.} \textbf{\bibinfo{volume}{D73}},
  \bibinfo{pages}{054019} (\bibinfo{year}{2006}), \eprint{hep-ph/0601245}.

\bibitem[{\citenamefont{Cacciari et~al.}(1998)\citenamefont{Cacciari, Greco,
  and Nason}}]{Cacciari:1998it}
\bibinfo{author}{\bibfnamefont{M.}~\bibnamefont{Cacciari}},
  \bibinfo{author}{\bibfnamefont{M.}~\bibnamefont{Greco}}, \bibnamefont{and}
  \bibinfo{author}{\bibfnamefont{P.}~\bibnamefont{Nason}},
  \bibinfo{journal}{JHEP} \textbf{\bibinfo{volume}{9805}}, \bibinfo{pages}{007}
  (\bibinfo{year}{1998}), \eprint{hep-ph/9803400}.

\bibitem[{\citenamefont{Forte et~al.}(2010)\citenamefont{Forte, Laenen, Nason,
  and Rojo}}]{Forte:2010ta}
\bibinfo{author}{\bibfnamefont{S.}~\bibnamefont{Forte}},
  \bibinfo{author}{\bibfnamefont{E.}~\bibnamefont{Laenen}},
  \bibinfo{author}{\bibfnamefont{P.}~\bibnamefont{Nason}}, \bibnamefont{and}
  \bibinfo{author}{\bibfnamefont{J.}~\bibnamefont{Rojo}},
  \bibinfo{journal}{Nucl.Phys.} \textbf{\bibinfo{volume}{B834}},
  \bibinfo{pages}{116} (\bibinfo{year}{2010}), \eprint{1001.2312}.

\bibitem[{\citenamefont{Ball et~al.}(2011)\citenamefont{Ball, Bertone, Cerutti,
  Del~Debbio, Forte et~al.}}]{Ball:2011mu}
\bibinfo{author}{\bibfnamefont{R.~D.} \bibnamefont{Ball}},
  \bibinfo{author}{\bibfnamefont{V.}~\bibnamefont{Bertone}},
  \bibinfo{author}{\bibfnamefont{F.}~\bibnamefont{Cerutti}},
  \bibinfo{author}{\bibfnamefont{L.}~\bibnamefont{Del~Debbio}},
  \bibinfo{author}{\bibfnamefont{S.}~\bibnamefont{Forte}},
  \bibnamefont{et~al.}, \bibinfo{journal}{Nucl.Phys.}
  \textbf{\bibinfo{volume}{B849}}, \bibinfo{pages}{296} (\bibinfo{year}{2011}),
  \eprint{1101.1300}.

\bibitem[{\citenamefont{Ball et~al.}(2013{\natexlab{a}})\citenamefont{Ball,
  Bertone, Carrazza, Deans, Del~Debbio et~al.}}]{Ball:2012cx}
\bibinfo{author}{\bibfnamefont{R.~D.} \bibnamefont{Ball}},
  \bibinfo{author}{\bibfnamefont{V.}~\bibnamefont{Bertone}},
  \bibinfo{author}{\bibfnamefont{S.}~\bibnamefont{Carrazza}},
  \bibinfo{author}{\bibfnamefont{C.~S.} \bibnamefont{Deans}},
  \bibinfo{author}{\bibfnamefont{L.}~\bibnamefont{Del~Debbio}},
  \bibnamefont{et~al.}, \bibinfo{journal}{Nucl.Phys.}
  \textbf{\bibinfo{volume}{B867}}, \bibinfo{pages}{244}
  (\bibinfo{year}{2013}{\natexlab{a}}), \eprint{1207.1303}.

\bibitem[{\citenamefont{Alekhin et~al.}(2010)\citenamefont{Alekhin, Blumlein,
  Klein, and Moch}}]{Alekhin:2009ni}
\bibinfo{author}{\bibfnamefont{S.}~\bibnamefont{Alekhin}},
  \bibinfo{author}{\bibfnamefont{J.}~\bibnamefont{Blumlein}},
  \bibinfo{author}{\bibfnamefont{S.}~\bibnamefont{Klein}}, \bibnamefont{and}
  \bibinfo{author}{\bibfnamefont{S.}~\bibnamefont{Moch}},
  \bibinfo{journal}{Phys.Rev.} \textbf{\bibinfo{volume}{D81}},
  \bibinfo{pages}{014032} (\bibinfo{year}{2010}), \eprint{0908.2766}.

\bibitem[{\citenamefont{Alekhin et~al.}(2012)\citenamefont{Alekhin, Blumlein,
  and Moch}}]{Alekhin:2012ig}
\bibinfo{author}{\bibfnamefont{S.}~\bibnamefont{Alekhin}},
  \bibinfo{author}{\bibfnamefont{J.}~\bibnamefont{Blumlein}}, \bibnamefont{and}
  \bibinfo{author}{\bibfnamefont{S.}~\bibnamefont{Moch}},
  \bibinfo{journal}{Phys.Rev.} \textbf{\bibinfo{volume}{D86}},
  \bibinfo{pages}{054009} (\bibinfo{year}{2012}), \eprint{1202.2281}.

\bibitem[{\citenamefont{Gluck et~al.}(2008{\natexlab{a}})\citenamefont{Gluck,
  Jimenez-Delgado, and Reya}}]{Gluck:2007ck}
\bibinfo{author}{\bibfnamefont{M.}~\bibnamefont{Gluck}},
  \bibinfo{author}{\bibfnamefont{P.}~\bibnamefont{Jimenez-Delgado}},
  \bibnamefont{and} \bibinfo{author}{\bibfnamefont{E.}~\bibnamefont{Reya}},
  \bibinfo{journal}{Eur.Phys.J.} \textbf{\bibinfo{volume}{C53}},
  \bibinfo{pages}{355} (\bibinfo{year}{2008}{\natexlab{a}}),
  \eprint{0709.0614}.

\bibitem[{\citenamefont{Jimenez-Delgado and
  Reya}(2009{\natexlab{a}})}]{JimenezDelgado:2008hf}
\bibinfo{author}{\bibfnamefont{P.}~\bibnamefont{Jimenez-Delgado}}
  \bibnamefont{and} \bibinfo{author}{\bibfnamefont{E.}~\bibnamefont{Reya}},
  \bibinfo{journal}{Phys. Rev.} \textbf{\bibinfo{volume}{D79}},
  \bibinfo{pages}{074023} (\bibinfo{year}{2009}{\natexlab{a}}),
  \eprint{0810.4274}.

\bibitem[{\citenamefont{Gluck et~al.}(2008{\natexlab{b}})\citenamefont{Gluck,
  Jimenez-Delgado, Reya, and Schuck}}]{Gluck:2008gs}
\bibinfo{author}{\bibfnamefont{M.}~\bibnamefont{Gluck}},
  \bibinfo{author}{\bibfnamefont{P.}~\bibnamefont{Jimenez-Delgado}},
  \bibinfo{author}{\bibfnamefont{E.}~\bibnamefont{Reya}}, \bibnamefont{and}
  \bibinfo{author}{\bibfnamefont{C.}~\bibnamefont{Schuck}},
  \bibinfo{journal}{Phys.Lett.} \textbf{\bibinfo{volume}{B664}},
  \bibinfo{pages}{133} (\bibinfo{year}{2008}{\natexlab{b}}),
  \eprint{0801.3618}.

\bibitem[{\citenamefont{Jimenez-Delgado and
  Reya}(2009{\natexlab{b}})}]{JimenezDelgado:2009tv}
\bibinfo{author}{\bibfnamefont{P.}~\bibnamefont{Jimenez-Delgado}}
  \bibnamefont{and} \bibinfo{author}{\bibfnamefont{E.}~\bibnamefont{Reya}},
  \bibinfo{journal}{Phys. Rev.} \textbf{\bibinfo{volume}{D80}},
  \bibinfo{pages}{114011} (\bibinfo{year}{2009}{\natexlab{b}}),
  \eprint{0909.1711}.

\bibitem[{\citenamefont{Thorne and Tung}(2008)}]{Thorne:2008xf}
\bibinfo{author}{\bibfnamefont{R.}~\bibnamefont{Thorne}} \bibnamefont{and}
  \bibinfo{author}{\bibfnamefont{W.}~\bibnamefont{Tung}}
  (\bibinfo{year}{2008}), \eprint{0809.0714}.

\bibitem[{\citenamefont{Olness and Schienbein}(2009)}]{Olness:2008px}
\bibinfo{author}{\bibfnamefont{F.}~\bibnamefont{Olness}} \bibnamefont{and}
  \bibinfo{author}{\bibfnamefont{I.}~\bibnamefont{Schienbein}},
  \bibinfo{journal}{Nucl.Phys.Proc.Suppl.} \textbf{\bibinfo{volume}{191}},
  \bibinfo{pages}{44} (\bibinfo{year}{2009}), \eprint{0812.3371}.

\bibitem[{\citenamefont{Andersen et~al.}(2010)}]{Binoth:2010ra}
\bibinfo{author}{\bibfnamefont{J.}~\bibnamefont{Andersen}} \bibnamefont{et~al.}
  (\bibinfo{collaboration}{SM and NLO Multileg Working Group}), pp.
  \bibinfo{pages}{21--189} (\bibinfo{year}{2010}), \eprint{1003.1241}.

\bibitem[{\citenamefont{Collins}(1998)}]{Collins:1998rz}
\bibinfo{author}{\bibfnamefont{J.~C.} \bibnamefont{Collins}},
  \bibinfo{journal}{Phys.Rev.} \textbf{\bibinfo{volume}{D58}},
  \bibinfo{pages}{094002} (\bibinfo{year}{1998}), \eprint{hep-ph/9806259}.

\bibitem[{\citenamefont{Amundson et~al.}()\citenamefont{Amundson, Olness,
  Schmidt, Tung, and Wang}}]{Amundson:1998zk}
\bibinfo{author}{\bibfnamefont{J.}~\bibnamefont{Amundson}},
  \bibinfo{author}{\bibfnamefont{F.~I.} \bibnamefont{Olness}},
  \bibinfo{author}{\bibfnamefont{C.}~\bibnamefont{Schmidt}},
  \bibinfo{author}{\bibfnamefont{W.}~\bibnamefont{Tung}}, \bibnamefont{and}
  \bibinfo{author}{\bibfnamefont{X.}~\bibnamefont{Wang}},
  \emph{\bibinfo{title}{{Theoretical description of heavy quark production in
  DIS} \em{(1998)}}},
  \bibinfo{howpublished}{\url{http://lss.fnal.gov/archive/1998/conf/Conf-98-15%
3-T.pdf}}.

\bibitem[{\citenamefont{Collins and Tung}(1986)}]{Collins:1986mp}
\bibinfo{author}{\bibfnamefont{J.~C.} \bibnamefont{Collins}} \bibnamefont{and}
  \bibinfo{author}{\bibfnamefont{W.-K.} \bibnamefont{Tung}},
  \bibinfo{journal}{Nucl.Phys.} \textbf{\bibinfo{volume}{B278}},
  \bibinfo{pages}{934} (\bibinfo{year}{1986}).

\bibitem[{\citenamefont{Olness and Scalise}(1998)}]{Olness:1997yn}
\bibinfo{author}{\bibfnamefont{F.~I.} \bibnamefont{Olness}} \bibnamefont{and}
  \bibinfo{author}{\bibfnamefont{R.~J.} \bibnamefont{Scalise}},
  \bibinfo{journal}{Phys.Rev.} \textbf{\bibinfo{volume}{D57}},
  \bibinfo{pages}{241} (\bibinfo{year}{1998}), \eprint{hep-ph/9707459}.

\bibitem[{\citenamefont{Nadolsky and Tung}(2009)}]{Nadolsky:2009ge}
\bibinfo{author}{\bibfnamefont{P.~M.} \bibnamefont{Nadolsky}} \bibnamefont{and}
  \bibinfo{author}{\bibfnamefont{W.-K.} \bibnamefont{Tung}},
  \bibinfo{journal}{Phys.Rev.} \textbf{\bibinfo{volume}{D79}},
  \bibinfo{pages}{113014} (\bibinfo{year}{2009}), \eprint{0903.2667}.

\bibitem[{\citenamefont{Olness et~al.}(1999)\citenamefont{Olness, Scalise, and
  Tung}}]{Olness:1997yc}
\bibinfo{author}{\bibfnamefont{F.~I.} \bibnamefont{Olness}},
  \bibinfo{author}{\bibfnamefont{R.}~\bibnamefont{Scalise}}, \bibnamefont{and}
  \bibinfo{author}{\bibfnamefont{W.-K.} \bibnamefont{Tung}},
  \bibinfo{journal}{Phys.Rev.} \textbf{\bibinfo{volume}{D59}},
  \bibinfo{pages}{014506} (\bibinfo{year}{1999}), \eprint{hep-ph/9712494}.

\bibitem[{\citenamefont{Kniehl et~al.}(2005{\natexlab{a}})\citenamefont{Kniehl,
  Kramer, Schienbein, and Spiesberger}}]{Kniehl:2004fy}
\bibinfo{author}{\bibfnamefont{B.}~\bibnamefont{Kniehl}},
  \bibinfo{author}{\bibfnamefont{G.}~\bibnamefont{Kramer}},
  \bibinfo{author}{\bibfnamefont{I.}~\bibnamefont{Schienbein}},
  \bibnamefont{and}
  \bibinfo{author}{\bibfnamefont{H.}~\bibnamefont{Spiesberger}},
  \bibinfo{journal}{Phys.Rev.} \textbf{\bibinfo{volume}{D71}},
  \bibinfo{pages}{014018} (\bibinfo{year}{2005}{\natexlab{a}}),
  \eprint{hep-ph/0410289}.

\bibitem[{\citenamefont{Kniehl et~al.}(2005{\natexlab{b}})\citenamefont{Kniehl,
  Kramer, Schienbein, and Spiesberger}}]{Kniehl:2005mk}
\bibinfo{author}{\bibfnamefont{B.}~\bibnamefont{Kniehl}},
  \bibinfo{author}{\bibfnamefont{G.}~\bibnamefont{Kramer}},
  \bibinfo{author}{\bibfnamefont{I.}~\bibnamefont{Schienbein}},
  \bibnamefont{and}
  \bibinfo{author}{\bibfnamefont{H.}~\bibnamefont{Spiesberger}},
  \bibinfo{journal}{Eur.Phys.J.} \textbf{\bibinfo{volume}{C41}},
  \bibinfo{pages}{199} (\bibinfo{year}{2005}{\natexlab{b}}),
  \eprint{hep-ph/0502194}.

\bibitem[{\citenamefont{Kniehl et~al.}(2008)\citenamefont{Kniehl, Kramer,
  Schienbein, and Spiesberger}}]{Kniehl:2008zza}
\bibinfo{author}{\bibfnamefont{B.~A.} \bibnamefont{Kniehl}},
  \bibinfo{author}{\bibfnamefont{G.}~\bibnamefont{Kramer}},
  \bibinfo{author}{\bibfnamefont{I.}~\bibnamefont{Schienbein}},
  \bibnamefont{and}
  \bibinfo{author}{\bibfnamefont{H.}~\bibnamefont{Spiesberger}},
  \bibinfo{journal}{Phys.Rev.} \textbf{\bibinfo{volume}{D77}},
  \bibinfo{pages}{014011} (\bibinfo{year}{2008}), \eprint{0705.4392}.

\bibitem[{\citenamefont{Lai et~al.}(2000)}]{Lai:1999wy}
\bibinfo{author}{\bibfnamefont{H.}~\bibnamefont{Lai}} \bibnamefont{et~al.}
  (\bibinfo{collaboration}{CTEQ Collaboration}), \bibinfo{journal}{Eur.Phys.J.}
  \textbf{\bibinfo{volume}{C12}}, \bibinfo{pages}{375} (\bibinfo{year}{2000}),
  \eprint{hep-ph/9903282}.

\bibitem[{\citenamefont{Martin et~al.}(2006)\citenamefont{Martin, Stirling, and
  Thorne}}]{Martin:2006qz}
\bibinfo{author}{\bibfnamefont{A.}~\bibnamefont{Martin}},
  \bibinfo{author}{\bibfnamefont{W.}~\bibnamefont{Stirling}}, \bibnamefont{and}
  \bibinfo{author}{\bibfnamefont{R.}~\bibnamefont{Thorne}},
  \bibinfo{journal}{Phys.Lett.} \textbf{\bibinfo{volume}{B636}},
  \bibinfo{pages}{259} (\bibinfo{year}{2006}), \eprint{hep-ph/0603143}.

\bibitem[{\citenamefont{Martin et~al.}(2010)\citenamefont{Martin, Stirling,
  Thorne, and Watt}}]{Martin:2010db}
\bibinfo{author}{\bibfnamefont{A.}~\bibnamefont{Martin}},
  \bibinfo{author}{\bibfnamefont{W.}~\bibnamefont{Stirling}},
  \bibinfo{author}{\bibfnamefont{R.}~\bibnamefont{Thorne}}, \bibnamefont{and}
  \bibinfo{author}{\bibfnamefont{G.}~\bibnamefont{Watt}},
  \bibinfo{journal}{Eur.Phys.J.} \textbf{\bibinfo{volume}{C70}},
  \bibinfo{pages}{51} (\bibinfo{year}{2010}), \eprint{1007.2624}.

\bibitem[{\citenamefont{Beringer et~al.}(2012)}]{Beringer:1900zz}
\bibinfo{author}{\bibfnamefont{J.}~\bibnamefont{Beringer}} \bibnamefont{et~al.}
  (\bibinfo{collaboration}{Particle Data Group}), \bibinfo{journal}{Phys.Rev.}
  \textbf{\bibinfo{volume}{D86}}, \bibinfo{pages}{010001}
  (\bibinfo{year}{2012}).

\bibitem[{\citenamefont{Aktas et~al.}(2006)}]{Aktas:2005iw}
\bibinfo{author}{\bibfnamefont{A.}~\bibnamefont{Aktas}} \bibnamefont{et~al.}
  (\bibinfo{collaboration}{H1 Collaboration}), \bibinfo{journal}{Eur.Phys.J.}
  \textbf{\bibinfo{volume}{C45}}, \bibinfo{pages}{23} (\bibinfo{year}{2006}),
  \eprint{hep-ex/0507081}.

\bibitem[{\citenamefont{Botje}(2011)}]{Botje:2010ay}
\bibinfo{author}{\bibfnamefont{M.}~\bibnamefont{Botje}},
  \bibinfo{journal}{Comput. Phys. Commun.} \textbf{\bibinfo{volume}{182}},
  \bibinfo{pages}{490} (\bibinfo{year}{2011}), \eprint{1005.1481}.

\bibitem[{\citenamefont{Buza et~al.}(1998)\citenamefont{Buza, Matiounine,
  Smith, and van Neerven}}]{Buza:1996wv}
\bibinfo{author}{\bibfnamefont{M.}~\bibnamefont{Buza}},
  \bibinfo{author}{\bibfnamefont{Y.}~\bibnamefont{Matiounine}},
  \bibinfo{author}{\bibfnamefont{J.}~\bibnamefont{Smith}}, \bibnamefont{and}
  \bibinfo{author}{\bibfnamefont{W.}~\bibnamefont{van Neerven}},
  \bibinfo{journal}{Eur.Phys.J.} \textbf{\bibinfo{volume}{C1}},
  \bibinfo{pages}{301} (\bibinfo{year}{1998}), \eprint{hep-ph/9612398}.

\bibitem[{\citenamefont{Thorne}(2012)}]{Thorne:2012az}
\bibinfo{author}{\bibfnamefont{R.}~\bibnamefont{Thorne}},
  \bibinfo{journal}{Phys.Rev.} \textbf{\bibinfo{volume}{D86}},
  \bibinfo{pages}{074017} (\bibinfo{year}{2012}), \eprint{1201.6180}.

\bibitem[{\citenamefont{Ball et~al.}(2013{\natexlab{b}})}]{Ball:2013gsa}
\bibinfo{author}{\bibfnamefont{R.~D.} \bibnamefont{Ball}} \bibnamefont{et~al.}
  (\bibinfo{collaboration}{The NNPDF Collaboration}),
  \bibinfo{journal}{Phys.Lett.} \textbf{\bibinfo{volume}{B723}},
  \bibinfo{pages}{330} (\bibinfo{year}{2013}{\natexlab{b}}),
  \eprint{1303.1189}.

\bibitem[{\citenamefont{Kovarik et~al.}(2011)\citenamefont{Kovarik, Schienbein,
  Olness, Yu, Keppel et~al.}}]{Kovarik:2010uv}
\bibinfo{author}{\bibfnamefont{K.}~\bibnamefont{Kovarik}},
  \bibinfo{author}{\bibfnamefont{I.}~\bibnamefont{Schienbein}},
  \bibinfo{author}{\bibfnamefont{F.}~\bibnamefont{Olness}},
  \bibinfo{author}{\bibfnamefont{J.}~\bibnamefont{Yu}},
  \bibinfo{author}{\bibfnamefont{C.}~\bibnamefont{Keppel}},
  \bibnamefont{et~al.}, \bibinfo{journal}{Phys.Rev.Lett.}
  \textbf{\bibinfo{volume}{106}}, \bibinfo{pages}{122301}
  (\bibinfo{year}{2011}), \eprint{1012.0286}.

\bibitem[{\citenamefont{Aaron et~al.}(2011)}]{Aaron:2011gp}
\bibinfo{author}{\bibfnamefont{F.}~\bibnamefont{Aaron}} \bibnamefont{et~al.}
  (\bibinfo{collaboration}{H1 Collaboration}), \bibinfo{journal}{Eur.Phys.J.}
  \textbf{\bibinfo{volume}{C71}}, \bibinfo{pages}{1769} (\bibinfo{year}{2011}),
  \eprint{1106.1028}.

\bibitem[{\citenamefont{{ATLAS~Collaboration}}(2013)}]{ATLAS-CONF-2013-017}
\bibinfo{author}{\bibnamefont{{ATLAS~Collaboration}}},
  \bibinfo{journal}{ATLAS-CONF-2013-017}  (\bibinfo{year}{2013}).

\bibitem[{\citenamefont{Chatrchyan et~al.}(2013)}]{Chatrchyan:2013qha}
\bibinfo{author}{\bibfnamefont{S.}~\bibnamefont{Chatrchyan}}
  \bibnamefont{et~al.} (\bibinfo{collaboration}{CMS Collaboration})
  (\bibinfo{year}{2013}), \eprint{1302.4794}.

\bibitem[{\citenamefont{Harris and Smith}(1998)}]{Harris:1997zq}
\bibinfo{author}{\bibfnamefont{B.}~\bibnamefont{Harris}} \bibnamefont{and}
  \bibinfo{author}{\bibfnamefont{J.}~\bibnamefont{Smith}},
  \bibinfo{journal}{Phys.Rev.} \textbf{\bibinfo{volume}{D57}},
  \bibinfo{pages}{2806} (\bibinfo{year}{1998}), \eprint{hep-ph/9706334}.

\bibitem[{\citenamefont{Harris and Smith}(1995)}]{Harris:1995tu}
\bibinfo{author}{\bibfnamefont{B.}~\bibnamefont{Harris}} \bibnamefont{and}
  \bibinfo{author}{\bibfnamefont{J.}~\bibnamefont{Smith}},
  \bibinfo{journal}{Nucl.Phys.} \textbf{\bibinfo{volume}{B452}},
  \bibinfo{pages}{109} (\bibinfo{year}{1995}), \eprint{hep-ph/9503484}.

\bibitem[{\citenamefont{Gluck and Reya}(2007)}]{Gluck:2006ju}
\bibinfo{author}{\bibfnamefont{M.}~\bibnamefont{Gluck}} \bibnamefont{and}
  \bibinfo{author}{\bibfnamefont{E.}~\bibnamefont{Reya}},
  \bibinfo{journal}{Mod.Phys.Lett.} \textbf{\bibinfo{volume}{A22}},
  \bibinfo{pages}{351} (\bibinfo{year}{2007}), \eprint{hep-ph/0608276}.

\bibitem[{\citenamefont{Berger et~al.}(2010)\citenamefont{Berger, Guzzi, Lai,
  Nadolsky, and Olness}}]{Berger:2010rj}
\bibinfo{author}{\bibfnamefont{E.~L.} \bibnamefont{Berger}},
  \bibinfo{author}{\bibfnamefont{M.}~\bibnamefont{Guzzi}},
  \bibinfo{author}{\bibfnamefont{H.-L.} \bibnamefont{Lai}},
  \bibinfo{author}{\bibfnamefont{P.~M.} \bibnamefont{Nadolsky}},
  \bibnamefont{and} \bibinfo{author}{\bibfnamefont{F.~I.}
  \bibnamefont{Olness}}, \bibinfo{journal}{Phys.Rev.}
  \textbf{\bibinfo{volume}{D82}}, \bibinfo{pages}{114023}
  (\bibinfo{year}{2010}), \eprint{1010.4315}.

\bibitem[{\citenamefont{Buza et~al.}(1996)\citenamefont{Buza, Matiounine,
  Smith, Migneron, and van Neerven}}]{Buza:1995ie}
\bibinfo{author}{\bibfnamefont{M.}~\bibnamefont{Buza}},
  \bibinfo{author}{\bibfnamefont{Y.}~\bibnamefont{Matiounine}},
  \bibinfo{author}{\bibfnamefont{J.}~\bibnamefont{Smith}},
  \bibinfo{author}{\bibfnamefont{R.}~\bibnamefont{Migneron}}, \bibnamefont{and}
  \bibinfo{author}{\bibfnamefont{W.}~\bibnamefont{van Neerven}},
  \bibinfo{journal}{Nucl.Phys.} \textbf{\bibinfo{volume}{B472}},
  \bibinfo{pages}{611} (\bibinfo{year}{1996}), \eprint{hep-ph/9601302}.

\end{thebibliography}

\end{document}